\documentclass[12pt]{report}
\usepackage[letterpaper,top=2.5cm,bottom=2.5cm,hmargin=2.5cm]{geometry}
\usepackage[T1]{fontenc}
\usepackage[utf8]{inputenc}
\usepackage{hyperref}
\usepackage[super,sort&compress]{natbib}
\usepackage{rotating}
\usepackage{graphicx}
\graphicspath{{figures/}}
\usepackage{paralist}
\usepackage{fancyhdr}
\usepackage{multirow}
\usepackage{collcell}
\usepackage{siunitx}
\usepackage{booktabs}
\usepackage[dvipsnames, table]{xcolor}
\usepackage{enumitem}
\usepackage{url}
\usepackage{verbatim}
\usepackage{amsmath}
\usepackage{amssymb}
\usepackage{float}

\begin{document}

\title{Considerations for an Integrated Detector Design at FCC-ee:
       A Human-AI Exploration}

\author{Charles Young \\[0.5em]
\small SLAC National Accelerator Laboratory \\[1em]
}

\date{\today}
\maketitle

\begin{abstract}
This report explores detector design considerations for the Future
Circular Collider in its electron-positron mode (FCC-ee) through an
extended dialogue between a physicist and an AI assistant. Starting
from initial ``prejudice'' detector concepts proposed by the AI assistant without explicit physicist input, each subsystem is
examined in detail, with the AI's assumptions challenged and revised
through the exchange. The discussion covers the full detector from
beam pipe to luminosity monitor, with particular attention to
the interplay between subsystem choices and the practical
considerations --- calibration, stability, and operational
simplicity --- that are essential for a fifteen-year precision
physics program. The narrative documents how the integrated
detector design evolved substantially from the starting point to revised ``prejudice'' detector concepts of the AI assistant. The focus of this report is on the process to 
illustrate both the potential and the limitations of human-AI
collaboration in experimental physics design, and the 
physics capabilities of any of the ``prejudice'' detector concepts
reamin to be explored.

\end{abstract}

\tableofcontents

% ======================================================================
% ======================================================================
\chapter{Introduction}
\label{ch:introduction}
% ======================================================================

\section{The FCC-ee Program}
\label{sec:fcc_program}

The Future Circular Collider in its electron-positron mode
(FCC-ee)~\cite{fcc_cdr, fcc_snowmass, fcc_epjst} is a proposed
circular $e^+e^-$ collider to be housed in a new $\sim$91~km tunnel
in the Geneva region at CERN. It is designed to operate at four
principal center-of-mass energy points, each targeting a distinct
set of precision measurements:

\begin{table}[htbp]
\centering
\caption{The four principal operating energy points of FCC-ee and their
primary physics goals. Approximate integrated luminosities and event
yields correspond to the baseline run plan as described in the
FCC-ee Conceptual Design Report~\cite{fcc_cdr}.}
\label{tab:energy_points}
\begin{tabular}{llll}
\toprule
$\sqrt{s}$ (GeV) & Program & Key measurements & Approximate yield \\
\midrule
91   & Z pole        & Electroweak precision,       & $10^{12}$ Z bosons \\
     &               & flavor physics, $R_b$,       & \\
     &               & $\alpha_s$, $N_\nu$          & \\
\addlinespace
160  & WW threshold  & $m_W$ from threshold scan,   & $10^{8}$ W pairs \\
     &               & W branching ratios           & \\
\addlinespace
240  & ZH production & Higgs couplings,             & $10^{6}$ ZH events \\
     &               & $\sigma_{\text{ZH}}$,        & \\
     &               & Higgs recoil mass            & \\
\addlinespace
350  & $t\bar{t}$    & Top mass, EW couplings,      & $10^{6}$ $t\bar{t}$ \\
     & threshold     & top Yukawa                   & \\
\bottomrule
\end{tabular}
\end{table}

The breadth of this program is both its greatest strength and its
most significant challenge for detector design. At the Z~pole, the
experiment is overwhelmingly statistics-rich --- with $10^{12}$
Z~bosons, many measurements of electroweak observables and flavor physics
processes will be limited by systematic uncertainties rather than
statistics, placing extraordinary demands on detector stability,
calibration, and acceptance control. At the ZH threshold, the Higgs
recoil mass measurement requires excellent tracking momentum
resolution. At the $t\bar{t}$ threshold, the reconstruction of
six-jet final states demands good jet energy resolution at the
highest particle energies accessible at FCC-ee. Energy scans 
to determine masses and line widths require unprecedented precision in
\textit{relative} luminosity measurements. No single
measurement drives the detector design; rather, the detector must
deliver competitive performance across this full spectrum.

FCC-ee benefits from a comparatively benign radiation
environment relative to hadron colliders. This has profound
implications for detector technology choices: it liberates the
designer from the radiation hardness requirements that dominate
technology selection at the LHC, and opens the door to technologies
that prioritize \emph{precision over survivability}. As will be seen
throughout this report, this observation recurs in nearly every
subsystem discussion.

\section{Physics Goals and Detector Requirements}
\label{sec:physics_goals}

The mapping from physics measurements to detector requirements is
not one-to-one. Different energy points stress different aspects of
the detector, and in some cases the requirements are in tension.
Table~\ref{tab:requirements} summarizes the principal connections.

\begin{table}[htbp]
\centering
\caption{Key physics measurements at each energy point and the
detector capabilities they primarily require.}
\label{tab:requirements}
\begin{tabular}{lll}
\toprule
Energy point & Key measurement & Driving detector requirement \\
\midrule
Z pole & EW precision ($m_W$, $\sin^2\theta_W$)     & Absolute luminosity ($10^{-4}$)\\
       &                           & Relaitve luminosity ($10^{-5}$)\\
       &          & acceptance control \\
\addlinespace
       & $R_b$, $R_c$             & $b$/$c$ tagging, vertexing \\
\addlinespace
       & Flavor physics            & Vertex resolution, PID \\
\addlinespace
       & $\alpha_s$               & Jet energy resolution \\
\midrule
WW threshold & $m_W$ from threshold & Tracking momentum resolution, \\
             & scan                 & lepton identification \\
\midrule
ZH & Higgs recoil mass         & momentum resolution $\sigma(p_T)/p_T$ for \\
   &                           & $Z\to\mu\mu$ \\
\addlinespace
   & Higgs couplings           & Jet energy resolution, \\
   & (H$\to b\bar{b}$,        & $b$/$c$ tagging, \\
   & $c\bar{c}$, $\gamma \gamma$)        & photon resolution \\
\midrule
$t\bar{t}$ & Top mass, couplings & Jet energy resolution \\
            &                     & at highest energies, \\
            &                     & $b$-tagging, forward \\
            &                     & coverage \\
\bottomrule
\end{tabular}
\end{table}

Several cross-cutting themes emerge from this mapping:

\begin{itemize}[itemsep=2pt]
\item \textbf{Vertexing and heavy-flavor tagging} are essential
  across all energy points, from $R_b$ at the Z~pole to Higgs
  coupling measurements at 240~GeV and top physics at 350~GeV. This
  places a premium on minimizing the material budget of the vertex
  detector and beam pipe.

\item \textbf{Tracking momentum resolution} is critical for the
  Higgs recoil measurement ($Z\to\mu^+\mu^-$ recoil mass), which
  requires $\sigma(p_T)/p_T^2 \lesssim 2\times 10^{-5}$~GeV$^{-1}$.
  This demands a large tracking volume, strong magnetic field,
  many precise measurement points, and low material budget
  so measurement performance is not compromised by multiple scattering.

\item \textbf{Jet energy resolution} matters at every energy point
  but becomes most demanding at the $t\bar{t}$ threshold where jet
  energies reach 100--175~GeV and the particle flow approach becomes
  susceptible to shower overlap.

\item \textbf{Systematic control and long-term stability} dominate
  at the Z~pole. With $10^{12}$~Z~bosons, statistical uncertainties
  on most observables are negligible; the experiment is limited by
  how well systematics --- calibration, acceptance, luminosity ---
  are controlled. Technologies that offer intrinsic uniformity and
  stability deserve particular attention.

\item \textbf{Particle identification} ($\pi/K/p$ separation)
  enhances the flavor physics program at the Z~pole and improves
  jet flavor tagging at all energies. This capability is not
  always given high priority in detector concept studies but can
  significantly extend the physics reach.

\item \textbf{Luminosity measurement} at the $10^{-4}$ level
  (absolute) and $10^{-5}$ level (relative) is essential for the
  Z~lineshape scan and threshold measurements. This is a
  metrology challenge as much as a detector challenge.
\end{itemize}

\section{Existing Detector Concepts}
\label{sec:existing_concepts}

Several detector concepts have been put forward for FCC-ee. Four
\footnote{This report is based on the AI's understanding 
of them at the time of the investigation and does not necessarily 
represent their current status.}
have been formally submitted as Expressions of
Interest~\cite{fcc_detector_eoi}:

\textbf{IDEA} (Innovative Detector for an Electron-positron
Accelerator)~\cite{idea_ref} features a large low-mass drift chamber as its
central tracker, providing both momentum measurement and particle
identification through cluster counting ($dN/dx$), complemented by 
silicon vertex detector and outer tracking layer. It proposes a
dual-readout fiber calorimeter for combined electromagnetic and
hadronic energy measurement, exploiting the ratio of scintillation
to \v{C}erenkov light to correct for event-by-event fluctuations in
the electromagnetic fraction of hadronic showers. A thin, low-mass
superconducting solenoid and a preshower detector complement the
calorimeter system.

\textbf{CLD} (CLic Detector adapted for
FCC-ee)~\cite{cld_ref} is based on detector concepts developed for
the Compact Linear Collider (CLIC) and adapted to FCC-ee
conditions. It features an all-silicon tracker, a silicon-tungsten
electromagnetic calorimeter with high transverse and longitudinal
granularity optimized for particle flow, and a scintillator-steel
hadronic calorimeter with similar granularity. The overall design
philosophy is driven by the particle flow paradigm, in which
individual particles in jets are reconstructed using the combination
of tracking and highly segmented calorimetry.

\textbf{ALLEGRO} (A Lepton coLlidEr detector with Granular
Read-Out)~\cite{allegro_ref} explores a noble liquid electromagnetic
calorimeter (liquid argon or liquid krypton with lead or tungsten
absorbers), combined with fine-granularity readout electrodes for
particle flow capability. It shares many elements with CLD for the
tracking and hadronic calorimeter systems, with the electromagnetic
calorimeter being the primary distinguishing feature.

\textbf{ILD-FCCee}~\cite{ild_fccee_ref} is an adaptation of the
International Large Detector (ILD) concept originally developed for
the International Linear Collider (ILC)~\cite{ild_ilc_ref}. Its
distinguishing feature is the use of a Time Projection Chamber (TPC)
as the main tracker, providing three-dimensional space points
with many measurements per track and only gaseous material in the
tracking volume. The TPC is complemented by silicon inner and outer
tracking layers. The calorimeter system follows the particle flow
approach with a silicon-tungsten electromagnetic calorimeter and a
scintillator-steel hadronic calorimeter.

These four concepts share many common elements --- notably MAPS-based
vertex detectors and a superconducting solenoid --- while differing
in their choices for the central tracker, electromagnetic
calorimeter, and hadronic calorimeter. They are explicitly designated
as ``concepts'' rather than final designs, and it is expected that
the eventual FCC-ee detector(s) will draw elements from multiple
concepts.

It is worth noting that similar detector studies are underway for
the Circular Electron-Positron Collider
(CEPC)~\cite{cepc_cdr, cepc_detector_ref} proposed in China. CEPC
and FCC-ee share remarkably similar accelerator designs, performance
parameters, and physics goals --- both are circular $e^+e^-$ colliders
designed to operate as Z, W, Higgs, and top factories at comparable
luminosities. The CEPC reference detector concept features a TPC
main tracker, a silicon-tungsten electromagnetic calorimeter based
on the particle flow approach, and a scintillator-steel hadronic
calorimeter, with extensive studies of TPC performance, ion back-flow
management, and particle flow optimization at Higgs factory
conditions~\cite{cepc_tpc_ref}. Given the close correspondence
between the two machines, the CEPC detector studies are directly
relevant to FCC-ee and vice versa. The CEPC reference detector
can be considered, in all but name, a fifth concept for an $e^+e^-$
Higgs factory detector.

This report takes the approach of examining each subsystem
independently rather than adopting any single concept, constructing
an integrated design based on the merits of the available technology
options. Where relevant, the strengths and limitations of choices
made by the existing concepts are discussed.

\section{Approach: Human-AI Collaborative Design Exploration}
\label{sec:approach}

This report documents an extended exploration of FCC-ee detector
design conducted through a dialogue between a physicist with
experience in collider detector design and an AI assistant
(Claude, developed by Anthropic). The methodology was as follows:

\begin{enumerate}[itemsep=2pt]
\item The AI was asked to propose initial detector concepts for
  FCC-ee, including a rationale for each subsystem choice and
  criteria for the integrated design.

\item Each subsystem was then examined in detail through iterative
  dialogue. The physicist posed questions, challenged assumptions,
  and introduced practical considerations based on experience. The
  AI responded with analysis, revised its positions where the
  arguments warranted, and maintained its positions where it judged
  the original reasoning sound.

\item Through this process, the integrated detector concept evolved
  substantially from its starting point, with changes in several
  major subsystem choices.
\end{enumerate}

The intent is twofold. First, the resulting detector concepts and the
technical analysis underlying them are of interest in their own right,
contributing to the ongoing discussion within the FCC-ee detector
community. However, the level of study and scrutiny is so far minimal, and these concepts should not be compared at face value with the existing FCC-ee detector concepts.  
Second, the process itself --- the strengths and
limitations of human-AI collaboration for complex physics detector design ---
is worth documenting as the field begins to explore such tools.

The report is structured to reflect this process. Chapter~\ref{ch:initial_concepts} presents the initial detector concepts
as proposed by the AI, including the criteria and reasoning behind
each choice. Chapter~\ref{ch:evolution} walks through the subsystem-by-subsystem examination, documenting how and why the design
evolved. Chapter~\ref{ch:revised} presents the final integrated
detector concept as proposed by the AI. Chapter~\ref{ch:reflections} reflects on the
collaborative process itself, and Chapter~\ref{ch:summary}
summarizes the conclusions.

It should be noted that the quantitative estimates throughout this
report are intended to be illustrative rather than definitive. They
are based on analytical calculations, scaling arguments, and
published performance data, not on dedicated Monte Carlo simulations
of the specific detector geometry proposed here. A number of
the conclusions reached in this study would benefit from validation
through full simulation, and these are identified as R\&D priorities
where appropriate.

The AI assistant used in this study was Claude, developed by
Anthropic, accessed through the Anthropic web interface (model
identifier \texttt{claude-opus-4}). A large part of the dialogue was
conducted in a single extended conversation, allowing the AI to
maintain context across all subsystem discussions. The dialogue
took place in April 2026.

The interaction was conducted in a direct dialogue (chatbot) mode
in which the AI operated solely from its training data and from
information explicitly provided by the physicist in the text of
the conversation or through file uploads. Notably, when the
physicist provided URLs to external resources, the AI could not
access or read those resources --- only information directly
pasted or uploaded into the conversation was available. The AI
did not have access to web search, database queries, or any
external information retrieval during the dialogue. This is a
significant limitation: for a rapidly evolving field, the AI's
training data inevitably misses recent developments, and the
burden of supplying current information falls entirely on the
human contributor. More capable configurations exist in which
the AI can retrieve and read documents from the web in real
time; such tool-augmented modes would likely improve the
efficiency and completeness of a study of this kind.

% ======================================================================
\chapter{Initial Detector Concepts}
\label{ch:initial_concepts}
% ======================================================================

\noindent\textit{This chapter presents the initial detector concepts
as proposed by the AI at the outset of the study, before the
detailed subsystem examination described in
Chapter~\ref{ch:evolution}. The criteria, reasoning, and specific
technology choices reflect the AI's starting positions, with no guidance from the physicist. Several of
these positions were subsequently revised through the collaborative
discussion, leading to the updated concepts presented
in Chapter~\ref{ch:revised}.}

\vspace{0.5em}
\noindent\textbf{Naming convention.} The detector concepts developed
in this report are labeled as follows:
\begin{itemize}[itemsep=2pt]
\item \textbf{CL1a} --- the initial primary detector concept
  (this chapter);
\item \textbf{CL1b} --- the initial complementary detector concept
  (this chapter);
\item \textbf{CL2a} --- the revised primary detector concept, after
  the subsystem examination of Chapter~\ref{ch:evolution} (presented
  in Chapter~\ref{ch:revised});
\item \textbf{CL2b} --- the revised complementary detector concept
  (presented in Chapter~\ref{ch:revised}).
\end{itemize}

\section{Criteria for an Integrated Detector}
\label{sec:criteria}

Before specifying technology choices, it is useful to articulate the
criteria that should guide the overall detector design. The following
were identified as the principal considerations, roughly in order of
priority.

\subsection{Physics Performance Across All Energy Points}

The FCC-ee program spans a factor of nearly four in center-of-mass
energy, from 91 to 350~GeV, as well as significant differences in event rate. The detector cannot be optimized for a
single energy point. As discussed in
Section~\ref{sec:physics_goals}, the Z~pole program demands
vertexing, particle identification, high beam crossing rate, and systematic control; the ZH
program demands tracking momentum resolution and jet energy
resolution; and the $t\bar{t}$ program demands jet resolution at the
highest energies. A detector optimized purely for the Higgs recoil
measurement at 240~GeV might not be optimal for the Z~pole flavor
physics program, and vice versa. The integrated design must make
thoughtful compromises.

\subsection{Systematic Robustness and Long-Term Stability}

For a facility expected to operate for fifteen or more years and to
produce measurements at the permille level or below, systematic
control is paramount. This criterion favors technologies that offer:
\begin{itemize}[itemsep=2pt]
\item \textbf{Intrinsic uniformity} --- detector response that is
  uniform by construction, rather than requiring extensive
  calibration to achieve uniformity.
\item \textbf{Stability over time} --- technologies that do not
  degrade, or degrade predictably, over years of operation.
\item \textbf{Redundancy} --- the ability to cross-check critical
  measurements using independent subsystems or methods.
\item \textbf{Insensitivity to environmental conditions} ---
  minimal dependence on temperature, pressure, humidity, and other
  external parameters.
\end{itemize}

\subsection{Particle Flow vs.\ Intrinsic Calorimetric Resolution}

Two fundamentally different philosophies exist for measuring jet
energies at $e^+e^-$ colliders:

\textbf{Particle flow} reconstructs every particle in the event
individually. Charged particles ($\sim$65\% of jet energy) are
measured by the tracker with excellent resolution
($\sigma/E \sim 10^{-4}$--$10^{-3}$). Photons ($\sim$25\% of jet energy) are
measured by the electromagnetic calorimeter. Only neutral hadrons
($\sim$10\% of jet energy) rely on the hadronic calorimeter. This approach
requires high-granularity calorimetry for pattern recognition and
a strong magnetic field to separate charged and neutral particles.

\textbf{Calorimetric approaches}, such as dual-readout, measure jet
energy primarily from the calorimeter with event-by-event corrections
for the different response of the calorimeter to the varying
electromagnetic fraction of hadronic showers. This is less
dependent on tracking perfection and can provide robust jet energy
measurement even when particle flow breaks down.

The optimal strategy may involve elements of both approaches, and the
choice of electromagnetic calorimeter technology has significant
implications for which approach is more effective.

\subsection{Hermeticity and Acceptance}

Missing energy signatures (e.g., $H\to\text{invisible}$,
$Z\to\nu\bar{\nu}$) require near-$4\pi$ coverage with minimal
cracks and dead material. The transition region between barrel and
endcap deserves particular attention, as does the forward region
where beam pipe and machine-detector interface elements create
unavoidable gaps.

\subsection{Technical Maturity and Risk}

Some technologies are well-proven at collider scale while others
remain at the prototype stage. A prudent design balances ambition
with risk. The FCC-ee timeline, with
first collisions anticipated in the mid-2040s, provides a finite
window for R\&D, and technologies requiring extensive
development may not be viable.

\subsection{Cost and Schedule}

Detector cost for a flagship facility is typically in the 1000M~CHF range. 
Technology choices have significant cost implications.
Calorimeters are often the most expensive subsystem due to 
material cost and cryogenic infrastructure in the case of noble liquid calorimetry. 
Magnets are also expensive and their costs are driven by field strength and volume.

\section{The Case for Two Complementary Detectors}
\label{sec:complementarity_case}

The current FCC-ee plan foresees four interaction points and detectors. This presents a strategic opportunity
that goes beyond simple redundancy.

If the two detectors are designed with \emph{maximally different}
technology choices, they provide:

\begin{enumerate}[itemsep=2pt]
\item \textbf{Independent systematic checks.} Every major physics
  measurement is performed twice, with different detector
  technologies and different systematic uncertainties. Agreement
  between the two results provides powerful validation; disagreement
  reveals problems that might otherwise go undetected.

\item \textbf{Complementary performance.} Different technologies
  excel in different regimes. A particle-flow-optimized detector may
  outperform at lower jet energies, while a dual-readout calorimeter
  may be superior at higher energies. Together, the two detectors
  offer excellent
  coverage over the full performance space.

\item \textbf{Risk mitigation.} If a novel technology encounters
  unforeseen limitations in operation, the other detector with its
  different technology provides a fallback for the physics program.

\item \textbf{Healthy competition.} The history of collider physics
  shows that having two experiments analyzing the same data leads to
  more careful work and better results from both.
\end{enumerate}

The decisions for the two detectors should therefore be made jointly,
with the goal of maximizing complementarity rather than duplicating
the same design. In what follows, the AI proposed two initial detector
concepts --- CL1a and CL1b --- designed with this principle in mind.

\section{CL1a: The Initial Primary Detector}
\label{sec:initial_concept_1}

Concept CL1a was constructed by selecting, for each subsystem, the
technology judged by the AI assistant to offer the best overall performance across the
full FCC-ee program, with emphasis on systematic robustness and
long-term stability. The initial choices and their rationale are
summarized below and in Table~\ref{tab:cl1a}.

\begin{table}[htbp]
\centering
\caption{CL1a: initial primary detector concept as proposed at the
start of the study. Entries marked with $\dagger$ were subsequently
revised through the collaborative discussion
(Chapter~\ref{ch:evolution}), leading to the updated concept CL2a
(Chapter~\ref{ch:revised}).}
\label{tab:cl1a}
\begin{tabular}{lll}
\toprule
Subsystem & Technology & Key feature \\
\midrule
Beam pipe          & Beryllium, locally thinned
                   & Minimal material \\
\addlinespace
Vertex detector    & MAPS, 65~nm CMOS,
                   & $\sim$0.02\% $X_0$/layer \\
                   & curved shells & \\
\addlinespace
Inner silicon$^\dagger$ & Additional MAPS layers
                   & Bridge to main tracker \\
\addlinespace
Main tracker       & Drift chamber (He-based gas)
                   & Low material, $dN/dx$ PID \\
\addlinespace
Outer silicon$^\dagger$ & Si strips + LGAD
                   & Momentum + TOF \\
\addlinespace
ECAL$^\dagger$     & Noble liquid (LAr)
                   & Uniformity, stability \\
                   & + Pb/W absorber & \\
\addlinespace
HCAL               & Scintillator-steel tiles
                   & PFA granularity \\
\addlinespace
Solenoid$^\dagger$ & 2~T, outside HCAL
                   & Uninterrupted calorimetry \\
\addlinespace
Muon system$^\dagger$ & $\mu$-RWELL
                   & Position resolution \\
\addlinespace
Luminometer        & Si-W calorimeter
                   & Precision acceptance \\
\bottomrule
\end{tabular}
\end{table}

\subsection{Vertex Detector}

The vertex detector was envisioned as an array of ultra-thin
Monolithic Active Pixel Sensors (MAPS) in 65~nm CMOS technology,
building on the ALICE ITS3 development~\cite{alice_its3}. The key
design choices were:
\begin{itemize}[itemsep=2pt]
\item Silicon thickness of 20--30~$\mu$m per layer, thinned from
  standard wafer thickness.
\item Curved, (partially) self-supporting cylindrical shells, exploiting the
  geometric rigidity of the curved shape to minimize support
  material.
\item Air cooling, enabled by the low power density of 65~nm MAPS
  ($\sim$20--50~mW/cm$^2$), eliminating coolant pipes and associated
  material from the active volume.
\item First layer as close as possible to the interaction point
  ($R \approx 12$--13~mm), with subsequent layers extending to
  $R \approx 30$--35~mm.
\end{itemize}

The target material budget was $\sim$0.02--0.03\% $X_0$ per layer,
or less than 0.1\% $X_0$ for the three innermost layers combined.
At FCC-ee energies, where multiple scattering dominates the impact
parameter resolution for tracks below $\sim$10~GeV, the material
budget of the vertex detector is arguably as important a
design parameter as its hit resolution. 

\subsection{Inner Silicon Tracker}

In CL1a, a set of additional MAPS layers at intermediate radii
($R \approx 80$--300~mm) was proposed as a distinct subsystem
bridging the gap between the vertex detector and the main tracker.
These layers would provide precision space points for pattern
recognition seeding and momentum resolution, using the same MAPS
technology as the vertex detector but with relaxed pixel pitch
($\sim$20--25~$\mu$m). The separation of this into a distinct
subsystem was subsequently reconsidered
(Section~\ref{sec:evo_vertex_silicon}).

\subsection{Main Tracker: Drift Chamber}

A large drift chamber was chosen as the central tracker, similar in
concept to the IDEA proposal. The principal arguments were:
\begin{itemize}[itemsep=2pt]
\item \textbf{Low material budget.} A helium-based gas mixture
  (He:iC$_4$H$_{10}$ 90:10) contributes negligibly to the material
  budget ($\sim$0.03\% $X_0$ radially). The wires add
  $\sim$0.1--0.2\% $X_0$.
\item \textbf{Many measurements per track.} With $\sim$100+ layers,
  each track receives many independent position measurements,
  providing robust pattern recognition and good momentum resolution
  ($\sim$100~$\mu$m per point).
\item \textbf{Particle identification.} Cluster counting ($dN/dx$)
  potentially provides better particle identification than
  traditional $dE/dx$ truncated mean, with a theoretical resolution
  of $\sim$2\% for a 2~m path length.
\item \textbf{Proven technology.} Drift chambers have been
  successfully operated at LEP, SLD, BaBar, Belle, KLOE, and MEG, among others, 
  though cluster counting needs further study. 
\end{itemize}

\subsection{Outer Silicon Wrapper}

A silicon layer at the outer tracker radius ($R \approx 2000$~mm)
was proposed to provide:
\begin{enumerate}[itemsep=2pt]
\item A precision $r$-$\phi$ space point for momentum resolution
  (the outermost measurement has the largest lever arm and therefore
  the greatest impact on curvature determination).
\item Longitudinal position measurement for acceptance control.
\item Time-of-flight measurement for $\pi/K$ separation using
  Low-Gain Avalanche Detectors (LGADs).
\end{enumerate}

In CL1a, this was envisioned as two layers: a fine-pitch strip layer
for position and a separate LGAD layer for timing. This was
subsequently reconsidered in favor of a single multi-function layer
(Section~\ref{sec:evo_wrapper}).

\subsection{Electromagnetic Calorimeter: Noble Liquid}

The initial choice for the electromagnetic calorimeter was a liquid
argon sampling calorimeter with lead or tungsten absorber plates,
offering:
\begin{itemize}[itemsep=2pt]
\item Expected energy resolution of $\sim$7--10\%/$\sqrt{E}$.
\item Intrinsic uniformity --- the liquid medium is homogeneous by
  construction, providing stable and uniform response without
  channel-by-channel calibration.
\item Proven technology at scale (ATLAS LAr calorimeter, NA48 LKr
  calorimeter).
\item Fine granularity with user-defined geometry readily achievable with modern multi-layer PCB
  readout electrodes.
\item Multiple longitudinal samplings for shower profiling.
\end{itemize}

This was one of the subsystem choices in CL1a that underwent the
most significant revision through the subsequent discussion
(Section~\ref{sec:evo_ecal}).

\subsection{Hadronic Calorimeter: Scintillator-Steel Tiles}

A scintillator-steel sampling calorimeter with high granularity was
chosen, optimized for particle flow reconstruction:
\begin{itemize}[itemsep=2pt]
\item Steel absorber plates ($\sim$20~mm) interleaved with
  scintillator tiles ($\sim$3~mm thick, $\sim$30$\times$30~mm$^2$).
\item Each tile read out by an individual SiPM.
\item $\sim$50 layers for $\sim$6 nuclear interaction lengths
  ($\lambda_I$).
\item The design prioritizes spatial granularity for shower imaging
  over intrinsic energy resolution.
\end{itemize}

\subsection{Solenoid}

A 2~T superconducting solenoid was proposed, placed outside the
hadronic calorimeter to avoid introducing dead material between the
electromagnetic and hadronic calorimeters. The field strength of 2~T
was chosen as a compromise between tracking performance (which
benefits from higher field) and accelerator compatibility (the
solenoid field must be compensated in the interaction region optics,
and compensation becomes more difficult at lower beam energies).
Both the field strength and the placement were subsequently revised
(Section~\ref{sec:evo_solenoid}).

\subsection{Muon System}

The muon system was initially proposed as 3--4 stations of
micro-Resistive WELL ($\mu$-RWELL) detectors in the iron return
yoke, providing $\sim$200~$\mu$m position resolution. This choice
was subsequently reconsidered
(Section~\ref{sec:evo_muon}).

\subsection{Luminosity Monitor}

A precision silicon-tungsten sampling calorimeter at small angles
($\theta \approx 60$--100~mrad) was proposed for luminosity
measurement through Bhabha event counting, targeting $10^{-4}$
absolute and $10^{-5}$ relative precision.

\section{CL1b: The Initial Complementary Detector}
\label{sec:initial_concept_2}

Following the principle of maximal complementarity articulated in
Section~\ref{sec:complementarity_case}, concept CL1b was proposed
with deliberately different technology choices for the major
subsystems. The intent was that each major physics measurement would
be performed with two different detector technologies, providing
independent systematic cross-checks.

\begin{table}[htbp]
\centering
\caption{CL1a and CL1b compared: technology choices selected to
maximize complementarity. Shared elements are listed below.}
\label{tab:cl1a_cl1b}
\begin{tabular}{lll}
\toprule
Subsystem & CL1a & CL1b \\
\midrule
Main tracker   & Drift chamber      & Full silicon tracker \\
ECAL           & Noble liquid (LAr)  & Si-W (particle flow) \\
HCAL           & Scint.-steel tiles  & Dual-readout fibers \\
\midrule
\multicolumn{3}{l}{\textit{Shared elements in CL1a and CL1b:}} \\
\addlinespace
Vertex         & \multicolumn{2}{l}{MAPS, 65~nm CMOS, curved shells} \\
Solenoid       & \multicolumn{2}{l}{Superconducting, $\sim$2~T} \\
Muon system    & \multicolumn{2}{l}{$\mu$-RWELL in return yoke} \\
Luminometer    & \multicolumn{2}{l}{Si-W calorimeter} \\
\bottomrule
\end{tabular}
\end{table}

The key complementary choices in CL1b were:

\begin{itemize}[itemsep=2pt]
\item \textbf{Tracker:} A full silicon tracker provides excellent
  spatial resolution at discrete layers but with more material than
  a gaseous tracker. It offers a fundamentally different systematic
  profile --- silicon tracking is limited by alignment and material
  knowledge, while drift chamber tracking is limited by
  calibration of the time-to-distance relationship and wire
  positions.

\item \textbf{Electromagnetic calorimeter:} A silicon-tungsten
  sampling calorimeter with 5$\times$5~mm$^2$ pads and $\sim$30
  longitudinal layers, specifically optimized for the particle flow
  approach. This provides different shower imaging capability
  compared to the noble liquid option in CL1a.

\item \textbf{Hadronic calorimeter:} A dual-readout fiber
  calorimeter, exploiting the scintillation-to-\v{C}erenkov ratio
  for event-by-event correction of electromagnetic fraction
  fluctuations. This attacks the fundamental problem of hadronic
  calorimetry through a completely different approach than the
  particle flow strategy of CL1a.
\end{itemize}

Subsystems where the technology choice is clearer ---
the vertex detector, solenoid, and luminometer --- were kept the
same in both concepts. The vertex detector technology (MAPS) is
clearly optimal for FCC-ee regardless of other choices. The
luminometer design is dictated by the measurement requirements rather
than by detector philosophy.

\vspace{1em}
\noindent\rule{\textwidth}{0.4pt}
\vspace{0.5em}

\noindent Concepts CL1a and CL1b served as the starting point for
the detailed subsystem examination that follows in
Chapter~\ref{ch:evolution}. As will be seen, several of the
technology choices --- including some of the most consequential
ones --- were revised through the subsequent discussion. The
resulting concepts, CL2a and CL2b, are presented in
Chapter~\ref{ch:revised}. The revisions affected CL1a more
significantly than CL1b, reflecting the fact that the detailed
examination focused primarily on the technologies chosen for the
primary detector.

% ======================================================================
\chapter{Subsystem Examination and Evolution}
\label{ch:evolution}
% ======================================================================

% This is the heart of the report: each subsystem examined in detail,
% documenting what changed and why. The narrative follows the
% conversation flow.

\noindent\textit{This chapter documents the detailed examination of
each detector subsystem through iterative dialogue between the
physicist and the AI. For each subsystem, the AI typically presented
an initial technical assessment, the physicist posed questions or
challenges drawing on practical experience and physical insight, and
the AI revised its analysis in response. Physicist questions are 
often open-ended to allow the AI to respond; they are occasionally 
leading when the physicist sees the need. The resulting conclusions
reflect this collaborative process. The written account is primarily
drafted by the AI, but the intellectual content --- particularly the
key questions and insights that redirected the analysis --- is
jointly developed. Where a specific insight originated clearly from
one contributor, this is noted in the text.}

\section{Beam Pipe and Vertex Region}
\label{sec:evo_vertex}

\subsection{Beam Pipe: Material Budget Dominance}
\label{sec:evo_beampipe}

The beam pipe is the first material encountered by particles
emerging from the interaction point, and its contribution to the
material budget sets a floor on the impact parameter resolution that
no amount of vertex detector optimization can overcome. A particle
originating at the interaction point scatters in the beam pipe wall
\emph{before} reaching the first tracking layer, and this scattering
cannot be corrected because there is no measurement between the
interaction point and the beam pipe.

The beam pipe design for FCC-ee is still under active study. Current expectations are for a
liquid-cooled, double-layer structure with a total thickness
corresponding to $\sim$0.6\% $X_0$~\cite{fccee_detector_requirements}.
This is a substantial amount of material --- roughly three times
the total material budget of the entire six-layer vertex detector
described in Section~\ref{sec:evo_vertex_silicon}.

The beam pipe inner radius is typically $\sim$10--12~mm. There is an intimate coupling between
the beam pipe design (a machine element), the vertex detector
design (an experiment element) and shielding (machine element for the benefit of experimental detector); the machine-detector interface
working group treats these as a single optimization problem.
Reducing the beam pipe material even modestly would have a direct and
significant impact on the vertex detector physics performance,
particularly for heavy-flavor tagging at the Z~pole where
multiple scattering dominates the impact parameter resolution
for tracks below $\sim$10~GeV.

\subsection{The Idea of a Sensor Inside the Beam Pipe}
\label{sec:evo_sensor_inside}

Given that the beam pipe dominates the material before the first
measurement, the physicist raised a deliberately provocative
question: could a sensor layer be placed \emph{inside} the beam
pipe, between the interaction point and the beam pipe wall?

% In Section 3.1.2, update the physics motivation:

\subsubsection{Physics motivation}

A MAPS layer inside the beam pipe at
$R \approx 8$--10~mm would measure the track position before
scattering in the beam pipe wall. With measurements on both
sides of the beam pipe, the scattering can be partially measured
and corrected, rather than contributing as an irreducible smearing.
For impact parameter resolution at low momenta (1--5~GeV), where
multiple scattering dominates, this could yield a significant
improvement --- potentially larger than the 20--40\% estimated
for a thinner beam pipe, given that the current design has more
material to mitigate.

\subsubsection{Engineering challenges}

The challenges are formidable:

\begin{itemize}[itemsep=2pt]
\item \textbf{Vacuum compatibility.} The sensor would operate in
  the ultra-high vacuum of the beam pipe ($10^{-9}$--$10^{-10}$~Torr).
  All materials must be UHV-compatible with low outgassing, ruling
  out standard PCB substrates and adhesives. The sensor must survive
  bakeout at 150--200$^\circ$C.

\item \textbf{Beam backgrounds.} The region inside the beam pipe is
  exposed to greater synchrotron radiation from the final-focus quadrupoles,
  beamstrahlung pairs, and off-momentum particles. A sensor at
  $R \approx 8$--10~mm would need to cope with these backgrounds
  without being overwhelmed.

\item \textbf{Beam impedance.} Any structure inside the beam pipe
  affects the beam coupling impedance. With FCC-ee's high stored
  beam current ($\sim$1.4~A at the Z~pole), this is a serious
  concern requiring detailed study.

\item \textbf{Power and data.} Electrical connections must cross the
  beam pipe wall through UHV-compatible feedthroughs, constraining the mechanical design.

\item \textbf{Mechanical constraints.} The sensor area is small
  ($\sim$50~cm$^2$ at $R = 10$~mm), but alignment to $\mu$m
  precision inside a $\sim$20~mm diameter tube is extremely
  challenging.
\end{itemize}

\subsubsection{Assessment}

The sensor-inside-beam-pipe concept is aggressive but not
implausible. The FCC-ee environment --- little radiation damage and moderate
backgrounds compared to hadron colliders, and small sensor area
required --- makes it more feasible here than at any previous or
planned collider. A pragmatic approach would be:

\begin{enumerate}[itemsep=2pt]
\item \textbf{Baseline:} Minimize beam pipe thickness (0.2--0.3~mm
  Be) and place the first MAPS layer at the smallest possible
  radius, essentially touching the outer beam pipe surface. This
  captures most of the benefit without the vacuum complications.

\item \textbf{R\&D path:} Pursue a dedicated program including
  UHV compatibility testing of thinned MAPS, beam background
  simulation, impedance studies, and thermal management in vacuum,
  with a decision point in the early 2030s.

\item \textbf{Upgrade option:} Design the beam pipe and vertex
  detector to be mechanically compatible with a future inner sensor
  installation, so that if the R\&D succeeds, the sensor can be
  added without redesigning the surrounding systems.
\end{enumerate}

\subsection{Unified Vertex and Inner Silicon Tracker}
\label{sec:evo_vertex_silicon}

\subsubsection{The initial separation}

In CL1a, the vertex detector (three layers at $R \approx 12$--35~mm)
and the inner silicon tracker (layers at $R \approx 80$--300~mm) were
treated by the AI as separate subsystems. This separation was inherited from
LHC detector designs, where the two subsystems genuinely employ
different technologies: hybrid pixels at small radii (for radiation
hardness and rate capability) and silicon strips at larger radii
(for cost-effective coverage of larger areas).

\subsubsection{The challenge}

The physicist questioned this separation: at FCC-ee, what is the
physical basis for distinguishing the vertex detector from the inner
silicon tracker?

The question is penetrating because the conditions that motivate
the separation at the LHC do not exist at FCC-ee:
\begin{itemize}[itemsep=2pt]
\item The radiation environment is benign at all radii --- there is
  no threshold radius beyond which a less radiation-hard technology
  becomes viable.
\item The same MAPS technology (65~nm CMOS) works at all radii from
  12~mm to 300~mm.
\item The power density decreases with radius as hit rates drop off, so air cooling becomes easier, not harder, at larger
  radii.
\item The design philosophy of curved (partially) self-supporting shells with
  minimal material applies equally at all radii, though the
  engineering details differ with scale.
\end{itemize}

\subsubsection{Resolution: one unified subsystem}

The vertex detector and inner silicon tracker in CL1a are unified
into a single subsystem: a set of MAPS layers extending continuously
from the beam pipe to the inner wall of the main tracker, using the
same base technology throughout. The layer configuration is:

\begin{table}[H]
\centering
\caption{Unified vertex and inner silicon tracker: layer
configuration. All layers use 65~nm CMOS MAPS on curved
self-supporting shells.}
\label{tab:unified_silicon}
\begin{tabular}{ccccc}
\toprule
Layer & Radius (mm) & Pixel pitch ($\mu$m) & Thickness ($\mu$m)
      & $x/X_0$ (\%) \\
\midrule
1 & $\sim$12  & 10--15 & $\sim$20 & $\sim$0.02 \\
2 & $\sim$20  & 10--15 & $\sim$20 & $\sim$0.02 \\
3 & $\sim$35  & 15     & $\sim$25 & $\sim$0.03 \\
4 & $\sim$90  & 20     & $\sim$25 & $\sim$0.03 \\
5 & $\sim$160 & 20--25 & $\sim$30 & $\sim$0.03 \\
6 & $\sim$300 & 25     & $\sim$30 & $\sim$0.03 \\
\bottomrule
\end{tabular}
\end{table}

The only variation across layers is in pixel pitch (which can be
relaxed at larger radii where hit densities are lower and the
multiple scattering contribution from inner layers already limits
extrapolation precision) and silicon thickness (slightly thicker
at larger radii, where the larger physical size of the shells
makes handling and assembly more demanding). The specific
thicknesses shown in Table~\ref{tab:unified_silicon} are
illustrative; the actual values will be determined by mechanical
prototyping and engineering studies.

The key design features that enable this ultra-low material budget
are:

\begin{itemize}[itemsep=2pt]
\item \textbf{Curved (partially) self-supporting shells.} A flat thin membrane
  is mechanically floppy, but a curved cylindrical shell of the 
  same thickness has
  significant rigidity against transverse loads. This geometric
  stiffness, demonstrated by the ALICE ITS3
  program~\cite{alice_its3}, potentially eliminates the need for
  carbon fiber staves or ladder structures in the active volume.

\item \textbf{Air cooling.} At power densities of
  $\sim$20--50~mW/cm$^2$ (achievable in 65~nm CMOS with rolling
  shutter readout), forced air flow through the gaps between
  cylindrical shells can remove the dissipated heat. A rough thermal
  estimate for the innermost layer ($R = 15$~mm, power density
  50~mW/cm$^2$) gives a temperature rise of $\sim$17$^\circ$C with
  modest air flow (1--2~m/s), which is acceptable in the
  radiation-benign FCC-ee environment where there is no need to
  operate at low temperature for leakage current control.

  The viability of air cooling eliminates coolant pipes and their
  associated material from the active volume. This single design
  choice removes what is typically a major material contribution
  in vertex detectors at hadron colliders.

\item \textbf{Integrated signal routing.} Electrical services
  (power delivery and data readout) are routed via ultra-thin
  aluminum flex cables, potentially printed directly on the MAPS
  sensor surface, exiting at the barrel ends. This avoids routing
  cables through the active tracking volume.
\end{itemize}

Several aspects require further R\&D to validate the design:
\begin{itemize}[itemsep=2pt]
\item \textbf{Air cooling validation.} A realistic prototype with
  representative power density and geometry should be tested,
  measuring thermal performance, vibration amplitudes (air flow
  can excite mechanical resonances in thin shells), and thermal
  gradients (which cause differential expansion and position shifts).

\item \textbf{65~nm MAPS at target thickness.} Sensors thinned to
  20--30~$\mu$m must be characterized for detection efficiency,
  noise, and long-term stability. While 50~$\mu$m has been
  demonstrated by ALICE ITS3, pushing to 20~$\mu$m is more
  aggressive.

\item \textbf{Self-supporting shells at larger radii.} The curved
  shell approach is demonstrated at small radii
  ($R \sim 18$--30~mm). At $R = 300$~mm and beyond, multiple wafers 
  are likely necessary to span the length. 
  Whether the self-supporting
  concept scales to these radii, or whether minimal support
  structures are needed, requires mechanical prototyping.

\item \textbf{Beam pipe co-design.} The combined material budget of
  beam pipe plus first sensor layer should be minimized as a system,
  in collaboration with the accelerator team.
\end{itemize}

This unification of vertex and inner silicon into a single subsystem
is the first example in this study of a design simplification that
emerged from questioning inherited assumptions. The separation was a
vestige of LHC thinking that does not survive scrutiny in the FCC-ee
context. This change is carried forward into the revised concept
CL2a (Chapter~\ref{ch:revised}).

\section{Main Tracker}
\label{sec:evo_tracker}

The main tracker occupies a large volume in the detector and
provides the primary momentum measurement for charged particles. In
CL1a, a drift chamber was chosen, similar in concept to the IDEA
proposal. The subsequent discussion examined both the drift chamber, straw tubes 
and the Time Projection Chamber (TPC) alternative in detail,
exploring mechanical design, particle identification capability, gas
choice, and the practical limitations that separate theoretical
performance from achievable performance.

\subsection{Drift Chamber: Mechanical Design}
\label{sec:evo_dc_mechanical}

The IDEA drift chamber concept proposes a large cylindrical chamber
with inner radius $\sim$350~mm, outer radius $\sim$2000~mm, and
full length $\sim$4~m, containing $\sim$56,000 wires (sense and
field) in a He:iC$_4$H$_{10}$ 90:10 gas mixture. The wires are
organized in 112 layers.

\subsubsection{The wire tension challenge}

The aggregate tension from tens of thousands of wires creates an
enormous mechanical load that must be borne by the end structures.
The IDEA collaboration estimates a total wire tension load of
$\sim$10$^5$~N~\cite{idea_wire_tension}, acting axially and
pulling the two ends of the chamber toward each other.

In a conventional drift chamber, this load is carried by rigid end
plates --- flat or slightly domed structures spanning the full
annular area of the chamber. For the IDEA geometry
($R_{\text{inner}} \approx 350$~mm, $R_{\text{outer}} \approx 2000$~mm),
a conventional aluminum end plate thick enough to resist this load
without excessive deflection would be $\sim$10--30~mm thick,
contributing $\sim$10--20\% $X_0$ of material, significantly degrading the measurement 
for any track passing through the forward region.

\subsubsection{Spokes and stays: the IDEA solution}

\begin{figure}[htbp]
\centering
\includegraphics[width=0.7\textwidth]{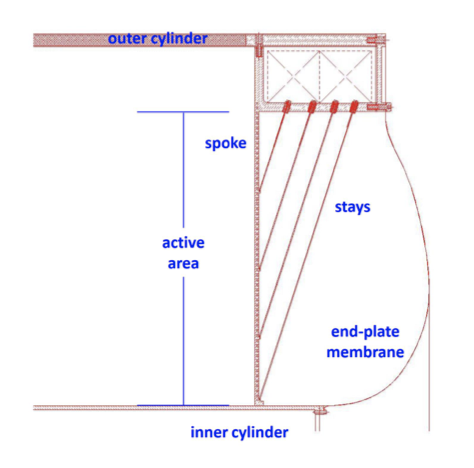}
\caption{Schematic layout of the tension recovery system in the IDEA 
drift chamber. The stays act as tension members connecting points along
the spoke to the outer cylinder, analogous to the cables of a
cable-stayed bridge. Figure adapted from
Ref.~\cite{idea_drift_chamber_2025}.}
\label{fig:spokes_stays}
\end{figure}

The IDEA collaboration proposes replacing the conventional end plates
with a system of \emph{spokes} and \emph{stays}, as illustrated
schematically in Figure~\ref{fig:spokes_stays}.\footnote{The discussion of the spoke and stay geometry benefited
from a schematic drawing from the IDEA
collaboration~\cite{idea_ref}, shared by the
physicist during the dialogue.}

The \textbf{spokes} are radial structural members at each end of the
chamber, running from the inner cylinder to the outer cylinder. They
replace the continuous end plate with a discrete set of narrow
members (typically 8--16 spokes per end), leaving most of the
azimuthal space open.

The \textbf{stays} are tension members --- essentially thin cables
or wires --- that fan from attachment points along each spoke outward
to the outer cylinder. They function precisely like the cables of a
cable-stayed bridge, with the spoke as the deck and the outer
cylinder as the anchor. Each spoke has $\sim$15 stays, corresponding
to attachment points at different radii where groups of wire layers
transfer their tension.

An \textbf{end-plate membrane} provides gas tightness without
carrying structural load. Because it is not structural, it can be
made very thin. Care must be taken to minimize diffusion of helium 
through this membrane.

\subsubsection{Material comparison: spokes vs.\ end plates}

The material advantage of spokes over end plates is substantial.
A conventional aluminum end plate spanning the full annular area
of the chamber would need to be $\sim$10--30~mm thick to resist
the wire tension without excessive deflection, contributing
$\sim$10--20\% $X_0$ of material for any track passing through the
forward region. In contrast, a set of $\sim$12 narrow spokes
occupies only a few percent of the azimuthal acceptance, and the
azimuth-averaged material is significantly less than a
continuous end plate. The detailed mechanical design, including
spoke and stay dimensions for the actual wire tension load
of $\sim$10$^5$~N~\cite{idea_wire_tension}, is being carried
out by the IDEA collaboration; the qualitative advantage of
the spoke approach over conventional end plates is decisive
regardless of the precise dimensions.

Since the total azimuthal obstruction is approximately independent
of the number of spokes --- fewer thick spokes or more thin spokes
give similar total filling fractions --- the choice of spoke count
can be driven by practical considerations such as manufacturing,
wire feedthrough hardware, and assembly.

\subsubsection{Longitudinal space and engineering considerations}

The stays fan from attachment points along each spoke to the outer
cylinder, with the innermost stays spanning the largest radial
distance and therefore requiring the most longitudinal space. The
stay angle represents a trade-off: steeper angles (more nearly
radial) reduce the longitudinal extent but require higher stay
tension to cancel the axial wire load; shallower angles are more
efficient structurally but consume more space. The detailed
optimization of stay geometry, spoke cross-sections, and the
resulting force balance is an engineering task that the IDEA
collaboration is actively pursuing~\cite{idea_wire_tension}.

The key implication for the overall detector layout is that the
stay system consumes significant longitudinal space beyond the
active wire volume --- perhaps up to several hundred millimeters
per end. This space is not
available for endcap tracking or calorimetry, and represents a
meaningful constraint on the detector layout
(Section~\ref{sec:cl2a_longitudinal}).

Other well-known engineering challenges --- wire gravitational
sag, electrostatic deflections, and the time-to-distance
calibration --- are standard issues in drift chamber design that
must be carefully modeled and corrected in reconstruction but do
not raise concerns specific to the FCC-ee application.

One consideration that deserves attention in the context of the
spoke-and-stay design is \textbf{electronics cooling}. In a
conventional drift chamber with solid end plates, the end plate
serves as a natural heat sink for the front-end electronics
mounted on its outer surface, and cooling pipes can be routed
along or integrated into the plate structure. With the
spoke-and-stay approach, this continuous thermal path is absent.
The front-end electronics --- particularly if waveform
digitization for cluster counting is pursued
(Section~\ref{sec:evo_cluster_counting}) --- can dissipate
substantial power that must be removed from a mechanically
sparse structure. Even with conventional $dE/dx$ readout, the
electronics cooling must be addressed through dedicated cooling
paths integrated into the spoke structure or through separate
cooling manifolds in the end region. An alternative approach is 
to locate electronics away from wire ends to a location with 
simpler thermal management. However, the cables needed to transmit
signals add to the material budget and can degrade electronic
performance. These are solvable
engineering problems; their costs and benefits should be 
examined carefully in
the overall design.

\subsection{Drift Chamber: Cluster Counting}
\label{sec:evo_cluster_counting}

One of the headline capabilities of the IDEA drift chamber is
particle identification through cluster counting ($dN/dx$) --- 
counting the individual primary ionization clusters along a track
rather than measuring the total ionization charge ($dE/dx$). The
projected performance is substantially better than conventional
$dE/dx$, but significant challenges have to be overcome for this projection.

\subsubsection{The theoretical promise}

The projected performance is based on a simple statistical argument.
A charged particle traversing the He:iC$_4$H$_{10}$ 90:10 gas
mixture produces an average of $\sim$12 primary ionizations per cm.
For a 200~cm path length (representative of a track at moderate
polar angle), the total number of primary clusters is:
\begin{equation}
N_{\text{total}} = 12 \times 200 = 2400
\end{equation}

Assuming Poisson statistics, the resolution on the cluster count is:
\begin{equation}
\frac{\sigma_N}{N} = \frac{1}{\sqrt{N}}
= \frac{1}{\sqrt{2400}} \approx 2.0\%
\end{equation}

This 2\% resolution, if achieved, would provide excellent $\pi/K$
separation: 3-$\sigma$ significance up to 30 GeV in momentum~\cite{elmetenawee_2024_cluster_counting}. 

The subsequent discussion identified several effects that degrade
this idealized performance. These are examined in turn below.

\subsubsection{Delta ray contamination}

Each primary ionization ejects an electron from a gas atom. The
energy transferred to this electron follows an approximately
$1/E^2$ distribution: most transfers are just above threshold,
but occasionally a large energy transfer produces an energetic
secondary electron (delta ray).

The consequences for cluster counting depend on the delta ray
energy and hence its range:

\begin{itemize}[itemsep=2pt]
\item \textbf{Low-energy secondaries} (the majority of cases):
  the ejected electron travels only a few microns or less and
  remains spatially coincident with the primary ionization. The
  result is a multi-electron cluster rather than multiple
  separate clusters. At the point of production, this does not
  increase the cluster count --- it simply makes the cluster
  have higher charge. However, as the multi-electron cluster drifts toward
  the sense wire, diffusion causes the individual electrons to
  separate. If the separation exceeds the resolving threshold
  of the electronics, what was produced as a single cluster may
  be \emph{detected} as multiple clusters. This effect is
  discussed further under diffusion
  (Section~\ref{sec:evo_diffusion}).

\item \textbf{Moderate-energy delta rays}
  ($\sim$300--1000~eV, range $\sim$50--200~$\mu$m): these
  travel far enough that their secondary ionization is spatially
  separated from the primary cluster even before drift. These create
  genuinely additional clusters that are counted separately,
  inflating the cluster count. Crucially, if the delta ray
  remains within the same drift cell as the primary ionization,
  there is no topological information to distinguish its
  clusters from genuine primary ionization clusters.

\item \textbf{High-energy delta rays} ($>$ a few keV): these
  have ranges of millimeters or more and may cross cell
  boundaries, where they can potentially be identified
  topologically as short track stubs at large angles to the
  primary track. These are in principle removable, though the
  pattern recognition is challenging for marginal cases.
\end{itemize}

The physicist identified the concern: for moderate-energy
delta rays that remain within the same drift cell,
\textbf{there is no way to distinguish the extra clusters from
genuine primary ionization clusters}. 
The contamination from moderate-energy delta rays necessarily
degrades the $dN/dx$ resolution: the measured cluster count is
the sum of the true primary count and the contamination count,
and the variance of the sum is the sum of the variances,
regardless of the shape of the contamination distribution. The
achievable resolution is therefore broader than the
$1/\sqrt{N}$ Poisson floor from primary ionization statistics
alone.

\subsubsection{Diffusion effects}
\label{sec:evo_diffusion}

As ionization electrons drift toward the sense wire, they undergo
random thermal diffusion. For He:iC$_4$H$_{10}$ at typical drift
fields, the longitudinal diffusion coefficient is
$D_L \approx 100$--200~$\mu$m/$\sqrt{\text{cm}}$, giving a spatial
spread of $\sigma_x \approx 60$--130~$\mu$m at the maximum drift
distance of $\sim$7~mm.

Diffusion causes two competing effects:

\textbf{Splitting of multi-electron clusters.} A primary ionization
that produces multiple electrons (e.g., a 3-electron cluster) sees
its electrons diffuse independently during drift. At maximum drift
distance, the separation between electrons from the same cluster
($\sigma_{\text{sep}} \approx \sqrt{2}\sigma_x \approx 130~\mu$m)
frequently exceeds the resolving threshold. Each electron then
appears as a separate arrival at the sense wire, inflating the
cluster count. The physicist identified the key point:
\textbf{once a multi-electron cluster has split, the individual
electrons are indistinguishable from genuine single-electron primary
clusters}. 

\textbf{Merging of separate clusters.} Conversely, electrons from
two genuinely separate primary ionizations can diffuse together and
arrive at overlapping times, appearing as a single cluster with higher charge and
reducing the count.

Both effects are drift-distance-dependent (worse at longer drift),
creating systematic position-dependent biases within each cell.
Even if the mean effects partially cancel, the variances from
splitting and merging add, degrading the resolution beyond the
Poisson floor.

Once a multi-electron cluster has split during drift, each
fragment is typically a single electron, identical in every
measurable respect to a genuine single-electron primary cluster.
No amplitude or timing analysis can identify it as a split
fragment --- \textbf{a single electron carries no memory of its
origin}.

For the converse problem --- merging, where two separate primary
clusters arrive at overlapping times and appear as a single
pulse --- one might hope to use amplitude to flag the merged
pair, since two electrons arriving together should produce a
larger signal than one. However, the gas amplification process
at the sense wire is inherently stochastic: for a single
electron, the avalanche size follows a
distribution with $\sim$100\% relative fluctuation. The
distributions for one-electron and two-electron pulses overlap
enormously, making reliable discrimination impractical.

Furthermore, the effective gain is not constant along the cluster
arrival time series. Positive ions from earlier avalanches create
space charge near the wire that reduces the effective field for
subsequent avalanches. This space charge effect is
position-dependent within each cell, event-dependent, and
track-angle-dependent (tilted tracks spread ionization along
the wire in $z$, reducing the space charge concentration). A
merged two-electron cluster arriving late in the sequence may
have a smaller amplitude than a single electron arriving early,
further undermining any amplitude-based discrimination.

\subsubsection{Electronics requirements}

The mean cluster arrival rate of $\sim$50~MHz (corresponding to
$\sim$20~ns mean spacing) sets the minimum bandwidth for the
front-end electronics. 
Modeling the time separation distribution as approximately
exponential with a mean of $\sim$20~ns, the required
electronics bandwidth depends on the target cluster-pair
resolving efficiency:

\begin{table}[htbp]
\centering
\caption{Required electronics bandwidth as a function of target
cluster-pair resolving efficiency.}
\label{tab:cluster_bandwidth}
\begin{tabular}{ccc}
\toprule
Target efficiency & Required $\delta_t$ (ns)
                  & Required bandwidth (MHz) \\
\midrule
50\% & 14.5 & $\sim$35  \\
70\% & 7.5  & $\sim$67  \\
80\% & 4.7  & $\sim$106 \\
90\% & 2.2  & $\sim$227 \\
95\% & 1.1  & $\sim$470 \\
\bottomrule
\end{tabular}
\end{table}

To resolve $\sim$90\% of cluster pairs, the front-end
electronics must have analog bandwidth of several hundred MHz.
The subsequent digitization must sample the waveform above
the Nyquist rate,
implying sampling rates in the GS/s range.
Compared to conventional drift chamber readout (where each
wire produces a drift time and an integrated charge per event),
cluster counting requires substantially greater electronics
resources across multiple dimensions: analog bandwidth must
increase from tens of MHz to hundreds of MHz; front-end power
per channel increases correspondingly; digitization rates move
from the MS/s range to the GS/s range; and the data volume per
hit wire increases by one to two orders of magnitude (from a
few tens of bits for a time and charge measurement to hundreds
or thousands of bits for a sampled waveform or a list of
individual cluster times and amplitudes). On-chip cluster
finding can reduce the data volume significantly, but at the
cost of 
precluding reprocessing of raw waveforms with improved
algorithms later.

These increased requirements cascade into higher total power
dissipation, greater data bandwidth, and more complex front-end
ASICs, all of which must be weighed against the improvement in
particle identification performance that cluster counting
provides over conventional $dE/dx$.

\subsubsection{Gas choice optimization}

The conventional wisdom for cluster counting is to maximize the
primary ionization rate $n_{\text{primary}}$, since the Poisson
resolution improves as $1/\sqrt{n_{\text{primary}} \times L}$. This
favors heavier gases (argon-based mixtures with $\sim$25/cm) over
helium-based ($\sim$12/cm).

The physicist suggested a counterintuitive alternative: could a gas
with \emph{lower} primary ionization actually perform better in
practice? The argument is that with
fewer clusters per unit time:
\begin{itemize}[itemsep=2pt]
\item The merging fraction decreases (clusters are more separated).
\item The electronics bandwidth requirements relax.
\end{itemize}

The trade-off is that the Poisson floor worsens (e.g., from 2.0\%
to 2.5\% for a reduction from 12/cm to 8/cm). But if the practical
degradations discussed above are large enough, there may be an optimum
$n_{\text{primary}}$ that minimizes the \emph{actual} resolution
rather than the theoretical floor.

Similarly, the physicist noted that a gas with lower drift
velocity would spread clusters further apart in the time
domain, potentially improving resolvability. However, the
benefit depends on the details of the gas mixture: changing
the drift field affects both drift velocity and diffusion,
and the two do not in general scale in the same way. The
improvement from lower drift velocity is most likely to be
realized when the electronics resolving time, rather than
diffusion, is the dominant limitation on cluster separation.
These considerations suggest that the gas optimization for
cluster counting deserves a systematic study, using both simulation
and empirical measurements,
that accounts for all practical effects simultaneously,
rather than optimizing solely for maximum
$n_{\text{primary}}$.

\subsubsection{Realistic performance assessment}

Taking the delta ray contamination, diffusion effects, left-right
folding\footnote{Clusters produced on the two sides of a sense wire share
a common drift time, doubling 
the effective cluster density in the time domain over a naive estimate. This
is an example where the physicist had to take decisive action to change the AI's
perspective. In this case, the consideration led to greater electronics bandwidth demands.}, and electronics limitations together, the achievable
$dN/dx$ resolution is likely to be worse than the theoretical floor of 2\%. The dominant degradations are:
\begin{itemize}[itemsep=2pt]
\item Delta ray contamination: adds non-Gaussian variance that
  cannot be removed for same-cell delta rays.
\item Diffusion-induced splitting and merging: adds drift-distance-dependent variance beyond Poisson.
 
\end{itemize}

This worsened resolution may still represent an improvement over
conventional $dE/dx$ truncated mean (typically $\sim$5--6\% in
drift chambers), though the improvement is \textbf{more modest than
the factor-of-two-or-more suggested by the idealized calculation}.
Particle identification at low momenta (up to a few GeV) is
provided by time-of-flight measurement from the outer silicon
wrapper (Section~\ref{sec:evo_wrapper}), independent of the
drift chamber readout approach. At higher momenta, where the
physics interest is greatest, the relevant question is whether
the improvement from cluster counting ($dN/dx$) over
conventional $dE/dx$ --- after accounting for the practical
limitations discussed above --- justifies the substantially
greater electronics resources and service requirements. This question deserves careful
study through detailed simulation with realistic cluster-finding
algorithms.

\subsection{Conventional End Plate Drift Chamber}
\label{sec:evo_conventional_dc}

Before the IDEA spoke-and-stay design, all large drift chambers
at $e^+e^-$ colliders --- including those at LEP (OPAL),
SLC (SLD), and the B-factories (BaBar, Belle) --- used
conventional end plates to carry the wire tension load. This
proven approach deserves consideration for FCC-ee as a lower-risk
alternative to the spoke-and-stay system.

\subsubsection{What stays the same}

In the barrel tracking volume, a conventional end plate drift
chamber with He:iC$_4$H$_{10}$ gas is essentially identical to
the IDEA concept: the same gas, similar wire configurations,
similar spatial resolution per layer, and similar material budget
from gas and wires ($\sim$0.2\% $X_0$ radially). The helium-based
mixture remains the gas of choice for its low material budget and
moderate primary ionization rate. The cluster counting capability
and its associated challenges
(Section~\ref{sec:evo_cluster_counting}) are unaffected by the
choice of end structure --- the ionization physics is identical.

\subsubsection{The end plate trade-off}

The key difference is at the chamber ends. A conventional end
plate is a continuous structural disc (or annular ring) spanning
the full chamber aperture, carrying the aggregate wire tension
of $\sim$10$^5$~N~\cite{idea_wire_tension}. End plate thickness
can be reduced through careful engineering --- for example, the
SLD drift chamber used a curved end plate geometry in which the
membrane tension reduced the bending stress
at the junction with the outer cylinder, achieving a thinner
plate than a flat design would require. Nevertheless, the end
plate material traversed by forward tracks is substantial ---
typically several 10's percent $X_0$ or more.

However, the total material in the end region is not dominated
by the end plate alone. Front-end electronics boards, power
cables, data cables, cooling manifolds, and cable routing
collectively contribute material that is likely comparable to the structural plate itself. Much of this service material
is present regardless of whether the end structure is a
conventional plate or a spoke-and-stay system --- the electronics
must be housed and cooled in either case.

\subsubsection{Advantages over the spoke-and-stay approach}

\begin{itemize}[itemsep=2pt]
\item \textbf{Proven engineering.} Conventional end plates have
  been designed, built, and operated successfully in multiple
  large drift chambers over decades. The engineering is
  well-understood with no unresolved R\&D questions.

\item \textbf{Longitudinal space.} The end plate is compact in
  $z$ --- typically a few centimeters of structural material
  plus the electronics layer immediately behind it. This is
  substantially less longitudinal space than the
  $\sim$500--800~mm consumed by the IDEA stay system
  (Section~\ref{sec:evo_dc_mechanical}), freeing space for
  endcap detectors.

\item \textbf{Electronics mounting and cooling.} The end plate
  provides a natural rigid surface for mounting front-end
  electronics and a thermal path for cooling. With the
  spoke-and-stay approach, the absence of a continuous end
  plate complicates electronics cooling, particularly if
  high-power waveform digitization for cluster counting is
  pursued (Section~\ref{sec:evo_cluster_counting}).

\item \textbf{Lower engineering risk.} No novel mechanical
  concepts are required. The design can draw directly on
  the extensive experience base from previous chambers.
\end{itemize}

\subsubsection{Disadvantages}

\begin{itemize}[itemsep=2pt]
\item \textbf{Forward region material.} The end plate and
  associated services represent several 10's percent $X_0$ of
  material at small polar angles. This has two consequences.
  First, if endcap tracking layers (silicon disks) are placed
  beyond the end plate, the intervening material degrades the
  combined track fit through multiple scattering. Second, and
  perhaps more importantly, this material sits directly in
  front of the endcap electromagnetic calorimeter, degrading
  photon energy resolution and $\pi^0$ reconstruction in the
  forward region through early shower initiation --- an effect
  analogous to the cryostat dead material discussed for noble
  liquid barrel calorimeters
  (Section~\ref{sec:evo_cryostat}).

\item \textbf{Non-uniform material distribution.} The material
  is concentrated in a narrow $z$ range (the end plate
  location), creating a sharp transition between the
  low-material barrel region and the high-material end plate.
  This complicates the acceptance modeling for precision
  measurements.
\end{itemize}

\subsubsection{Assessment}

The conventional end plate drift chamber is a viable and
low-risk option for FCC-ee. Its barrel performance is
essentially identical to the IDEA concept, and its engineering
is thoroughly proven. The trade-off is additional material in
the forward region, partially offset by the recovery of
longitudinal space and the practical advantage of a natural
electronics mounting and cooling surface.

The choice between conventional end plates and the IDEA
spoke-and-stay system depends on how heavily one weights
forward-region material (favoring IDEA) versus engineering
simplicity and longitudinal compactness (favoring conventional
end plates). For the CL2a concept, the spoke-and-stay approach
is adopted following the IDEA design philosophy, but the
conventional alternative remains a credible fallback with
lower engineering risk. A quantitative comparison ---
evaluating the impact of end plate material on specific
Z-pole physics measurements against the benefit of recovered
longitudinal space for endcap detectors --- would help
clarify this trade-off.

\subsection{Straw Tube Alternative}
\label{sec:evo_straw}

Straw tubes offer an alternative implementation of the drift
chamber concept in which each sense wire is enclosed in a thin
metallized tube that serves as both cathode and cell boundary,
replacing the field wires of a conventional drift chamber. This
approach has operational advantages: the cylindrical geometry
gives a clean $1/r$ electric field with a well-understood
time-to-distance relationship; a broken sense wire is contained
within its straw and cannot affect neighboring cells; and
individual straws can be tested before integration into the full
chamber.

However, for the FCC-ee application, straw tubes present several
challenges relative to the wire drift chamber:

\begin{itemize}[itemsep=2pt]
\item \textbf{Material budget.} Even with 12~$\mu$m wall
  Mylar tubes (at the frontier of what has been prototyped),
  the cumulative straw wall material for a radial track
  traversing $\sim$165 layers is $\sim$1.4\% $X_0$ from the
  tube walls alone --- roughly 7$\times$ the material of the
  IDEA drift chamber (gas + wires). With support structures
  included, the total material is comparable to that of a
  full silicon tracker ($\sim$1--2\% $X_0$). This does not
  meet the momentum resolution target at low momenta, though
  it is worth noting that silicon tracker concepts such as
  CLD has a similar material budget.

\item \textbf{Stereo geometry.} In a wire drift chamber,
  stereo wires run at a small angle to the beam axis, and the
  cell boundaries defined by the wire positions naturally adapt
  along $z$ --- the cell size varies smoothly from midplane to
  endplate. A straw tube has a fixed diameter along its length.
  Tubes that are close-packed and provide full azimuthal coverage
  at the end plates overlap and clash
  for smaller $\lvert z \rvert$, while 
  close-packed tubes that provide full coverage at the midplane
  develop gaps at the endplates when tilted for stereo. 
  This gap size grows with stereo angle; for typical
  angles of $\sim$50~mrad over a 2~m half-length, the
  displacement at the endplate ($\sim$100~mm) is many straw
  diameters, creating substantial coverage gaps. A possible
  mitigation is to use mostly axial straws with a few stereo
  layers at small angles, accepting moderate gaps and reduced
  $z$ resolution compared to a full-stereo wire chamber. 

\item \textbf{Mechanical support.} A 12~$\mu$m Mylar tube
  cannot withstand the axial tension of its sense wire ---
  the critical buckling load for such a thin-walled cylinder
  is far too low. The total wire tension load (comparable to
  that of a conventional drift chamber) must therefore be
  carried by end plates held apart by compression struts
  through the active volume. This reintroduces end plate
  material in the forward region that the IDEA spoke-and-stay
  design was specifically created to minimize.

\item \textbf{Fabrication complexity.} Since individual straws
  cannot support their own sense wire tension, each straw
  requires an external fixture to bear the wire load during
  fabrication and testing, until the straw is integrated into
  the full chamber with end plates providing the reaction
  force. This complicates the production process for tens of
  thousands of straws.
\end{itemize}

The particle identification capability ($dE/dx$ or $dN/dx$)
is not fundamentally different between straw tubes and a wire
drift chamber --- the ionization physics, delta ray production,
diffusion, and cluster counting challenges discussed in
Section~\ref{sec:evo_cluster_counting} apply equally to both.

The straw tube approach offers real operational advantages in
fault isolation and field geometry simplicity, and its material
budget, while substantially greater than the IDEA wire chamber,
is not greater than that of a silicon tracker. It is not adopted
for the CL2a concept, where the IDEA-style wire drift chamber
achieves lower material budget and better stereo $z$ coverage,
but it could merit consideration for applications where the
operational advantages outweigh the material penalty.

\subsection{Time Projection Chamber}
\label{sec:evo_tpc}

The TPC was considered as an alternative to the drift chamber for
the main tracker. The discussion assumed operation at atmospheric
pressure with modern MicroPattern Gaseous Detector (MPGD) such as GEM or Micromegas readout, 
similar to the ALICE TPC upgrade, the ILD-FCCee concept,
and the CEPC Reference Detector~\cite{cepc_detector_ref}.

\subsubsection{Advantages of true 3D reconstruction}

The TPC's defining feature is that every hit is a true
three-dimensional space point. The $r$-$\phi$ coordinates come from
the pad position on the endcap readout plane, and $z$ comes from the
drift time. This provides several concrete advantages over a drift
chamber:

\begin{itemize}[itemsep=2pt]
\item \textbf{Pattern recognition.} With 3D points, track finding
  is a search for helical patterns in a point cloud with no
  combinatorial ambiguities. There is no left-right ambiguity
  (present in drift chambers) and no stereo matching required for
  the $z$ coordinate. This makes pattern recognition more robust
  in complex event topologies --- jets, displaced vertices, kinks,
  and $V^0$s.

\item \textbf{$z$ resolution.} Every TPC hit provides a $z$
  coordinate directly from the drift time, with a resolution
  of $\sim$1~mm. This means every measurement point is a true
  3D space point, without requiring the combination of axial
  and stereo layers needed in a wire drift chamber to obtain
  $z$ information.

\item \textbf{Uniform coverage.} The TPC provides measurements
  throughout its volume with no dead regions from wire support
  structures, spokes, or stays. The active volume is purely gas.
\end{itemize}

\subsubsection{Gas choice}

The long drift distance in a TPC ($\sim$2~m per half-chamber)
makes transverse diffusion a critical design constraint.
Physicist pointed out that this requirement is in tension 
with the preference for a light gas to minimize material budget.
Adequate diffusion suppression requires two complementary
mechanisms: magnetic field suppression (effective in any gas
with a strong solenoidal magnetic field parallel to the drift direction)
and the Ramsauer-Townsend (RT) effect, a quantum mechanical
minimum in the electron-atom elastic scattering cross-section
that occurs in argon, krypton, and xenon but is absent in
helium and neon.

With both mechanisms active in an optimized argon-based mixture
at 2~T, transverse diffusion over 2~m of drift is suppressed to
the mm level, competitive with the per-point resolution of a
drift chamber.

\subsubsection{The material budget trade-off}

The Ramsauer-Townsend effect creates a tension between diffusion
performance and material budget. The best TPC gases for diffusion
suppression (argon-based) are significantly heavier than the best
gases for material budget (helium-based):

\begin{table}[htbp]
\centering
\caption{Material budget comparison for different tracker gas
options over a $\sim$1.65~m radial path length at atmospheric
pressure.}
\label{tab:gas_material}
\begin{tabular}{lSS}
\toprule
Gas option & {Density (mg/cm$^3$)} & {$x/X_0$ (\%)} \\
\midrule
He:iC$_4$H$_{10}$ 90:10 (drift chamber) & 0.26 & 0.03 \\
Ne:CO$_2$ 90:10 (TPC)                   & 0.88 & 0.48 \\
Ar:CH$_4$ 90:10 (TPC)                   & 1.56 & 1.5  \\
Ar:CF$_4$ 95:5 (TPC)                    & 1.70 & 1.5  \\
\bottomrule
\end{tabular}
\end{table}

The argon-based TPC gas contributes $\sim$1.5\% $X_0$ radially ---
a factor of $\sim$50 more than the helium-based drift chamber gas.
For low-momentum tracks from $B$ and $D$ hadron decays at the
Z~pole, where multiple scattering limits the impact parameter
resolution even after the vertex detector, this is a significant
disadvantage.

A helium-based TPC would avoid the material penalty but would not
benefit from the RT effect, relying solely on magnetic field
suppression for diffusion control. The resulting resolution at
maximum drift would be substantially worse than in an argon-based
TPC.

The drift chamber sidesteps this dilemma entirely: it uses
helium (lightest gas, minimal multiple scattering) and does
not need long-range diffusion suppression because the drift
distances are short ($\sim$7~mm per cell). It is worth noting
that the TPC gas material ($\sim$1.5\% $X_0$) is comparable
to that of a straw tube tracker or a full silicon tracker
(Section~\ref{sec:evo_straw}) --- the material penalty is a
common feature of all main tracker options other than the
helium wire drift chamber.

\subsubsection{Ion back-flow management}

Ion back-flow (IBF) --- the drift of positive ions from the
amplification stage back into the TPC drift volume --- creates space
charge that distorts the electric field and degrades spatial
resolution. This was historically the TPC's most serious operational
challenge.

The discussion confirmed that IBF is now considered manageable for
FCC-ee conditions through a two-pronged strategy:

\textbf{Passive suppression:} Multi-GEM stacks (as used in the
ALICE TPC upgrade~\cite{alice_tpc_upgrade}) achieve IBF $\sim$0.5--1\%
through geometric/electrostatic trapping of ions at each GEM stage.
This works continuously without active gating or dead time,
compatible with FCC-ee's continuous readout requirement.

\textbf{Calibration of residual distortions:} The remaining IBF
creates slowly-varying space charge distortions that can be measured
and corrected using:
\begin{itemize}[itemsep=2pt]
\item Laser calibration tracks at known positions, providing an
  absolute distortion map.
\item Track-based calibration using reconstructed tracks
  (which should be smooth helices in a uniform field),
  providing high-statistics fine-grained
  corrections~\cite{alice_tpc_calibration_2026}.
\end{itemize}

At FCC-ee's Z-pole event rate ($\sim$100~kHz), the IBF-induced
space charge density is much lower than in ALICE Pb-Pb collisions,
where the technique has been successfully demonstrated. The CEPC
collaboration has performed detailed studies reaching similar
conclusions for Higgs factory
conditions~\cite{cepc_tpc_ref}.

Active gating (opening and closing a wire grid to block ions) is
not viable for FCC-ee because it introduces dead time incompatible
with continuous readout at the $\sim$20~ns bunch spacing.

\subsubsection{Rate capability and pile-up}

The TPC's long drift time ($\sim$80~$\mu$s for 2~m drift at
$v_d \approx 2.5$~cm/$\mu$s) means multiple physics events are
simultaneously present in the drift volume. At the Z-pole rate
of $\sim$100~kHz:
\begin{equation}
N_{\text{pileup}} \approx 10^5 \times 80 \times 10^{-6} \approx 8~\text{events}
\end{equation}

These eight overlapping events produce $\sim$160 tracks spread
over $\sim$50,000 readout pads per row --- a manageable occupancy.
The tracks can be associated to the correct event using timing
information from the vertex detector and outer silicon wrapper,
which have nanosecond-level time resolution.

The drift chamber, with its $\sim$200--350~ns drift time per cell,
has essentially zero pile-up at any FCC-ee energy point. This is a
practical advantage, though not a decisive one given that TPC
pile-up is manageable.

\subsection{Drift Chamber vs.\ TPC: Comparative Assessment}
\label{sec:evo_tracker_choice}

Table~\ref{tab:tracker_comparison} summarizes the comparison.

\begin{table}[htbp]
\centering
\caption{Drift chamber vs.\ TPC: comparative assessment for FCC-ee.}
\label{tab:tracker_comparison}
\begin{tabular}{lll}
\toprule
Property & Drift chamber & TPC \\
\midrule
Material (gas)          & Excellent ($\sim$0.03\% $X_0$)
                        & $\sim$1.5\% $X_0$ \\
                        &
                        & (Ar-based, RT required) \\
\addlinespace
$r$-$\phi$ resolution   & $\sim$100~$\mu$m, uniform
                        & $\sim$100~$\mu$m, z-dependent \\
\addlinespace
$z$ information         & Requires axial + stereo
                        & Every hit, \\
                        & layer combination $\sim$1~mm 
                        & $\sim$1~mm \\
\addlinespace
3D reconstruction       & 2D per layer + stereo
                        & True 3D per hit \\
\addlinespace
Pattern recognition     & Left-right ambiguity,
                        & Unambiguous, \\
                        & stereo matching required
                        & robust \\
\addlinespace
Pile-up                 & Negligible
                        & $\sim$8 events at Z pole \\
\addlinespace
Calibration             & Local (per wire/cell)
                        & Global (drift field, \\
                        &
                        & space charge) \\
\addlinespace
PID                     & $dE/dx$; $dN/dx$
                        & $dE/dx$; $dN/dx$ \\
                        & under study
                        & to be investigated \\
\addlinespace
Maintainability  & Front-end boards       & Endcap modules \\
                 & replaceable;           & replaceable; \\
                 & wire repair difficult  & field cage monolithic \\
\addlinespace
Forward material        & Spokes (localized)
                        & Readout endcap \\
                        &
                        & (continuous) \\
\bottomrule
\end{tabular}
\end{table}

Both options are viable for FCC-ee, and neither is clearly superior
overall. The drift chamber's advantages in material budget and
calibration robustness are counterbalanced by the TPC's superior 3D
reconstruction and $z$ resolution.

The material budget argument favors the drift chamber,
particularly for low-momentum tracks at the Z~pole where
multiple scattering in the tracker volume degrades the
momentum resolution.

The $z$ resolution argument favors the TPC, particularly for
acceptance determination at the Z~pole where $\cos\theta$-dependent
systematics must be controlled at extreme precision.

For the CL2a concept (Chapter~\ref{ch:revised}), the drift chamber
is retained as the baseline, primarily for its lower material budget.
However, the TPC is recommended as the tracker for the complementary
detector CL2b, providing maximal complementarity in tracking
technology and systematic profile. The extensive TPC studies by the
CEPC collaboration and the ILD-FCCee concept provide a strong
foundation for this choice.

The main tracker technology remains, in our assessment, 
resolved and would benefit from
a direct simulation comparison under FCC-ee conditions, including
realistic material budgets, pattern recognition performance in
complex events, and the impact of gas material on vertexing
performance at the Z~pole.

\section{Outer Silicon Wrapper}
\label{sec:evo_wrapper}

The outer silicon wrapper sits at the boundary between the main
tracker and the electromagnetic calorimeter, at $R \approx 2000$~mm.
In CL1a, it was envisioned as two separate layers with distinct functionalities: silicon strips
to determine position for momentum measurement and LGADs for timing. The subsequent
discussion refined both the functional requirements and the
technology choice.

\subsection{Three Functions in Priority Order}
\label{sec:evo_wrapper_functions}

The physicist proposed a clear priority ordering for the wrapper's
functions, which shaped the subsequent technology discussion:

\begin{enumerate}[itemsep=4pt]
\item \textbf{Precision $r$-$\phi$ space point for momentum
  resolution.} The outermost measurement point has the largest lever
  arm and therefore the greatest weight in the track curvature
  determination. A single high-precision $r$-$\phi$ point at
  $R \approx 2000$~mm with $\sigma_{r\phi} \approx 7$--10~$\mu$m
  provides a constraint equivalent to many drift chamber points.
  This alone justifies the wrapper.

\item \textbf{Longitudinal ($z$) position for acceptance
  determination.} At the Z~pole with $10^{12}$~Z~bosons,
  statistical uncertainties are negligible and acceptance edges in
  $\cos\theta$ must be known at the $10^{-4}$ level. This requires
  good $z$ resolution at the outer radius. At $R = 2000$~mm, a
  $z$ resolution of $\sigma_z \sim 1$~mm gives
  $\sigma_\theta \sim 0.5$~mrad --- adequate but not generous.
  Better $z$ resolution is desirable.

  The physicist noted that since the $z$-resolution requirement is of order a few hundred
microns rather than $\sim$10 $\mu$m, strip sensors (as discussed below), arranged
as a stereo doublet, can provide adequate $z$ measurement
regardless of the specific sensor technology. This significantly
reduces the channel count compared to a pixel implementation at
$R = 2000$~mm.

\item \textbf{Time-of-flight for particle identification.} With a
  flight path of $\sim$2~m and TOF resolution of $\sim$30~ps,
  $\pi/K$ separation at $3 \sigma$ level extends to $\sim$3.5~GeV. 
  \footnote{A useful mnemonic: the time-of-flight difference
between a 1~GeV/$c$ pion and kaon over 2~m is $\sim$750~ps.
This time difference scales linearly with flight distance and inversely with $p^2$.}
  This complements the
  $dE/dx$ (or $dN/dx$) measurement from the main tracker, which
  provides separation at higher momenta. The combination gives
  continuous $\pi/K$ coverage over most of the relevant momentum
  range at the Z~pole.
\end{enumerate}

This priority ordering has a practical consequence: the technology
choice should be optimized primarily for the first function
(position) and must not compromise it in pursuit of the third
(timing).

\subsection{LGAD Technology Options}
\label{sec:evo_lgad_options}

The discussion examined several LGAD variants for their ability to
serve all three functions simultaneously in a single sensor layer,
minimizing the total material.

\subsubsection{Conventional LGAD}

Standard LGADs have pad sizes of $\sim$1$\times$1~mm$^2$ or larger,
limited by the junction termination structure between pads that
creates a dead no-gain region of $\sim$50--100~$\mu$m width. The
position resolution is $\sim$1~mm/$\sqrt{12} \approx 300$~$\mu$m
--- completely inadequate for the precision $r$-$\phi$ measurement.
Timing is excellent with $\sim$30~ps demonstrated, but conventional
LGADs cannot serve as the precision space point layer.

\subsubsection{AC-LGAD (Resistive Silicon Detectors)}

AC-LGADs use a continuous gain layer (no inter-pad dead zones) with
a resistive sheet and AC-coupled readout pads. The signal from any
point spreads through the resistive layer and is detected by
multiple neighboring pads, enabling position reconstruction from
the charge sharing pattern.

Demonstrated performance includes position resolution of
$\sim$5--15~$\mu$m (depending on pitch and resistivity) and timing
of $\sim$30--40~ps --- potentially meeting all three functional
requirements in a single sensor.

However, the physicist identified a potential vulnerability: the
charge sharing mechanism that enables position resolution is
compromised when multiple tracks hit the sensor within the signal
spreading distance ($\sim$1~mm). The overlapping charge clouds
produce a biased or ambiguous position reconstruction.

Studies of machine learning approaches to disentangle overlapping hits indicate that performance can be characterized in three regimes, based on the fundamental signal overlap geometry:
\begin{itemize}[itemsep=2pt]
\item Well-separated hits (separation $> 1.5\times$ spreading
  distance): both positions recovered with near-single-hit
  resolution.
\item Moderately close hits ($\sim$1$\times$ spreading distance):
  positions recoverable with degraded resolution ($\sigma \sim
  20$--50~$\mu$m).
\item Very close hits ($< 0.5\times$ spreading distance): genuinely
  degenerate; even ML cannot reliably separate them.
\end{itemize}

The physicist noted that while the average track density at
$R = 2000$~mm is low, the cases where two tracks are
close are precisely the cases where momentum resolution matters
most (e.g., the core of a boosted jet). Losing a
precision point at the end of a track seriously affects that
individual track's momentum measurement.

The ML disentangling techniques, while promising, are strongly
dependent on the specific sensor geometry, require very accurate
simulation for training, and have not been validated under real
detector conditions with non-uniformities and long-term variations.
The assessment is \textbf{promising but not yet proven} at the level
needed for a precision measurement.

\subsubsection{Trench-Isolated LGAD (TI-LGAD)}

TI-LGADs use deep narrow trenches between readout elements
(filled with insulating material) to provide electrical
isolation while maintaining high fill factor ($>$95\%) even
for fine pitch. Each readout element has its own independent
gain region, so there is no charge sharing between
neighboring elements.

Key properties:
\begin{itemize}[itemsep=2pt]
\item Position resolution determined by pitch:
$\sigma \sim \mathrm{pitch}/\sqrt{12}$. For a strip pitch
of 50~$\mu$m, this gives $\sigma \approx 15$~$\mu$m in
$r$-$\phi$.
\item Timing $\sim$30--40~ps, comparable to standard
LGADs~\cite{tilgad_fbk}.
\item \textbf{Multi-hit capability is inherently robust}
--- each readout element is independent, so two hits on
adjacent strips are cleanly separated with no charge
sharing ambiguity.
\end{itemize}

Compared to AC-LGADs, the TI-LGAD position resolution is
binary (hit or no hit per strip) rather than derived from
analog charge sharing. While the achievable resolution is
somewhat coarser for the same pitch ($\sim$15~$\mu$m vs.\
$\sim$5--15~$\mu$m for AC-LGADs), it is fully adequate for
the wrapper's $r$-$\phi$ requirement and has the significant
advantage of being immune to the multi-hit confusion
discussed above.

\subsubsection{Technology assessment}

For CL2a, the AC-LGAD is preferred as the baseline technology for a
single-layer outer wrapper, primarily because:
\begin{itemize}[itemsep=2pt]
\item The track density at $R = 2000$~mm is low --- the
  fraction of tracks affected by multi-hit confusion is small; however 
  this should be confirmed by detailed physics simulation of the
  applicable use cases. 
\item The material savings of a single thin layer (vs.\ separate
  strip + LGAD layers) for calorimeter performance.
\item Both AC-LGAD technology and the ML disentangling techniques
  will continue to mature before the FCC-ee construction timeline.
\end{itemize}

The TI-LGAD is maintained as a backup that could become preferred
if the technology matures sufficiently, particularly because it
eliminates the multi-hit concern entirely.

Separate strip and LGAD layers (as in the
original CL1a) is a conservative path guaranteed
to work, at the cost of approximately doubling the wrapper material
budget.

\subsection{Material Budget}
\label{sec:evo_wrapper_material}

The physicist pressed for a realistic accounting of the full wrapper
material budget, not just the sensor thickness. This exercise carried
out by the AI 
revealed that the sensor itself is a minor contributor:

\begin{table}[htbp]
\centering
\caption{Material budget for the outer silicon wrapper (single
AC-LGAD layer), showing optimistic and conservative estimates.}
\label{tab:wrapper_material}
\begin{tabular}{lSS}
\toprule
Component & {Optimistic $x/X_0$ (\%)} & {Conservative $x/X_0$ (\%)} \\
\midrule
LGAD sensor (50~$\mu$m Si) & 0.053 & 0.053 \\
Readout ASIC (75~$\mu$m Si) & 0.080 & 0.107 \\
Bump bonds / interconnect & 0.005 & 0.015 \\
Mechanical support & 0.15 & 0.40 \\
Power cables & 0.05 & 0.10 \\
Data cables & 0.03 & 0.08 \\
Cooling (air vs.\ pipes) & 0.00 & 0.25 \\
\midrule
\textbf{Total} & \textbf{0.37} & \textbf{1.0} \\
\bottomrule
\end{tabular}
\end{table}

The dominant contributions are the mechanical support structure
(which cannot use the self-supporting curved shell approach at
$R = 2000$~mm --- the structure is too large and gravity matters)
and potentially the cooling system (if air cooling proves
insufficient for the LGAD power dissipation).

At $\sim$0.4--1.0\% $X_0$, the outer wrapper is
\textbf{10--30 times heavier per layer than the MAPS vertex layers},
even though the sensors are comparably thin. This is driven by the
hybrid nature of LGADs (separate ASIC doubling the silicon
material), the structural requirements at large radius, and the
power delivery and cooling infrastructure.

This material sits directly in front of the ECAL, where every photon
from $\pi^0 \to \gamma\gamma$ must traverse it. The photon
conversion probability is $\sim$0.3--0.8\% per photon, which
degrades ECAL performance. This consideration
\textbf{strongly reinforces the case for a single AC-LGAD layer}
rather than the two-layer strip + LGAD configuration of CL1a, which
would roughly double the material. The physicist noted that the silicon
wrapper realized in a detector will need multiple sensor layers compared
with this conceptual layout to ensure robustness. 

\subsection{Start Time for Time-of-Flight}
\label{sec:evo_tof_start}

The time-of-flight measurement requires both a stop time (at the
outer wrapper) and a start time (at the interaction point). The
physicist raised the question of whether the nominal bunch crossing
time is adequate as the start time, or whether the bunch length
introduces a significant spread.

\subsubsection{Bunch length and start time spread}

The FCC-ee bunch length varies across the energy points, primarily due to
the interplay of RF voltage, synchrotron radiation, and
beamstrahlung:

\begin{table}[htbp]
\centering
\caption{FCC-ee bunch lengths and corresponding start time
spreads.}
\label{tab:bunch_length}
\begin{tabular}{lSS}
\toprule
Energy point & {$\sigma_z$ (mm)} &
{$\sigma_t$ (ps)} \\
\midrule
Z pole (91~GeV)         & \mbox{3{--}4} & \mbox{7{--}9} \\
WW threshold (160~GeV)  & \mbox{3{--}4} & \mbox{7{--}9} \\
ZH (240~GeV)            & \mbox{4{--}5} & \mbox{9{--}12} \\
$t\bar{t}$ (350~GeV)    & \mbox{5{--}6} & \mbox{12{--}14} \\
\bottomrule
\end{tabular}
\end{table}

The start time spread of $\sim$10~ps is not negligibly smaller than
the LGAD stop-time resolution of $\sim$25--30~ps. 
Using the nominal bunch crossing time as the start time
would degrade the effective TOF resolution especially at the
$t\bar{t}$ threshold. Two approaches to determine the start
on an event-by-event were examined.

\subsubsection{Method 1: Common vertex time fit}

Every track in the event originates from he same collision at a 
definite time $t_0$. The stop time $t_{stop,i}$ is measured for $N$ tracks, 
each with a known path length $L_i$ and momentum $p_i$. Their masses are
known for identified leptons and the remainder are assumed to be charged pions. 
These $N$ independent estimates of $t_0$ can be combined in a fit to determine
an overall vertex time. The vertex time fit is combined with bunch timing 
information for an improved start time, yielding $\sigma_{t_0} \approx 5$~ps for
high-multiplicity hadronic events.
Table~\ref{tab:tof_summary} summarizes the situation.

\begin{table}[htbp]
\centering
\caption{Start time determination: intrinsic collision
time spread from bunch length compared with vertex time
fit precision. In practice, their optimal combination is used.}
\label{tab:tof_summary}
\begin{tabular}{lcc}
\toprule
Scenario & \mbox{$\sigma_{t,\text{bunch}}$ (ps)} &
\mbox{$\sigma_{t_0,\text{fit}}$ (ps)} \\
\midrule
Z hadronic ($\sim$12 tracks) & \mbox{7--9} & $\sim$7 \\
Z $\to\tau\tau$ ($\sim$3 tracks) & \mbox{7--9} &
$\sim$14 \\
$t\bar{t}$ hadronic ($\sim$20 tracks) & \mbox{12--14} &
$\sim$6 \\
\bottomrule
\end{tabular}
\end{table}

\subsubsection{Method 2: Dedicated inner timing layer}

A timing-capable sensor layer at small radius
($R \sim 80$--100~mm) could provide a per-track start
time, eliminating the need for either bunch timing or a
vertex time fit. However, the achievable timing resolution
for a thin silicon sensor is $\sim$25~ps at best, which is
significantly worse than the $\sim$7--9~ps start time
already available from the bunch crossing timing. Adding the 
information from a dedicated inner timing layer does not improve the TOF
measurement in any meaningful way. 
And the physicist noted the physics cost of adding material
in the tracking volume ($\sim$0.1--0.3\%~$X_0$ for the
sensor alone, plus services).

An alternative would be to incorporate timing capability
directly into the MAPS vertex detector layers, avoiding
a dedicated additional layer. However, this timing capability 
must significantly out-perform the expectations of AC-LGAD 
to be meaningful. Furthermore, 
timing in MAPS requires additional
circuitry that increases power dissipation, which in turn
increases service material for power delivery and cooling,
and generates additional data volume --- cascading
consequences that undermine the air-cooled, ultra-thin
vertex detector design. If future MAPS technology
development were to provide adequate timing to improve on
the timing from bunch crossing time without these
penalties, it would be a welcome addition; at present,
however, the cost is not justified by the benefit.

\subsubsection{Assessment}

The common vertex time fit is recommended as a complement
to the bunch time determination. The two estimates are
independent and their combination yields
$\sigma_{t_0} \approx 5$~ps for high-multiplicity hadronic
events. Whether the improvement in start time precision
materially impacts $\pi/K$ separation across the physics
program is a question that deserves dedicated study.

\subsection{Summary of Changes from CL1a}
\label{sec:evo_wrapper_summary}

The outer silicon wrapper in CL2a differs from CL1a in two respects:

\begin{enumerate}[itemsep=2pt]
\item \textbf{Single layer replacing two.} The separate strip +
  LGAD configuration of CL1a is replaced by a single AC-LGAD layer
  serving all three functions (position, $z$, timing). This roughly
  halves the material budget, from $\sim$1.5--2\% $X_0$ to
  $\sim$0.4--1.0\% $X_0$, directly benefiting the ECAL performance
  behind it.

\item \textbf{Explicit start time strategy.} The common vertex time
  fit is adopted as the baseline method for TOF start time
  determination, with a dedicated inner timing layer identified as
  an upgrade path rather than a baseline requirement. This avoids
  adding material in the inner tracking volume while preserving the
  option for the future.
\end{enumerate}

\section{Electromagnetic Calorimeter}
\label{sec:evo_ecal}

The electromagnetic calorimeter underwent the most dramatic
revision of any subsystem in this study. The initial choice
in CL1a --- a liquid argon sampling calorimeter --- was changed
to a PbWO$_4$ crystal calorimeter through a sequence of
arguments that illustrate how practical considerations can
overturn choices that appear sound in the abstract. The
discussion proceeded through several stages of physicist-AI discussion: liquid argon vs.\
liquid krypton, the cryostat dead material problem, the
reassessment of crystals in the FCC-ee context, longitudinal
segmentation and dual-readout, SiPM readout challenges, and
finally the compatibility of crystals with particle flow
reconstruction.

\subsubsection{From Liquid Argon to Liquid Krypton}
\label{sec:evo_lar_to_lkr}

The physicist's first challenge was straightforward: why liquid
argon rather than liquid krypton? Everything else being equal,
liquid krypton has higher density (2.41 vs.\ 1.40~g/cm$^3$),
higher $dE/dx$ for minimum ionizing particles ($\sim$3.4 vs.\
$\sim$2.1~MeV/cm), shorter radiation length (4.7 vs.\
14.0~cm), and smaller Moli\`ere radius (4.7 vs.\ 7.2~cm).
These translate directly into higher sampling fraction and
hence better energy resolution for the same absorber geometry,
as summarized in Table~\ref{tab:noble_liquid_comparison}.

The principal argument against krypton is cost. Argon is
$\sim$1\% of the atmosphere and essentially free; krypton is
$\sim$1~ppm and must be extracted and purified at significant
expense ($\sim$\$1000--3000/liter at research quantities,
likely lower at industrial scale). For a full $4\pi$ sampling
calorimeter requiring $\sim$19~m$^3$ of liquid krypton
(estimated from a barrel geometry with 22~$X_0$ depth at inner
radius $\sim$2.1~m), the krypton cost alone would be of order
\$20--60M.

Against this, krypton operates at a higher boiling point
(120~K vs.\ 87~K for argon), making the cryogenics somewhat
simpler. Purity requirements are similar for both liquids.
The krypton is not consumed and can be recovered.

\begin{table}[htbp]
\centering
\caption{Comparison of noble liquid electromagnetic calorimeter
configurations and their approximate stochastic energy resolution
terms.}
\label{tab:noble_liquid_comparison}
\begin{tabular}{lS[table-format=2.0]}
\toprule
Configuration & {Stochastic term (\%/$\sqrt{\text{GeV}}$)} \\
\midrule
LAr + Pb (ATLAS-like) & 10{--}12 \\
LKr + Pb (similar geometry) & 7{--}9 \\
LKr + Pb (optimized) & 5{--}7 \\
LKr homogeneous (NA48-style) & 3.5 \\
\bottomrule
\end{tabular}
\end{table}

On physics grounds, liquid krypton is clearly preferred over
liquid argon. In response to the physicist's challenge, the AI
acknowledged that the initial choice of argon was insufficiently
justified --- inherited from the dominance of the ATLAS LAr
calorimeter in the literature rather than optimized for FCC-ee.
This was the first revision: if a noble liquid calorimeter were
to be built, it should use krypton.

\subsection{The Cryostat Problem}
\label{sec:evo_cryostat}

The physicist then posed two questions that proved decisive
for the entire ECAL technology choice: how deep is the
calorimeter in radiation lengths, and how thick is the
cryostat?

\paragraph{Calorimeter depth.}
Electromagnetic shower containment for photons of order
($E \sim 50$--100~GeV)
(as produced in Higgs decays at FCC-ee)
requires 22--25 radiation lengths for 99\% containment. For
a sampling calorimeter with 2~mm Pb absorber and 4~mm LKr
gaps, the effective radiation length of the sandwich is
$X_0^{\mathrm{eff}} \approx 13.6$~mm, giving a total depth
of $\sim$300~mm for 22~$X_0$.

\paragraph{Cryostat thickness.}
Any noble liquid calorimeter requires a thermally insulated
cryostat --- a vacuum-insulated cold vessel that every
particle must traverse before reaching the active
calorimeter. A realistic inner cryostat wall consists of
a warm wall, vacuum insulation and a cold wall with the 
material budget estimated in Table~\ref{tab:cryostat_material}. 
This is a substantial amount of dead material before the
first active calorimeter layer.

\begin{table}[htbp]
\centering
\caption{Cryostat inner wall material budget. The range
reflects material choices (aluminum vs.\ stainless steel
for the cold wall) and structural requirements. Wall
thicknesses reflect realistic engineering for a barrel
cryostat at $R \sim 2.1$~m, consistent with ATLAS LAr
calorimeter experience.}
\label{tab:cryostat_material}
\begin{tabular}{llc}
\toprule
Component & Material & $x/X_0$ \\
\midrule
Warm wall & \mbox{3--5}~mm Al &
\mbox{3--6\%} \\
Vacuum + MLI & \mbox{25--30}~mm &
$\sim$0.6\% \\
Cold wall & \mbox{4--5}~mm Al (or 2~mm SS) &
\mbox{4--6\%} (or 11\%) \\
\midrule
Total (Al cold wall) & & $\sim$10\%~$X_0$ \\
Total (SS cold wall) & & $\sim$15\%~$X_0$ \\
\bottomrule
\end{tabular}
\end{table}

\paragraph{Impact on calorimeter performance.}
The cryostat dead material degrades performance through
several mechanisms:

\textbf{Stochastic term.} The dead material introduces
additional energy loss fluctuations that degrade the
stochastic resolution term.\footnote{A useful rule of
thumb for the impact of dead material on the stochastic
energy resolution term is to add in quadrature a
contribution $\Delta a \approx (0.05\text{--}0.08)
\times \sqrt{x/X_0}$~(\%/$\sqrt{\mathrm{GeV}}$), where
$x/X_0$ is the dead material thickness in radiation
lengths. For $x/X_0 = 0.1$ (i.e., 10\%~$X_0$), this gives $\Delta a \approx
2\%$, a modest degradation of the bulk stochastic term
(from 7\% to $\sim$7.3\% for LKr sampling). This
parametrization is broadly consistent with the OPAL
lead-glass test beam measurements reported in Fabjan
and Gianotti~\cite{fabjan_gianotti_2003}.}
For a cryostat of $\sim$10--15\%~$X_0$, the bulk
stochastic term degradation is modest. However, the
more significant effects are the non-Gaussian tails
and constant term contributions discussed below.

\textbf{Non-Gaussian tails from photon conversions.}
Photons traversing $\sim$10--15\%~$X_0$ of dead
material have a conversion probability of
$\sim$8--11\%. For the
majority of photons that do not convert, the cryostat has
little effect on the energy measurement. However, photons
that do convert produce $e^+e^-$ pairs that deposit some
energy in the cryostat walls and enter the calorimeter
with degraded spatial coherence. These events populate
non-Gaussian tails of the energy response that are
difficult to model and correct --- a particular concern
for the precision measurements at FCC-ee.

\textbf{Constant term contribution.} Non-uniformity of
the cryostat thickness (at joints, welds, feedthroughs),
energy-dependent shower initiation probability, and
angular dependence of the effective thickness
($t/\sin\theta$) all contribute to the constant term at
the $\sim$0.2--0.5\% level. For high-energy photons from
Higgs decays ($E \sim 50$--120~GeV), where the stochastic
term is already small, the constant term may dominate the
resolution.

\textbf{Position and angular resolution.} Photons
converting in the cryostat produce $e^+e^-$ pairs that
are spread by the magnetic field before reaching the
calorimeter, degrading the photon position and pointing
resolution --- precisely the capabilities that the
longitudinal segmentation of a noble liquid calorimeter
is designed to provide.

A presampler (a thin active layer before the calorimeter,
as used in ATLAS) can partially correct the mean energy
loss in the cryostat walls. For the cryostat thicknesses
discussed here ($\sim$0.1--0.15~$X_0$), presampler
corrections can recover much of the average energy loss.
However, it does not eliminate the event-by-event
fluctuations that produce non-Gaussian tails, nor do they
address the position resolution degradation from pair
spreading in the magnetic field.

\paragraph{The decisive comparison.}
The cryostat dead material problem is fundamental to any
noble liquid calorimeter and cannot be eliminated --- a
vacuum-insulated cold vessel is required by the physics of
cryogenic containment. While a presampler can partially
mitigate the average energy loss, the combination of
non-Gaussian tails from photon conversions, constant term
contributions from thickness variations, and position
resolution degradation from pair spreading represents a
significant and irreducible performance penalty.

The crystal calorimeter maintains near-intrinsic resolution
because virtually no photons interact before reaching the
active material. The noble liquid options suffer not only
from the intrinsically worse stochastic term of a sampling
calorimeter, but from the additional and irreducible
penalties of the cryostat --- penalties that are
particularly damaging for the precision photon measurements
central to the FCC-ee physics program.

This was the turning point in the ECAL discussion. The
physicist's questions about cryostat thickness exposed a
fundamental limitation that the AI had not raised
unprompted. The AI agreed that the argument was sound ---
if the comparison includes the cryostat (as any realistic
design must), the noble liquid approach loses much of its
appeal relative to crystals.

\begin{table}[htbp]
\centering
\caption{Comparison of electromagnetic calorimeter
technologies including the impact of material before the
active calorimeter. The crystal calorimeter has only a
thin mechanical support structure ($\sim$0.5\%~$X_0$)
before the first active material. ``Effective $a$'' is
estimated using the rule of thumb $\Delta a \approx
(0.05\text{--}0.08) \times \sqrt{x/X_0}$, added in
quadrature to the intrinsic term; this captures only the
bulk Gaussian stochastic degradation, not the non-Gaussian
tails from photon conversions. ``Additional constant
term'' refers to the contribution from upstream material
only, not the total constant term of the calorimeter.}
\label{tab:effective_resolution}
\begin{tabular}{lccccc}
\toprule
Technology & Intrinsic $a$ & Dead material &
Effective $a$ & Conversion & Additional \\
& (\%/$\sqrt{\mathrm{GeV}}$) & before active &
(\%/$\sqrt{\mathrm{GeV}}$) & probability &
constant term \\
\midrule
LKr sampling & 7 & \mbox{10--15\%}~$X_0$ &
\mbox{7.3--7.4} & \mbox{8--11\%} &
\mbox{0.2--0.5\%} \\
LKr homogeneous & 3.5 & \mbox{10--15\%}~$X_0$ &
\mbox{4.1--4.3} & \mbox{8--11\%} &
\mbox{0.2--0.5\%} \\
PbWO$_4$ crystal & 2.8 & $\sim$0.5\%~$X_0$ &
$\sim$2.8 & $<$0.5\% & negligible \\
\bottomrule
\end{tabular}
\end{table}

\subsection{Crystal Calorimetry Reconsidered}
\label{sec:evo_crystals}

With the cryostat problem established, the case for crystal
calorimetry was reassessed. The AI's initial dismissal of crystals
had been based largely on the LHC experience, where
radiation-induced transparency loss is the dominant challenge. The
physicist pointed out that this experience is not applicable at
FCC-ee.

\subsubsection{Radiation damage: a non-issue at FCC-ee}

The radiation environment at FCC-ee is orders of magnitude more
benign than at the LHC:
\begin{itemize}[itemsep=2pt]
\item LHC: $\sim$1--10~Gy/hour in the ECAL barrel, (varying with $\eta$), 
accumulating to tens to
hundreds of kGy over the HL-LHC program.
\item FCC-ee: $\sim$mGy/hour at most --- essentially negligible for
  crystal damage.
\end{itemize}

The CMS PbWO$_4$ calorimeter experience --- transparency loss,
radiation-induced color centers, continuous laser monitoring, complex
correction procedures, and the resulting contribution to the
constant term --- is driven entirely by the LHC radiation
environment. At FCC-ee, crystals would maintain their initial
optical quality throughout the entire 15-year program. This removes
what has been the dominant systematic limitation of crystal
calorimeters in the current era.

\paragraph{Calibration and monitoring.}
Calibration of a crystal calorimeter is generally more
challenging than for a noble liquid calorimeter, where the
response is determined primarily by the stable and uniform
ionization properties of the liquid. In a crystal
calorimeter, the full response chain --- crystal light
yield, light transport, photodetector gain, and electronics
response --- must be monitored and calibrated across
$\sim$80,000 channels over the lifetime of the experiment.
This is typically addressed through a layered calibration
strategy, combining hardware monitoring systems (light
pulsers, charge injection, temperature sensors) for
short-timescale stability with physics-based methods for
absolute scale and intercalibration.

FCC-ee provides abundant clean electromagnetic samples at
all running points --- electrons and photons from
well-understood processes at known energies --- for
establishing and maintaining the absolute energy scale.
Compared to the LHC, the calibration environment is
qualitatively more favorable: the physics samples are
larger and cleaner, and the low-radiation environment
means the time-dependent corrections that drive much of
the calibration complexity at hadron colliders are far
smaller.

\subsubsection{Other advantages}

Beyond energy resolution, crystals offer:
\begin{itemize}[itemsep=2pt]
\item \textbf{Compactness.} PbWO$_4$ achieves 22~$X_0$ in
only $\sim$20~cm depth ($X_0 = 8.9$~mm), compared to
$\sim$37~cm total for a LKr sampling calorimeter
including its cryostat walls. This allows for a more
compact solenoidal coil.

\item \textbf{Small Moli\`ere radius.} PbWO$_4$ has
$R_M = 22$~mm, providing good transverse shower
separation. While larger than tungsten's 9~mm (used in
Si-W calorimeters), it is significantly smaller than the
$\sim$35~mm effective Moli\`ere radius of a Pb/LKr
sampling calorimeter.

\item \textbf{No cryogenic infrastructure.} Crystals operate at or
  near room temperature, eliminating the cryostat, cryogenics plant,
  liquid purity monitoring, and associated services. This
  simplifies the detector significantly.

\item \textbf{Cost.} PbWO$_4$ crystals at $\sim$\$200--500 each for
  $\sim$80,000 crystals give a crystal cost of $\sim$\$16M--40M ---
  comparable to the krypton cost alone for a noble liquid
  calorimeter, before accounting for the cryostat and cryogenics
  infrastructure.
\end{itemize}

\subsubsection{Longitudinal Segmentation}
\label{sec:evo_crystal_design}

A known limitation of crystal calorimeters relative to
noble liquid or Si-W designs is the lack of longitudinal
segmentation --- a crystal is typically read out as a
single unit, providing total energy but no information
about the shower development profile. The physicist raised
this as a concern following the shift to crystals. In
response, the AI proposed a design that addresses this
limitation through physical segmentation and multiple
SiPM readout points.

\subsubsection{Two-segment design}

The crystal is physically divided into two segments:
\begin{itemize}[itemsep=2pt]
\item \textbf{Front segment:} $\sim$4--6~$X_0$ ($\sim$35--55~mm
  of PbWO$_4$), read out by a SiPM on the particle entrance face.
\item \textbf{Back segment:} $\sim$16--18~$X_0$ ($\sim$140--160~mm),
  read out by SiPM(s) on the exit face.
\end{itemize}

The front SiPM sits on the entrance face of the front segment,
with its active surface facing into the crystal. It collects
primarily scintillation light from the front segment, with
little \v{C}erenkov contamination because \v{C}erenkov light is
emitted predominantly in the forward direction (away from this
SiPM, deeper into the crystal). The front SiPM and its readout substrate add a small
amount of material ($\sim$0.1--0.2\%~$X_0$) on the
entrance face, which must be included in the upstream
material budget.

The ratio of front to total energy provides a measure of the
longitudinal shower profile, enabling:
\begin{itemize}[itemsep=2pt]
\item Photon/hadron discrimination (electromagnetic showers deposit
  a large fraction in the front segment; hadronic showers penetrate
  deeper).
\item $\pi^0/\gamma$ discrimination (overlapping photons from
  high-energy $\pi^0$ decay produce a broader longitudinal profile
  than a single photon).
\item Longitudinal leakage correction (deeper showers have more
  energy in the back segment and are more likely to leak).
\end{itemize}

The front segment depth of $\sim$4--6~$X_0$ is a compromise: deep
enough that electromagnetic showers have begun developing
significantly, but shallow enough that the back segment contains the
majority of the energy and shower maximum ($\sim$7--9~$X_0$) falls
near the segment boundary, maximizing the sensitivity of the
front/back ratio to shower depth.

The physical interface between segments introduces a small dead
layer (optical isolation film or air gap, $\sim$50--200~$\mu$m ($\sim$50--200~$\mu$m, corresponding to
$\sim$0.05--0.2\%~$X_0$)).
This is near shower maximum for medium-energy photons, where energy
deposition per $X_0$ is highest, and contributes to the stochastic
term. The optimization of segment interface material is an important
engineering detail.

\subsubsection{Dual-readout in the ECAL: considered and
not adopted}

Hadrons interacting in the ECAL produce showers whose
electromagnetic fraction $f_{em}$ fluctuates event by
event, degrading the energy resolution. Dual-readout
calorimetry --- simultaneous measurement of \v{C}erenkov
and scintillation light --- can correct for these
fluctuations, since \v{C}erenkov light is produced almost
exclusively by the electromagnetic shower component. This
technique has been studied and developed over some three
decades~\cite{akchurin_2005_hadron,
akchurin_2014_em}, and
represents a well-researched if not frequently adopted  approach to improving
hadronic energy resolution.

With a depth of $\sim$20~cm, corresponding to
approximately one nuclear interaction length
($\lambda_I \approx 20$~cm for PbWO$_4$), a significant
fraction of hadron interactions are initiated in the ECAL, making
dual-readout correction potentially valuable.

In the context of the segmented PbWO$_4$ design discussed
above, dual readout could be implemented by adding a
third SiPM to the back face, equipped with an optical
filter that preferentially transmits \v{C}erenkov light
(UV/blue) while attenuating scintillation light (peaking
at $\sim$440~nm for PbWO$_4$). The C/S ratio would be extracted from the two back-face
SiPMs (filtered and unfiltered), enabling an event-by-event
correction of the measured energy for fluctuations in
$f_{em}$ in the back segment. Combined with the front SiPM --- which is
optically separate and provides longitudinal profile
information --- the three-SiPM configuration would yield
both a shower depth profile and an $f_{em}$-corrected
energy measurement.

The physicist identified two problems with this approach
in the FCC-ee context.

First, for purely electromagnetic showers --- the ECAL's
primary mission --- the dual-readout technique provides
no additional information. In an electromagnetic shower,
$f_{em} = 1$ by definition: there are no hadronic
fluctuations to correct for. Both the \v{C}erenkov and
scintillation signals are proportional to the total
deposited energy, with a fixed ratio that carries no
event-by-event discriminating power. The \v{C}erenkov
measurement is therefore redundant with the scintillation
measurement for electromagnetic showers --- it measures
the same quantity with fewer photons.

Meanwhile, implementing the filtered readout carries a
direct cost: the optical filter attenuates scintillation
photons, reducing the light available for the primary
energy measurement. This cost is compounded by the poor
spectral separation between \v{C}erenkov light and
PbWO$_4$ scintillation --- they overlap significantly in
the blue region, making clean filtering impractical
without substantial loss of scintillation signal.

Second, while the dual-readout correction would genuinely
help for hadrons interacting in the ECAL, the benefit is
limited: the ECAL sees only the beginning of hadronic
showers, with the bulk of the hadronic energy measured in
the HCAL behind. Improving the ECAL's hadronic
measurement at the cost of degrading its electromagnetic
measurement is trading performance on the primary mission
for a partial gain on a secondary one.

A quantitative estimate of the cost: with the back face
shared between two SiPMs (one filtered, one unfiltered),
the total detected photoelectrons drop by roughly a
third, degrading the photostatistical contribution to the
energy resolution by $\sim$20\%.

The dual-readout option in the ECAL is not adopted for
this design. The recommended configuration is two SiPMs
(front and back) without filtering, to provide
longitudinal shower profile information and the best
electromagnetic energy resolution. This does not preclude
dual-readout techniques in the HCAL, where the
cost-benefit trade-off is fundamentally different.

\subsection{SiPM Readout and Dynamic Range}
\label{sec:evo_sipm}

% NOTE FOR FUTURE COHERENCE PASS:
% - Light yield: 200 photons/MeV used here; may need
%   revision to ~100. Affects N_scint, N_pe, and
%   saturation estimates throughout.
% - Maximum energy: 120 GeV per crystal and 150 GeV
%   in opening may be too high given corrected photon
%   energy (~100 GeV max from H->gammagamma). Revisit.
% - Emission peak: 440 nm per CMS TDR Table 1.1,
%   not 420 nm as stated in dual readout section.
% - Decay time: TDR says 5 ns (39%), 15 ns (60%),
%   100 ns (1%). Current text says "fast component
%   (~60% of light, ~6 ns)" which conflates the two
%   fast components.
% - Moliere radius: use 2.2 cm consistently
%   (TDR gives 2.19 cm).

The discussion then turned to a practical challenge of crystal
calorimetry with SiPM readout at FCC-ee energies: the dynamic range
required to measure electromagnetic showers from $\sim$100~MeV to
$\sim$150~GeV without saturation.

\subsubsection{Maximum photon count}

For PbWO$_4$ at room temperature (scintillation yield
$\sim$200~photons/MeV)\footnote{Published values for
PbWO$_4$ light yield range from $\sim$100 to
$\sim$200~photons/MeV at room temperature, depending
on crystal quality, doping, and the precise definition
of ``room temperature'' (the yield has a steep
temperature coefficient of $\sim$$-$2\%/$^\circ$C).
We use 200~photons/MeV throughout as a representative
value; qualitative conclusions are not sensitive to
this choice.}, the maximum energy deposit in a single
crystal ($\sim$120~GeV, accounting for shower sharing between
crystals) produces:
\begin{equation}
N_{\text{scint}} = 200 \times 120{,}000 = 24 \times 10^6
~\text{photons}
\end{equation}

With $\sim$15\% light collection efficiency at the back face and
$\sim$25--40\% SiPM photon detection efficiency, the detected
photoelectrons of an idealized back SiPM are:
\begin{equation}
N_{\text{pe}} \approx 24 \times 10^6 \times 0.15 \times 0.3
\approx 10^6
\end{equation}

The scintillation light is emitted with three decay
components: $\sim$39\% with $\tau \approx 5$~ns,
$\sim$60\% with $\tau \approx 15$~ns, and $\sim$1\%
with $\tau \approx 100$~ns. Of the $\sim$10$^6$ total
photoelectrons at maximum energy, the peak arrival rate
is $\sim$10$^5$ per nanosecond, with the majority
delivered within $\sim$30~ns.

Internal reflections within the high-refractive-index crystal
($n \approx 2.3$) cause extensive light mixing as photons bounce
multiple times off the reflective wrapping before reaching the back
face. As a result, the photon density at the SiPM surface is
approximately uniform regardless of the transverse shower profile
--- there is no localized hot spot at the shower core. The
saturation assessment can therefore use the average photon density
across the SiPM face rather than the peak energy deposition density
within the crystal.

\subsubsection{SiPM saturation}

An SiPM consists of an array of microcells operating in Geiger mode.
Each cell fires once and recovers with a time constant
$\tau_{\text{recovery}} \sim 10$--50~ns. When many photons arrive
within the recovery time, some cells may not have recovered from an 
earlier arriving photon. This is a saturation effect that can be 
corrected on average; however, event-to-event fluctuations lead to
poorer resolution. The average number of cells expected to
fire is:
\begin{equation}
\label{eq:sipm_saturation}
N_{\text{fired}} = N_{\text{cells}} \times
\left(1 - \exp\left(-\frac{N_{\text{pe}}}{N_{\text{cells}}}\right)\right)
\end{equation}

For 1\% nonlinearity at maximum signal, one needs
$N_{\text{cells}} > N_{\text{pe,fast}} / 0.02 \approx 30 \times 10^6$
cells\footnote{For small saturation, expanding
Eq.~\ref{eq:sipm_saturation} to second order gives
$N_{\text{fired}} \approx N_{\text{pe}}
(1 - N_{\text{pe}}/(2 N_{\text{cells}}))$, where the second term
$N_{\text{pe}}/(2 N_{\text{cells}})$ is
the fractional nonlinearity.
Requiring this
to be less than 1\% gives
$N_{\text{cells}} > N_{\text{pe}} / 0.02$.
With $N_{\text{pe,fast}} \approx 600{,}000$, this
requires $\sim$30~million cells.}. Current SiPM technology at 10~$\mu$m cell pitch provides
$\sim$10,000 cells/mm$^2$; covering a 20$\times$20~mm$^2$ crystal
face ($\sim$320~mm$^2$ active area) gives only $\sim$3.2 million
cells --- roughly an order of magnitude too few.

\subsubsection{Fill factor as sampling fraction}

The physicist reframed the saturation problem in illuminating terms:
pushing to smaller cell pitch to increase cell count necessarily
reduces the fill factor, since each cell has a fixed-width border
region ($\sim$2--3~$\mu$m) for the quenching resistor and guard
ring. At 5~$\mu$m pitch, the fill factor drops to $\sim$16\%.

This is another form of ``sampling fraction'' --- whether photons
are lost in absorber plates (as in a conventional sampling
calorimeter) or in dead area between SiPM cells, the physics impact is
identical. The mean can be corrected, but the fluctuation in the
number of detected photons is irreducibly larger. The stochastic
resolution degrades as $1/\sqrt{\text{fill factor}}$.

Among common scintillating crystals, PbWO$_4$ is the
only one for which the SiPM saturation challenge is
even manageable. Higher-yield crystals such as LYSO
($\sim$33,000~photons/MeV), BGO
($\sim$9,000~photons/MeV), or CsI(Tl)
($\sim$54,000~photons/MeV) would produce
100--500$\times$ more light, making SiPM saturation
catastrophic at any energy above $\sim$1~GeV regardless
of cell pitch or fill factor. The low light yield of
PbWO$_4$, usually considered a disadvantage, is in this
context essential.

\subsubsection{Nonlinearity as a separate effect}
\label{sec:nonlinearity}

Even with 100\% fill factor, the Geiger-mode saturation compresses
the response. The correction $N_{\text{pe}} =
-N_{\text{cells}}\ln(1 - N_{\text{fired}}/N_{\text{cells}})$
amplifies the statistical variance:
\begin{equation}
\sigma_{\text{corrected}}^2 = \sigma_{\text{raw}}^2 \times
\left(\frac{N_{\text{cells}}}{N_{\text{cells}} -
N_{\text{fired}}}\right)^2
\end{equation}

At 20\% saturation ($N_{\text{fired}}/N_{\text{cells}} = 0.2$), the
variance amplification is $(1/0.8)^2 = 1.56$, degrading the
effective resolution by $\sim$25\%. At 50\% saturation, the
amplification is 4$\times$, doubling the resolution. These are
separate from and in addition to fill factor effects.

These estimates treat the scintillation light as arriving
instantaneously. In practice, the light is spread over the
scintillation decay profile (with components at $\sim$5, 15,
and 100~ns for PbWO$_4$), and cells that fire early in the pulse
may recover in time to detect later-arriving photons. This partial
recovery effectively increases the dynamic range beyond the
instantaneous estimate. However, the recovery process is
stochastic: which cells recover and re-fire fluctuates event by
event, introducing an additional variance that cannot be removed
by correcting the mean response. The estimates above are therefore
indicative of the scale of the saturation challenge rather than
precise predictions; quantifying the net effect of partial recovery
and its associated fluctuations would require detailed simulation
of the interplay between the scintillation time profile and cell
recovery characteristics.

\subsubsection{Digital SiPM}

A digital SiPMs (dSiPMs) is defined by the integration of 
per-cell digital electronics directly on the same substrate as the 
photo-sending cells. This electronics usually 
integrates active quenching and
recharge circuitry into each cell, reducing the recovery
time from $\sim$10--50~ns (passive) to $\sim$1--5~ns.
Cells can therefore fire multiple times during the
scintillation pulse, increasing the effective dynamic
range. The electronics can have additional logic 
to process individual micro-cell signals to produce a
custom output signal. 

The benefit is real but bounded. The electromagnetic
shower deposits all its energy within $\sim$1~ns, and
the scintillation photon rate is highest immediately
afterward --- precisely when cells have not yet had
time to recover. During the first recharge cycle, the
detector response is identical to that of an analog
SiPM. The improvement from active recharge grows as
the scintillation pulse decays and the photon arrival
rate drops below the cell recovery rate. Active
recharge therefore mitigates saturation --- substantially
for the later portion of the scintillation pulse, but
not for the initial burst\footnote{While scintillation light 
is governed by the decay time constants of the fluor, which can 
range from few nsec to few 100s of nsec, \v{C}erenkov light 
is prompt and does not benefit from active recharge}.

The power implications depend critically on what
functionality is implemented per cell. If the on-sensor
circuitry is limited to active quenching, recharge,
masking of noisy cells, 
and counting of cell firings within a time window, the
additional power dissipation is negligible. If per-cell
timing information is desired, the power requirements
increase substantially and depend on implementation
details that are not yet settled: the timing resolution,
the number of cells sharing a timing circuit, the
occupancy, and the CMOS technology node. Subject matter
experts consulted during this study were unable to
provide even order-of-magnitude power estimates without
a specific circuit design.

For the baseline calorimetric measurement (total energy
from photon counting), per-cell timing is not required.
A counting-only dSiPM with active recharge offers lower
system power than an analog SiPM (which requires a
front-end amplifier and ADC per channel), while
providing meaningful improvement in dynamic range
through cell recovery during the scintillation pulse.

\subsubsection{Preferred approach: digital SiPM in
counting mode}

The key R\&D items are the development and test beam
validation of a counting-mode dSiPM matched to
PbWO$_4$ scintillation properties: cell pitch, recharge
time, counter depth, and integration window must be
optimized for the FCC-ee energy range. This work should
be performed before committing to the readout technology.

\subsubsection{Fallback: analog SiPM with saturation
correction}

If counting-mode dSiPM proves impractical or
insufficiently mature on the FCC-ee timescale, the
fallback is a conventional analog SiPM. The analog
sensor draws essentially no quiescent power, though the
per-channel front-end electronics (amplifier and
digitizer) located behind the crystal add significant
power --- more than the counting-mode dSiPM alternative.

The saturation response of an analog SiPM is a known
function of $N_{\text{pe}}/N_{\text{cells}}$
(Eq.~\ref{eq:sipm_saturation}), and can be corrected
using the measured signal and a calibration of the
saturation curve. This corrects the mean response; the
residual resolution degradation from variance
amplification (Section~\ref{sec:nonlinearity}) remains,
but is a calculable and bounded effect.

It has been suggested that digitizing the scintillation
waveform and exploiting its time structure could improve
the saturation correction beyond the simple calibration
approach. However, with $\sim$10$^5$ photoelectrons per
nanosecond at the highest energies, the practical benefit 
of waveform-based correction over
simple calibration curve correction remains to be
demonstrated.

A test beam validation analogous to that described for
dSiPM --- measuring energy resolution across the FCC-ee
energy range with saturation correction applied ---
would establish the achievable performance of the analog
approach.

\subsubsection{Operating temperature}

PbWO$_4$ scintillation yield increases significantly at lower
temperatures: roughly $\times$1.5 at 0$^\circ$C and $\times$2.5 at
$-$25$^\circ$C, relative to room temperature. CMS operates at
18$^\circ$C as a compromise.

The physicist noted an ironic trade-off: lower temperature increases
the light yield (improving photostatistics and the stochastic term)
but also increases the number of photons arriving at the SiPM,
\emph{worsening} the saturation problem. Additionally, the
temperature coefficient of PbWO$_4$ light yield is steep
($\sim$$-$2\%/$^\circ$C near room temperature), requiring tight
thermal control ($\sim$0.05$^\circ$C for 0.1\% energy stability).

The optimal operating temperature balances photostatistics against
saturation and thermal stability demands. A moderate temperature
of $\sim$10--15$^\circ$C may be optimal, but this requires
dedicated study with the actual SiPM readout configuration.

\subsection{Crystal ECAL and Particle Flow Compatibility}
\label{sec:evo_ecal_pfa}

A concern with the crystal ECAL choice is its compatibility with
particle flow reconstruction, which has been the dominant paradigm
for $e^+e^-$ Higgs factory detectors. The AI initially expressed
skepticism about this compatibility, but the subsequent discussion
substantially revised this assessment.

\subsubsection{The initial concern}

The worry was that crystal granularity
(20$\times$20~mm$^2$ crystal face) is much coarser than the
Si-W pads (5$\times$5~mm$^2$) for which particle flow algorithms
were optimized. With only $\sim$1 crystal across a Moli\`ere radius
(22~mm), compared to $\sim$2 pads across the tungsten Moli\`ere
radius (9~mm), the ability to resolve nearby showers from different
particles appeared significantly compromised.

Furthermore, the crystal ECAL provides only two longitudinal
segments (front and back), compared to the $\sim$30 longitudinal
layers available in a Si-W ECAL. The 3D shower imaging capability
seemed fundamentally different.

\subsubsection{The reassessment}

Closer examination revealed that the concern, while not unfounded,
was overstated in some respects and genuine in others.

\textbf{What matters in the PFA context.} Particle flow jet energy
reconstruction uses tracker momentum for charged particles, ECAL
energy for photons, and HCAL energy for remaining neutral hadrons.
The ECAL's role is to provide the best possible measurement of the
photon component, which constitutes $\sim$25\% of jet energy. Any
ECAL deep enough to contain electromagnetic showers fulfills this
role; the crystal and Si-W options are equivalent in this regard.
The relevant question is how the crystal's superior photon energy
resolution trades off against Si-W's finer granularity and
longitudinal segmentation for resolving nearby clusters and for
identifying charged particle deposits.

\textbf{Granularity relative to shower width.} The relevant
comparison is not raw cell size but cell size relative to shower
width. Si-W has $\sim$1.8 cells per Moli\`ere radius; the crystal
has $\sim$1.1. The difference is real but less dramatic than the
raw 5~mm vs.\ 20~mm ratio suggests, because the shower widths scale
differently in the two materials.

\textbf{Position resolution through energy sharing.} In a crystal
calorimeter, the position of an electromagnetic shower is
reconstructed from the energy sharing among neighboring crystals.
The position resolution is approximately:
\begin{equation}
\sigma_{\text{pos}} \sim
\frac{R_M}{\sqrt{E/\text{GeV}}}
\sim \frac{22}{\sqrt{10}} \approx 7~\text{mm for a 10~GeV photon}
\end{equation}
This is comparable to the Si-W pad size, showing that the crystal
ECAL achieves good single-shower position resolution through energy
sharing even though its cells are large. However, position
resolution for a single isolated shower is distinct from the ability
to resolve two nearby overlapping showers, which is the more
relevant capability for particle flow. For two-shower separation,
the cell size relative to the shower width is the more directly
relevant quantity, where Si-W retains an advantage.

\textbf{Close pairs are rare.} At Z-pole jet energies
($\sim$45~GeV), the typical particle separation at the ECAL face
is $\sim$200~mm --- roughly 9 Moli\`ere radii, easily resolved in
any calorimeter. Only a few percent of photon-hadron pairs are
close enough for granularity to matter, and the magnetic field
bends charged particles away from neutrals, further reducing the
overlap.

\textbf{Photon energy resolution directly improves jets.} Photons
constitute $\sim$25\% of jet energy. The crystal's substantially
better photon energy resolution directly improves this component.
For a 10~GeV photon, the crystal gives $\sigma_E \sim 95$~MeV
vs.\ $\sim$350~MeV for Si-W --- a significant improvement that
partially compensates any loss from coarser granularity.

\textbf{Longitudinal segmentation: an honest limitation.} The
two-segment crystal design provides far less longitudinal
information than the $\sim$30 layers of a Si-W ECAL. This is a
genuine limitation for particle flow reconstruction. In PFA, the
ECAL measures not only photons but also receives energy deposits
from all charged particles --- electrons, muons, and charged
hadrons --- whose contributions must be identified and subtracted
so that the remaining energy accurately represents the neutral
(photon) component. Fine longitudinal and transverse segmentation
aids this subtraction by providing geometric discrimination between
a tracked particle's deposit and any overlapping neutral energy.
For example, a muon traversing a crystal deposits MIP energy that
cannot easily be disentangled from photon energy in the same
crystal; in Si-W, the muon's narrow trail across $\sim$30 small
pads is geometrically distinct from a photon shower. This is a
genuine advantage of Si-W that the crystal option cannot match
with only two longitudinal segments.

\subsubsection{The role of the magnetic field}

The particle flow jet energy resolution numbers commonly quoted in
the literature (e.g., $\sim$3--4\% at 45~GeV from ILD studies)
assumed a 3.5~T magnetic field and Si-W ECAL. These numbers cannot
be directly applied to a detector with 2~T field and crystal ECAL.

The magnetic field strength --- which determines how effectively
charged particles are separated from neutrals at the calorimeter
face --- is a significant driver of PFA performance. At the 2~T
field adopted for this design, the charged-neutral separation is
reduced compared to the 3.5~T assumed in ILD studies, and this
affects PFA performance regardless of the ECAL technology choice.
The ECAL technology and the magnetic field strength both contribute
to PFA performance, and the two effects should not be conflated
when comparing detector concepts.

\subsubsection{Revised assessment}

The crystal ECAL's compatibility with particle flow reconstruction
is a mixed picture:

\begin{itemize}[itemsep=2pt]
\item \textbf{Advantages:} substantially better photon energy
resolution, improving the $\sim$25\% photon component of jet
energy; adequate single-shower position resolution through energy
sharing; sufficient granularity for the majority of particle
separations at Z-pole jet energies.

\item \textbf{Limitations:} coarser transverse granularity reduces
two-shower separation capability for the small fraction of close
pairs; far less longitudinal segmentation limits the ability to
geometrically identify and subtract charged particle deposits from
overlapping neutral energy.
\end{itemize}

The crystal ECAL is not optimal for particle flow in the same way
that Si-W was specifically designed for it. The longitudinal
segmentation limitation, in particular, is genuine and cannot be
fully mitigated by the two-segment design. However, the crystal's
advantages in photon energy resolution, absence of cryostat dead
material, and operational simplicity are significant, and the
granularity concern is less severe than the raw cell size
comparison suggests. A definitive comparison would require full
simulation of jet reconstruction with both ECAL options in the
same detector geometry and magnetic field --- a study that has not
yet been performed and that this report recommends.

\subsection{Summary of ECAL Evolution}
\label{sec:evo_ecal_summary}

The electromagnetic calorimeter choice evolved through several
stages of physicist-AI discussion:
\begin{enumerate}[itemsep=2pt]
\item \textbf{LAr $\to$ LKr:} Liquid krypton preferred over
  liquid argon on physics grounds (higher sampling fraction,
  better resolution, smaller Moli\`ere radius), with cost as
  the main counterargument.

\item \textbf{Noble liquid $\to$ crystal:} The cryostat dead
  material ($\sim$10--15\%~$X_0$) degrades the noble liquid
  calorimeter performance through multiple mechanisms:
  non-Gaussian tails from photon conversions ($\sim$8--11\%
  conversion probability), constant term contributions from
  cryostat non-uniformities, and position resolution
  degradation from pair spreading in the magnetic field.
  Crystals achieve $\sim$3\%/$\sqrt{E}$ with negligible dead
  material upstream. The benign radiation environment at
  FCC-ee removes the historical weakness of crystals, and the
  abundant physics samples at all running points provide
  powerful calibration capability.

\item \textbf{Longitudinal segmentation:} A two-segment crystal
  design (front $\sim$4--6~$X_0$, back $\sim$16--18~$X_0$) with
  front and back SiPM readout providing longitudinal shower
  profile information to partially address a known limitation
  of crystal calorimeters.

\item \textbf{Dual readout in ECAL not adopted:} While the
  technique would genuinely help for hadrons interacting in the
  ECAL ($\sim$1~$\lambda_I$ depth), for electromagnetic showers
  the separate \v{C}erenkov measurement is redundant ($f_{em} = 1$
  always) and the optical filter degrades the scintillation
  light collection. The cost to the primary EM measurement
  outweighs the secondary hadronic benefit.

\item \textbf{Readout approach:} Counting-mode digital SiPM
  with active quenching and recharge preferred over analog
  SiPM, offering lower system power and improved dynamic
  range through cell recovery during the scintillation pulse.
  Analog SiPM with saturation curve correction is maintained
  as a fallback. Both options require test beam validation.

\item \textbf{PFA compatibility:} The crystal ECAL is not
  optimal for particle flow in the way Si-W was designed for
  it --- the coarser granularity and limited longitudinal
  segmentation are genuine limitations for identifying and
  subtracting charged particle energy deposits. However, the
  substantially better photon energy resolution partially
  compensates, and the granularity concern is less severe
  than the raw cell size comparison suggests. A full
  simulation comparison is recommended.
\end{enumerate}

This sequence illustrates a recurring theme of this study:
practical considerations (cryostat dead material, radiation
environment, power management) can decisively shift the
technology choice away from what appears optimal in the abstract.

\section{Solenoid}
\label{sec:evo_solenoid}

The solenoid underwent two significant changes from CL1a,
both emerging from the physicist's questions about cost,
accelerator coupling, and the magnetic flux return path.
First, the field strength evolved from a fixed 2~T to a
variable 1.5--3~T range matched to the beam energy at
each operating point. Second, the placement moved from
outside the HCAL to between the ECAL and HCAL, with
significant consequences for the overall detector layout.
The field strength is discussed in this subsection; the
placement is discussed in Section~\ref{sec:solenoid_placement}.

\subsection{Field Strength and Accelerator Coupling}
\label{sec:evo_variable_field}

\subsubsection{The beam-solenoid interaction}

The solenoid field at the interaction point couples to
the circulating beams, and this coupling must be
compensated by anti-solenoid coils or corrector magnets
in the interaction region. The compensation becomes more
demanding at lower beam energies. FCC-ee accelerator
studies indicate that solenoid fields above $\sim$2~T
become increasingly difficult to compensate at Z-pole
energies ($E_{\text{beam}} = 45.6$~GeV), while fields up
to 3~T are manageable at the $t\bar{t}$ threshold
($E_{\text{beam}} = 175$~GeV).

\subsubsection{Variable field strategy}

This accelerator constraint aligns naturally with the
physics requirements: lower-energy running (Z~pole)
involves softer jets where particle flow works well even
at moderate field, while higher-energy running
($t\bar{t}$) produces more collimated jets that benefit
most from strong magnetic separation. The proposed
operating scheme is:

\begin{table}[htbp]
\centering
\caption{Variable solenoid field strategy across the
FCC-ee energy points.}
\label{tab:variable_field}
\begin{tabular}{llll}
\toprule
Energy point & $E_{\text{beam}}$ (GeV) & $B$ (T) &
Rationale \\
\midrule
Z pole      & 45.6 & 1.5--2.0 & Accelerator
compatibility most demanding \\
\addlinespace
WW threshold & 80  & 2.0--2.5 & Relaxed accelerator
constraint \\
\addlinespace
ZH          & 120  & 2.5--3.0 & Best momentum resolution
for Higgs recoil \\
\addlinespace
$t\bar{t}$  & 175  & 3.0      & Maximum field;
highest-energy jets \\
\bottomrule
\end{tabular}
\end{table}

The solenoid is designed for the maximum field of 3~T but
operated at reduced current for lower-energy running.
This is technically straightforward --- a superconducting
solenoid designed for 3~T operates safely at any lower
field. The design choice to support 3~T rather than a
fixed 2~T carries costs: a thicker coil (more conductor
and structural material for the higher magnetic pressure,
which scales as $B^2$) increases both the construction
cost and the material budget before the HCAL. These are
meaningful but justified by the physics benefit of
matching the field to the beam energy across the full
FCC-ee program.

The implications for other subsystems are manageable:
\begin{itemize}[itemsep=2pt]
\item Track reconstruction, particle flow clustering,
  and calibrations must be validated across the
  1.5--3~T range.
\item The drift chamber is the most sensitive subsystem
  to field variations: the Lorentz angle affects drift
  paths over the $\sim$cm-scale drift distances, curving
  trajectories that are straight in the absence of a
  magnetic field. Cell geometry can be designed to
  optimize drift paths for a given field strength and
  gas mixture, but this optimization is valid for only
  one operating point. Operating over a 1.5--3~T range
  requires a cell design and gas choice that perform
  adequately across the full range, accepting some
  compromise at each individual field setting. This is
  a known and understood challenge with established
  mitigation strategies.
\item The track momentum resolution scales as $1/B$,
  so Z-pole running at 2~T has 50\% worse momentum
  resolution than 3~T operation. This is acceptable
  because the Z~pole physics does not require the
  extreme momentum resolution needed for the Higgs
  recoil measurement at 240~GeV.
\end{itemize}

\subsection{Solenoid Placement: Inside the HCAL}
\label{sec:solenoid_placement}

CL1a placed the solenoid outside the HCAL to avoid introducing dead
material between the ECAL and HCAL. The physicist challenged this
with a series of arguments about cost, magnetic flux return, and
the interplay with other subsystems.

\subsubsection{Cost scaling}

The cost of a superconducting solenoid scales steeply with bore
radius. The key quantities --- conductor volume, stored energy, and
return yoke mass --- all scale as $R^2 L$ or steeper. For the two
placement options:

\begin{table}[htbp]
\centering
\caption{Solenoid parameters for the two placement options,
assuming 3~T design field.}
\label{tab:solenoid_placement}
\begin{tabular}{lSS}
\toprule
Parameter & {Inside HCAL} & {Outside HCAL} \\
          & {($R = 2.5$~m)} & {($R = 4$~m)} \\
\midrule
Bore radius (m) & 2.5 & 4.0 \\
Length (m)       & {5--6} & {7--9} \\
Flux $\Phi$ (Wb) & 59 & 151 \\
Stored energy ratio & 1 & {$\sim$3.7} \\
Yoke mass ratio   & 1 & {$\sim$3.7} \\
Approximate cost (MCHF) & {25--35} & {80--120} \\
\bottomrule
\end{tabular}
\end{table}

The larger solenoid costs roughly \textbf{$\sim$4$\times$ more},
following the $(R_{\text{out}}/R_{\text{in}})^3 \approx
(4/2.5)^3 \approx 4$ scaling identified by the physicist. The cost
difference of $\sim$50--90M~CHF is a substantial fraction of the
total detector budget.

\subsubsection{The HCAL as flux return}

The physicist made a further observation that strengthens the case
for the inside placement: if the HCAL uses steel absorber plates
(as in the scintillator-steel tile design), this steel can carry a
substantial fraction of the magnetic return flux.

For a barrel HCAL with $\sim$40 layers of 20~mm steel plates
spanning from $R \approx 2.5$~m to $R \approx 4.2$~m, the total
steel cross-sectional area available for flux return is of order
15--20~m$^2$. At an average field near saturation
($B_{\text{sat}} \approx 1.6$~T), the HCAL steel can carry a
substantial fraction of the total 59~Wb return flux.

The remainder requires an additional return yoke outside the HCAL,
but this is significantly thinner than would be needed without the
HCAL contribution. Compare this to the solenoid-outside case, where
the full 151~Wb requires a massive dedicated return yoke at
$R \approx 4.5$--6~m, adding several thousand tonnes of iron.

The physicist summarized the argument concisely: with the solenoid
inside, every tonne of steel serves multiple purposes ---
calorimeter absorber, flux return, and muon filter. With the
solenoid outside, a massive dedicated flux return is needed whose
thickness is driven by magnetic requirements that may over- or
under-shoot the muon physics needs.

\subsubsection{The cost of coil dead material}

The price paid for the inside placement is dead material between the
ECAL and HCAL. The solenoid coil --- superconducting cable plus
aluminum stabilizer and structural support --- contributes:

\begin{table}[htbp]
\centering
\caption{Solenoid coil material budget for different design
aggressiveness levels, at 3~T design field and $R = 2.5$~m.}
\label{tab:coil_material}
\begin{tabular}{lSS}
\toprule
Design approach & {Thickness (mm)} & {$x/\lambda_I$} \\
\midrule
Conventional (CMS-like) & 250 & 0.64 \\
Aggressive              & 150 & 0.38 \\
Very aggressive (HTS)   & 100 & 0.26 \\
\bottomrule
\end{tabular}
\end{table}

The target is $\lesssim$150~mm, giving $\sim$0.4~$\lambda_I$ of
dead material. The structural feasibility at 3~T requires careful
engineering: the magnetic pressure is $\sim$3.6~MPa, giving hoop
stress of $\sim$60~MPa in a 150~mm wall at $R = 2.5$~m. This is
within the capability of high-strength aluminum alloy but leaves
modest safety margins. At 2~T (Z-pole operation), the stress drops
to $\sim$27~MPa, well within comfortable limits.

\subsubsection{Impact on hadronic energy measurement}

The coil dead material of $\sim$0.4~$\lambda_I$ means that
$\sim$33\% of hadrons entering the HCAL interact in the coil
first, with the resulting energy deposit unmeasured. This degrades
the hadronic energy resolution, primarily for neutral hadrons whose
energy depends entirely on the calorimetric measurement.

However, in the particle flow framework, charged hadron energies
are measured by the tracker and photon energies by the ECAL. The
neutral hadron component --- which is what the coil dead material
primarily affects --- constitutes only $\sim$10--15\% of jet energy.
Furthermore, the particle flow confusion term dominates jet energy
resolution over the HCAL intrinsic resolution at all FCC-ee
energies (Section~\ref{sec:evo_hcal}).

The coil dead material degrades the neutral hadron energy
resolution, contributing modestly to the jet energy resolution in
quadrature --- a small effect compared to the $\sim$2--3\%
confusion term that dominates.

\subsubsection{Interplay with the HCAL absorber material}

If the HCAL steel serves as flux return, the absorber
\textbf{must be ferromagnetic} --- low-carbon steel or iron. This
rules out non-magnetic absorber materials such as brass (used in
the CMS HCAL) or copper. Standard low-carbon structural steel
(e.g., S235) serves both purposes well, with saturation
magnetization of $\sim$1.6--1.8~T and adequate mechanical
properties for calorimeter construction.

The steel is magnetized at $\sim$1.6~T in operation. Charged shower
particles traversing a 20~mm steel plate experience both multiple
Coulomb scattering (RMS lateral displacement $\sim$1~mm for a
typical 200~MeV shower particle, scaling as $1/p$) and magnetic
bending ($\sim$0.5~mm at the same momentum). Multiple scattering
is the larger effect by roughly a factor of two, and both are small
compared to the characteristic transverse scale of hadronic showers
($\sim$10--20~cm). Neither materially affects the calorimeter
energy response.

\subsection{Field Uniformity}
\label{sec:evo_field_uniformity}

The placement of ferromagnetic steel (the HCAL) immediately
outside the solenoid coil changes the magnetic boundary
conditions and creates some field non-uniformity in the
tracking volume.

The physicist noted that this is not a practical concern.
The tracking volume extends to $R \approx 2000$~mm, while
the solenoid coil is at $R \approx 2500$~mm, with the
crystal ECAL (non-magnetic PbWO$_4$) providing
$\sim$500~mm of separation. Field perturbations from the
magnetized HCAL steel fall off with distance and are
expected to be small at the outer tracker radius. However,
even substantially larger perturbations would not be
problematic: every major collider detector maps its
magnetic field to $\sim$10$^{-4}$ precision before
operation, and track reconstruction integrates the
equations of motion through the measured 3D field map
rather than assuming a uniform field. What matters is not
uniformity but \emph{knowledge} of the field at every
point. A 2~T field with 1\% non-uniformity mapped to
$10^{-4}$ is far more useful for tracking than a
perfectly uniform field known only to 1\%.

Furthermore, at FCC-ee, abundant $Z \to \mu^+\mu^-$
events provide continuous in-situ field calibration
through the known Z~mass constraint.

Field non-uniformity is therefore not a factor in the
solenoid placement decision.

\subsection{Summary of Solenoid Evolution}
\label{sec:evo_solenoid_summary}

The solenoid design evolved in two important ways from
CL1a:
\begin{enumerate}[itemsep=2pt]
\item \textbf{Fixed 2~T $\to$ variable 1.5--3~T.} The
  field strength is matched to the beam energy at each
  operating point, naturally aligning the physics needs
  (higher field benefits higher-energy jets) with the
  accelerator constraints (higher beam energy tolerates
  more solenoid field). The solenoid is designed for 3~T
  maximum but operated at reduced current for
  lower-energy running.
\item \textbf{Outside HCAL $\to$ inside HCAL.} The
  inside placement reduces the solenoid cost by
  $\sim$4$\times$, enables the HCAL steel to serve
  double duty as flux return (substantially reducing the
  dedicated return yoke mass and cost), and allows the
  muon absorber to be optimized for physics rather than
  dictated by magnetic requirements. The cost is
  $\sim$0.4~$\lambda_I$ of dead material between ECAL
  and HCAL, which has a modest impact on jet energy
  resolution: in the particle flow framework, the coil
  material primarily affects the neutral hadron component
  ($\sim$10--15\% of jet energy), and the confusion term
  dominates the jet energy resolution at all FCC-ee
  energies.
\end{enumerate}

The solenoid discussion illustrates how cost and
engineering considerations can shift a design choice that
initially appears to be driven by physics alone. The
``obvious'' choice of placing the solenoid outside the
HCAL for uninterrupted calorimetry does not survive
scrutiny of the full cost-performance trade-off.

\section{Hadron Calorimeter}
\label{sec:evo_hcal}

\subsection{The Problem and the Particle Flow Solution}
\label{sec:evo_hcal_problem}

\subsubsection{The standalone problem}

Hadronic energy resolution is fundamentally limited by
event-by-event fluctuations in the electromagnetic fraction
$f_{\text{em}}$ of hadronic showers. In each nuclear interaction
within the shower, a fraction of the energy goes into $\pi^0$
production, while the remainder goes into charged hadron cascades,
nuclear breakup, and binding energy losses. The ratio of
electromagnetic to hadronic response ($e/h$) is typically 1.2--1.5
for non-compensating calorimeters, so fluctuations in
$f_{\text{em}}$ translate directly into energy resolution
degradation and non-Gaussian response tails.

\subsubsection{The particle flow approach to jet energy}

The physicist pointed out that for jet energy measurement, the
particle flow algorithm (PFA) largely avoids this problem by using
the best-suited detector component for each particle type:
\begin{itemize}[itemsep=2pt]
\item Charged particles ($\sim$65\% of jet energy) are measured
  by the tracker with excellent resolution
  ($\sigma/E \sim 10^{-3}$).
\item Photons and $\pi^0$s ($\sim$25\%) are measured by the ECAL.
\item Only neutral hadrons ($\sim$10\%) genuinely rely on the HCAL.
\end{itemize}

Since only $\sim$10\% of jet energy depends on the HCAL
measurement, even moderate intrinsic HCAL resolution has limited
impact on overall jet energy resolution. On the other hand, the
HCAL in a particle flow detector has the additional role to
\textbf{identify and locate} neutral hadron showers separately
from charged hadron showers, enabling correct energy assignment.

A known limitation of the PFA approach is that performance
degrades at higher jet energies, as showers become more collimated
and the confusion between charged and neutral deposits increases.
Published studies show jet resolution degrading from $\sim$3\% at
45~GeV to $\sim$5--6\% at 175~GeV, though these specific numbers
assume a 3.5~T field and Si-W ECAL and are not directly applicable
to CL2a.

This framing applies to any detector with a deep ECAL that
contains electromagnetic showers. The hadronic showers within the
HCAL still contain significant electromagnetic content from
$\pi^0$ production in the cascade; the PFA simplification is that
\emph{incoming} photons have already been measured by the ECAL.

\subsubsection{Two approaches to the HCAL}

The two HCAL technologies under consideration address the
remaining task through fundamentally different strategies:
the scintillator-steel tile calorimeter emphasizes high
granularity for pattern recognition and shower separation,
while the dual-readout fiber calorimeter attacks the
$f_{\text{em}}$ fluctuation problem directly to achieve
superior standalone hadron resolution.

\subsection{Scintillator-Steel Tile Calorimeter}
\label{sec:evo_tile_hcal}

\subsubsection{Design}

The particle flow approach drives the tile HCAL design
toward high granularity for pattern recognition and
shower separation:
\begin{itemize}[itemsep=2pt]
\item Steel absorber plates ($\sim$20~mm) interleaved
  with scintillator tiles ($\sim$3~mm thick,
  $\sim$30$\times$30~mm$^2$ area).
\item Each tile read out by an individual SiPM.
\item $\sim$50 layers for $\sim$6~$\lambda_I$ depth.
\item Total channel count: $\sim$8 million for a full
  4$\pi$ HCAL.
\end{itemize}

\subsubsection{Strengths}

\begin{itemize}[itemsep=2pt]
\item \textbf{Proven particle flow performance.}
  Extensively simulated for ILC (ILD, SiD) and FCC-ee
  (CLD), with jet energy resolution of $\sim$3--4\%
  demonstrated at Z-pole jet energies in full simulation
  with PandoraPFA reconstruction.

\item \textbf{3D shower imaging.} The fine granularity
  (30$\times$30~mm$^2$ tiles in $\sim$50 longitudinal
  layers) provides detailed shower topology. This is
  valuable beyond jet energy measurement: $\tau$
  reconstruction, missing energy vetoes, muon
  identification through minimum-ionizing signatures,
  and software compensation using shower shape
  information.

\item \textbf{Mature technology.} The CALICE
  collaboration has built and tested large prototypes
  in beam tests~\cite{calice_ahcal}. SiPM mass
  production is established. Tile production and
  assembly have been industrialized.

\item \textbf{Straightforward calibration.} Each
  tile/SiPM combination is calibrated with minimum
  ionizing particles (muons traversing the HCAL). The
  MIP calibration is robust, well-understood, and
  transfers reliably from test beam to in-situ
  operation.

\item \textbf{Compatibility with magnetized steel.}
  With the solenoid inside the HCAL
  (Section~\ref{sec:solenoid_placement}), the steel
  absorber carries return flux and is magnetized at
  $\sim$1.5--1.6~T. This is fully compatible with
  scintillator tile readout: SiPMs are insensitive to
  magnetic field, and the magnetized steel has
  negligible effect on hadronic shower development.
\end{itemize}

\subsubsection{Limitations}

\begin{itemize}[itemsep=2pt]
\item \textbf{Non-compensating.} The intrinsic resolution
  for neutral hadrons is $\sigma(E)/E \approx
  55$--60\%/$\sqrt{E}$, limited by the fundamental
  $e/h \neq 1$ of a steel-scintillator sampling
  calorimeter. The particle flow approach bypasses rather than solves
this problem. In kinematic regions where charged-neutral
separation is less effective, the jet energy measurement
depends more heavily on this raw non-compensating
response.

\item \textbf{Large channel count.} $\sim$8 million
  SiPM channels require massive quality
  control, and calibration effort. SiPM gain drift and
  dark count evolution over 15~years demand continuous
  monitoring.
\end{itemize}

\subsection{Dual-Readout Fiber Calorimeter}
\label{sec:evo_dr_hcal}

\subsubsection{Design}

The dual-readout approach attacks the $f_{\text{em}}$ fluctuation
problem directly. Two types of fibers are interleaved in a copper
or brass absorber matrix:
\begin{itemize}[itemsep=2pt]
\item \textbf{Scintillating fibers (S):} respond to all ionizing
  energy deposition, giving a signal proportional to total deposited
  energy with $e/h \neq 1$.
\item \textbf{Clear (\v{C}erenkov) fibers (C):} respond only to
  relativistic charged particles ($\beta > 1/n$), predominantly the
  $e^+e^-$ from the electromagnetic shower component, giving a
  signal approximately proportional to $f_{\text{em}}$.
\end{itemize}

From the two signals, $f_{\text{em}}$ can be extracted event by
event and the energy corrected:
\begin{equation}
E_{\text{corrected}} = \frac{S - \chi C}{1 - \chi}
\end{equation}
where $\chi$ is a calibration constant determined by the $e/h$
ratios for the two signal types.

RD52 test beam results~\cite{rd52_results} have demonstrated
hadronic energy resolution of $\sigma(E)/E \approx
30$--35\%/$\sqrt{E}$ after dual-readout correction, with more
Gaussian response than conventional non-compensating calorimeters.

\subsubsection{Strengths}

\begin{itemize}[itemsep=2pt]
\item \textbf{Addresses the fundamental physics problem.} The
  $f_{\text{em}}$ correction removes the dominant source of
  hadronic resolution degradation, achieving effective
  compensation event by event.

\item \textbf{Superior standalone hadron resolution.}
  $\sim$30--35\%/$\sqrt{E}$ after correction, compared to
  $\sim$55--60\%/$\sqrt{E}$ for the tile calorimeter.

\item \textbf{More Gaussian response.} The correction removes the
  non-Gaussian tails caused by $f_{\text{em}}$ fluctuations, which
  is important for avoiding systematic biases in mass
  reconstruction and jet energy measurements.

\item \textbf{Robust in challenging kinematic regions.} In
  conditions where particle flow performance is degraded, the
  dual-readout calorimeter provides a good standalone energy
  measurement --- a safety net that the tile approach lacks.

\item \textbf{Fewer channels.} With readout segmented into towers
  of $\sim$20--30~mm laterally and two channels per tower (S and
  C), the total channel count is $\sim$80,000 --- two orders of
  magnitude fewer than the tile approach.
\end{itemize}

\subsubsection{Limitations}

\begin{itemize}[itemsep=2pt]
\item \textbf{Limited granularity.} The effective transverse
  granularity is set by the tower size ($\sim$20--30~mm), and the
  standard design has \textbf{no longitudinal segmentation} ---
  fibers run the full depth with readout only at the back. This
  limits 3D shower imaging capability and makes it harder to
  separate overlapping showers, identify shower start points, and
  track muons through the calorimeter.

\item \textbf{\v{C}erenkov light yield is low.} Clear fibers
  collect only \v{C}erenkov photons, which are much less abundant
  than scintillation photons. The \v{C}erenkov channel has worse
  photostatistics, limiting the precision of the $f_{\text{em}}$
  measurement. This is particularly important at low hadron
  energies (few GeV) where the \v{C}erenkov signal may be too
  small for a reliable correction, and the dual-readout technique
  provides diminishing benefit.

\item \textbf{Technology maturity.} While RD52 has demonstrated
  the principle convincingly in test beams, significant engineering
  challenges remain for a full collider detector. The most serious
  is fiber routing: in test beam configurations, scintillating and
  clear fibers extend $\sim$1~m beyond the absorber to be grouped
  onto separate photodetectors. Scaling this approach to a full
  $4\pi$ detector with realistic space constraints is an unsolved
  problem. Assembly of the fiber/absorber matrix at the required
  scale and the long-term mechanical integrity of fibers under
  operational stresses are additional concerns.

\item \textbf{ECAL-HCAL shower splitting.} With a crystal ECAL
  in front, hadrons that begin showering in the ECAL have their
  energy split between two very different detector technologies.
  The dual-readout correction in the HCAL applies only to the HCAL
  portion of the shower, not the total. For neutral hadrons, the
  nuclear interaction length of PbWO$_4$ is $\sim$22~cm ---
  comparable to the crystal depth of $\sim$20~cm --- so
  $\sim$60\% of neutral hadrons interact in the ECAL, creating
  split showers for which the dual-readout correction is
  incomplete.
\end{itemize}

\subsection{Combining High Granularity and Dual Readout}
\label{sec:evo_hcal_combination}

The tile and fiber approaches each have distinct strengths ---
high granularity for pattern recognition, and dual readout for
standalone hadron resolution. It is natural to ask whether these
can be combined in a single detector.

\subsubsection{Adding dual-readout capability to
scintillator-steel tiles}

Each scintillator tile in the HCAL is replaced by a pair of
optically separate tiles: a front scintillator tile (producing
the S signal from all ionizing energy deposition) and a back
clear tile (producing only \v{C}erenkov light, predominantly
from the electromagnetic shower component). This achieves the
dual-readout S/C measurement within the existing high-granularity
tile geometry.

There are challenges:
\begin{itemize}[itemsep=2pt]
\item \textbf{Doubled channel count.} The number of readout
  channels increases from $\sim$8~million to $\sim$16~million,
  with corresponding increases in SiPM procurement, quality
  control, and calibration effort.
\item \textbf{\v{C}erenkov light yield.} The photostatistics
  from a single clear tile of $\sim$3~mm thickness are low.
  However, since the \v{C}erenkov light is detected by a
  separate photosensor from the scintillation light, that sensor
  can be independently optimized for this low-light regime
  (e.g., larger area, higher gain, or cooled operation).
\end{itemize}

\subsubsection{Increasing granularity in the dual-readout fiber
calorimeter}

Greater physical segmentation --- both transverse (smaller towers)
and longitudinal (multiple readout depths) --- could in principle
bring the fiber calorimeter's pattern recognition closer to that
of the tile approach. The challenges are primarily practical:
\begin{itemize}[itemsep=2pt]
\item \textbf{Higher channel count.} Achieving granularity
  comparable to the tile calorimeter implies a similar total
  channel count, eliminating one of the fiber approach's
  principal advantages.
\item \textbf{Increased complexity.} The already-unsolved fiber
  routing problem becomes substantially harder with longitudinal
  segmentation, as fibers must be extracted and grouped at
  multiple depths rather than only at the back.
\end{itemize}

An alternative is virtual longitudinal segmentation 
(and without improving transverse segmentation) through
precision timing of photon arrival at the photosensor: photons
produced at different depths travel different distances along the
fiber and arrive at slightly different times. However, this
approach faces fundamental limitations.

For scintillation light, whose emission time constants
($\sim$few~ns or longer) are comparable to or larger than the
photon transit time along the entire fiber, the measured arrival times
reflect primarily the scintillator decay characteristics rather than the
emission depth. Virtual segmentation cannot be applied to the
scintillation channel. This is a significant drawback because 
scintillation signal dominates over \v{C}erenkov signal.  

For the \v{C}erenkov channel, light emission is prompt,
but the time measured at the photosensor carries no
direct information on shower depth without additional
assumptions. The measured time is:
\begin{equation}
t_{\text{meas}} = t_{\text{emit}} + \frac{L}{v}
\end{equation}
where $t_{\text{emit}}$ is the time at which the charged
particle produced \v{C}erenkov light at a point inside
the calorimeter, $L$ is the distance from that point to
the photosensor, and $v$ is the speed of light
propagation in the fiber. The depth information is
contained in $L$, but extracting it requires knowledge
of $t_{\text{emit}}$ --- and this is a single equation
with two unknowns. Additional constraints must be
applied: assumptions about when the initiating particle
entered the calorimeter, how quickly the shower
propagates in depth, and at what depth the \v{C}erenkov
emission occurred. For an isolated single hadron entering
at a known time, these assumptions may be tractable. For
overlapping showers or secondary particles produced deep
in the cascade, they are not. It is essential to
understand the influence of such assumptions on the
extracted depth information and its associated
uncertainty.

\subsubsection{Assessment}

The benefits of combining high granularity with dual readout are
attractive, and simulation studies should be carried out to
quantify the potential gains. Of the approaches considered:

\begin{itemize}[itemsep=2pt]
\item \textbf{Adding dual readout to scintillator-steel tiles}
  appears most promising: the channel count increase (factor of 2)
  is bounded and technically manageable, the high-granularity
  pattern recognition is preserved, and the dual-readout correction
  addresses the standalone resolution limitation directly.

\item \textbf{Physical segmentation of the fiber calorimeter} is
  technically more challenging (fiber routing at multiple depths)
  and undermines the approach's inherent simplicity
  advantage. Approaching the granularity achievable with
  scintillator-steel is challenging.

\item \textbf{Virtual depth segmentation through timing} in the
  fiber calorimeter is limited by the fundamental mismatch between
  scintillator decay times and fiber transit times. It could
  provide depth information for the \v{C}erenkov channel only,
  which gives the longitudinal profile of the electromagnetic
  shower component but not the hadronic component --- precisely
  the part where depth information would be most valuable for
  pattern recognition.
\end{itemize}

\subsection{Quantitative Impact on Jet Energy Resolution}
\label{sec:evo_hcal_quantitative}

\subsubsection{Particle flow performance with crystal ECAL}

The commonly quoted PFA jet energy resolution numbers
($\sim$3--4\% at 45~GeV) originate from ILD studies that
assumed a 3.5~T magnetic field and a Si-W ECAL with
5$\times$5~mm$^2$ pads and $\sim$30 longitudinal layers.
These assumptions differ significantly from the CL2a
concept (2--3~T field, crystal ECAL with
20$\times$20~mm$^2$ granularity and 2 longitudinal
segments).

The PFA jet energy resolution can be approximately
decomposed as (treating each component as independent,
which is illustrative rather than rigorous):
\begin{equation}
\left(\frac{\sigma_E}{E}\right)^2_{\text{jet}} \approx
\left(\frac{0.65 \cdot \sigma_p/p}
{\text{tracker}}\right)^2 +
\left(\frac{0.25 \cdot \sigma_E^\gamma/E^\gamma}
{\text{ECAL}}\right)^2 +
\left(\frac{0.10 \cdot \sigma_E^{nh}/E^{nh}}
{\text{HCAL}}\right)^2 +
\sigma^2_{\text{confusion}}
\label{eq:pfa_decomposition}
\end{equation}

For a 45~GeV jet (average photon energy $\sim$5~GeV,
average neutral hadron energy $\sim$4.5~GeV):

\begin{table}[htbp]
\centering
\caption{Estimated contributions to jet energy resolution
at 45~GeV, comparing the tile and dual-readout HCAL
options with a crystal ECAL. The confusion term is
extrapolated from published ILD studies at 3.5~T and has
not been simulated for the CL2a configuration. The
tracker term is negligible in both cases.}
\label{tab:jet_resolution}
\begin{tabular}{lSS}
\toprule
Component & {Tile HCAL (\%)} &
{Dual-readout HCAL (\%)} \\
\midrule
Photon term (crystal ECAL) & 0.34 & 0.34 \\
Neutral hadron term & 2.6 & 1.4 \\
Confusion term (estimated) & {2--3} & {2--3} \\
\midrule
\textbf{Total (quadrature)} & {$\sim$3.6} &
{$\sim$2.9} \\
\bottomrule
\end{tabular}
\end{table}

The dual-readout HCAL improves the neutral hadron
contribution from $\sim$2.6\% to $\sim$1.4\%, yielding
a total jet resolution improvement of $\sim$0.7
percentage points. This is meaningful but not dominant
--- the \textbf{confusion term controls the overall jet
resolution} regardless of HCAL technology.

\subsubsection{The magnetic field as a significant driver}

The confusion term is driven significantly by the
magnetic field strength, which determines how effectively
charged particles are separated from neutrals at the
calorimeter face. The variable field strategy
(Section~\ref{sec:evo_variable_field}) --- running at
higher field for higher-energy physics --- addresses the
confusion term directly, independent of the HCAL
technology choice.

\subsection{The HCAL as Flux Return}
\label{sec:evo_hcal_flux}

With the solenoid placed inside the HCAL
(Section~\ref{sec:solenoid_placement}), the HCAL steel
absorber serves as partial magnetic flux return, carrying
a substantial fraction of the return flux. This has
specific implications for the HCAL technology choice:

\begin{itemize}[itemsep=2pt]
\item The absorber \textbf{must be ferromagnetic}
  (steel/iron), ruling out non-magnetic materials such
  as brass or copper. Standard low-carbon structural
  steel works well for both calorimetry and magnetics.

\item The steel is magnetized at $\sim$1.5--1.6~T, which
  is fully compatible with scintillator tile readout
  (SiPMs are field-insensitive) and has negligible
  impact on shower physics.

\item For the dual-readout fiber option, the absorber is
  typically copper or brass in the RD52 design. Using
  magnetized steel would require redesigning the
  fiber/absorber geometry. While not impossible in
  principle, this has not been studied and represents an
  additional engineering challenge.
\end{itemize}

The flux return consideration provides a mild preference
for the tile approach, which naturally uses steel
absorber, over the dual-readout approach, which would
need to adapt its absorber material.

\subsection{Technology Assessment}
\label{sec:evo_hcal_assessment}

The choice between scintillator-steel tiles and dual-readout
fibers is the most difficult technology decision in the entire
detector design. Neither option is clearly superior; each has
genuine strengths that the other lacks.

The arguments favoring \textbf{scintillator-steel tiles} for the
CL2a baseline are:

\begin{enumerate}[itemsep=2pt]
\item The FCC-ee physics program is dominated by jet measurements
  at 91--240~GeV where particle flow works well and the confusion
  term, rather than HCAL intrinsic resolution, limits performance.

\item The presence of a separate deep ECAL, which measures all
  incoming photons before they reach the HCAL, reduces the cluster
  multiplicity and simplifies the pattern recognition task in the
  HCAL. This benefits the tile approach in particular, though it
  also reduces the demands placed on it.

\item The 3D shower imaging from 50 longitudinal layers of
  30$\times$30~mm$^2$ tiles is valuable for many analyses beyond
  jet energy: $\tau$ reconstruction, missing energy, muon
  identification, and software compensation.

\item The technology is mature, with extensive prototyping and beam
  test validation by the CALICE collaboration.

\item Steel absorber naturally serves as magnetic flux return,
  integrating cleanly with the inside-HCAL solenoid placement.

\item MIP-based calibration is robust and well-understood.
\end{enumerate}

The arguments favoring \textbf{dual-readout fibers} are also
compelling:

\begin{enumerate}[itemsep=2pt]
\item Superior standalone hadron resolution
  ($\sim$30\%/$\sqrt{E}$ vs.\ $\sim$55\%/$\sqrt{E}$) with more
  Gaussian response.

\item Robust performance in kinematic regions where particle flow
  is degraded --- a safety net at the highest jet energies and in
  forward regions.

\item Far fewer readout channels ($\sim$80,000 vs.\
  $\sim$8 million), with corresponding savings in QC and
  calibration effort.

\item Addresses the fundamental $f_{\text{em}}$ fluctuation problem
  rather than bypassing it.
\end{enumerate}

For CL2a, the \textbf{scintillator-steel tile HCAL is adopted} as
the baseline, on the grounds that the particle flow approach is
well-matched to the dominant Z-pole and ZH physics programs, the
technology is mature, and the steel absorber integrates naturally
with the solenoid flux return.

However, the dual-readout fiber calorimeter is \textbf{strongly
recommended for the complementary detector CL2b}. This provides:
\begin{itemize}[itemsep=2pt]
\item Independent systematic checks on all jet energy measurements,
  with fundamentally different calorimetric approaches.
\item A direct experimental comparison of particle flow vs.\
  dual-readout at a Higgs factory --- data that would be invaluable
  for the design of future collider detectors.
\item Coverage of the high-energy regime ($t\bar{t}$ threshold)
  where dual-readout may outperform particle flow.
\item Risk mitigation: if either approach has unforeseen limitations
  in operation, the other provides a fallback.
\end{itemize}

This is perhaps the strongest case for the two-detector
complementarity strategy advocated throughout this report. The tile
vs.\ dual-readout question may ultimately be answerable only with
collider data, and having both technologies deployed simultaneously
would settle it definitively.

\subsection{Summary of HCAL Discussion}
\label{sec:evo_hcal_summary}

The HCAL technology choice did not change from CL1a to CL2a ---
scintillator-steel tiles remain the baseline. However, the
\emph{reasoning} evolved substantially through the discussion:

\begin{enumerate}[itemsep=2pt]
\item The presence of a separate deep ECAL was recognized as
  simplifying the HCAL's pattern recognition task by measuring
  incoming photons before they reach the HCAL, reducing both the
  demands on the HCAL and the performance difference between the
  two technologies.

\item The confusion term was identified as the dominant limitation
  on jet energy resolution, making the HCAL intrinsic resolution
  (and hence the tile vs.\ dual-readout difference) a second-order
  effect.

\item The magnetic field strength was recognized as a significant
  driver of jet resolution through its effect on the confusion
  term, reinforcing the variable field strategy.

\item The solenoid placement inside the HCAL created a mild
  preference for steel absorber (compatible with flux return),
  favoring the tile approach.

\item The case for two-detector complementarity was strengthened
  --- this is the subsystem where having both technologies deployed
  simultaneously offers the greatest scientific return.
\end{enumerate}

The PFA performance numbers frequently cited in the literature
(from ILD studies at 3.5~T with Si-W ECAL) were found to be
\textbf{not directly applicable} to the CL2a geometry. A proper
evaluation of jet energy resolution for the CL2a concept ---
crystal ECAL, 2--3~T variable field, scintillator-steel HCAL ---
requires dedicated full simulation, which is identified as a
priority R\&D activity.

\section{Muon System}
\label{sec:evo_muon}

The muon system underwent a complete technology change from CL1a to
CL2a --- from micro-Resistive WELL ($\mu$-RWELL) gaseous detectors
to scintillator bars with wavelength-shifting fiber and SiPM readout.
This change, prompted by the physicist's challenge, illustrates a
broader lesson about matching technology to actual requirements
rather than selecting the most technically impressive available
option.

\subsection{Requirements at FCC-ee}
\label{sec:evo_muon_requirements}

The starting point is a clear definition of what the muon system
must actually do. At FCC-ee, the muon system is an
\textbf{identifier}, not a \textbf{measurer}: its job is to confirm
that a particle penetrating the full calorimeter and solenoid coil
is a muon, not to independently measure the muon momentum (which
the tracker does far better).

The identification relies on the penetrating nature of muons:
multiple detector stations are interleaved with absorber (iron)
layers, and a particle reaching the outer stations through
$\sim$7--8 nuclear interaction lengths of material is almost
certainly a muon. Multi-station coincidence provides further
rejection of hadronic punch-through. The requirements on each
station are therefore:
\begin{itemize}[itemsep=2pt]
\item High detection efficiency ($>$95\%) for minimum ionizing
  particles.
\item Sufficient position resolution ($\sim$1~cm) for track
  matching.
\item Timing resolution better than $\sim$10~ns for bunch crossing
  identification.
\item Coverage of $\sim$1000--1400~m$^2$ total area across 3--4
  stations.
\item Reliable operation over 15~years with minimal maintenance.
\end{itemize}

\subsubsection{Position resolution}

The position resolution required for track matching is
set by the uncertainty in extrapolating a track from the
tracker through the calorimeter and coil to the
muon stations. This extrapolation uncertainty is
dominated by multiple scattering, giving approximately
$\sim$20~mm for a 10~GeV muon and scaling as $1/p$.

A muon detector with position resolution of $\sim$1~cm
provides a matching window of a few cm$^2$. Whether this
is adequate depends on the background hit rate in the
muon stations --- primarily from machine-related sources
rather than physics. For the background levels expected
at FCC-ee, this is likely sufficient, though the exact
requirement depends on detailed background simulations
that have not yet been performed for this configuration.
The size (and hence resolution) can be adjusted as
a straightforward design optimization once background
levels are better characterized.

What is clear is that the $\sim$100~$\mu$m resolution of
$\mu$-RWELL provides no benefit: even in the most
demanding background scenario, a matching window of
$\sim$1~mm$^2$ at the muon stations is far smaller than
any conceivable need.

\subsubsection{Rate capability and radiation hardness}

At FCC-ee, the muon stations at $R \approx 4$--5~m see hit rates of
$\sim$Hz/cm$^2$ from physics events, with modest additional
contributions from beam backgrounds. This is negligible compared to
the CMS endcap environment where $\mu$-RWELL is being deployed
(MHz/cm$^2$). Similarly, the integrated radiation dose over 15~years
of FCC-ee operation is negligible for any detector technology.

\subsubsection{The mismatch}

\begin{table}[htbp]
\centering
\caption{Muon system requirements at FCC-ee compared to the
capabilities of $\mu$-RWELL, illustrating the mismatch between
the initial technology choice and the actual needs.}
\label{tab:muon_requirements}
\begin{tabular}{llll}
\toprule
Property & FCC-ee need & $\mu$-RWELL & Matched? \\
\midrule
Position resolution & $\sim$1~cm & $\sim$100~$\mu$m &
Overkill \\
Rate capability & $\sim$Hz/cm$^2$ & MHz/cm$^2$ &
Overkill \\
Radiation hardness & Negligible & Excellent & Overkill \\
Detection efficiency & $>$95\% & $>$97\% & Yes \\
Timing & $<$10~ns & $\sim$5~ns & Yes \\
\bottomrule
\end{tabular}
\end{table}

The $\mu$-RWELL meets the efficiency and timing requirements but
massively exceeds what is needed in position resolution, rate
capability, and radiation hardness. This excess capability comes
at a cost in operational complexity (gas system, high voltage,
environmental sensitivity) without providing any physics benefit.

\subsection{Technology Choice: Scintillator Bars}
\label{sec:evo_scintillator_bars}

The physicist proposed scintillator bars with embedded
wavelength-shifting (WLS) fibers and dual-end SiPM readout as a
technology that precisely matches these requirements. This is the
approach being adopted by the Belle~II experiment for its muon
system upgrade~\cite{belle2_muon} and is used in the Mu2e cosmic
ray veto at Fermilab~\cite{mu2e_crv}, among other applications.

\subsubsection{Design}

Each detector element is an extruded plastic scintillator bar:
\begin{itemize}[itemsep=2pt]
\item Typical dimensions: $\sim$30~mm wide $\times$ $\sim$10~mm
  thick $\times$ $\sim$2--4~m long.
\item Co-extruded TiO$_2$ reflective coating for light containment.
\item Embedded wavelength-shifting fiber ($\sim$1~mm diameter,
  Y-11 or similar) running the length of the bar.
\item SiPM readout on both ends of each bar.
\end{itemize}

Each muon station consists of two layers of bars in orthogonal
orientations (X and Y views), providing two-coordinate position
measurement.

\subsubsection{Performance}

\textbf{Detection efficiency.} A minimum-ionizing muon traversing
$\sim$10~mm of plastic scintillator produces $\sim$20--40
photoelectrons at each SiPM (after WLS fiber collection and SiPM
photon detection efficiency). This gives essentially 100\% detection
efficiency with any reasonable threshold. The double-ended readout
provides coincidence-based noise rejection and redundancy --- if one
SiPM fails, the other end still detects the muon.

\textbf{Position resolution.} Transverse to the bar, the resolution
is determined by the bar width:
$\sigma_{\text{trans}} = 30/\sqrt{12} \approx 9$~mm. Along the bar,
the position is obtained from the time difference between the two
SiPM signals. With $\sim$200--300~ps timing per SiPM (achievable
with modern devices) and signal propagation speed of $\sim$17~cm/ns
in the WLS fiber:
\begin{equation}
\sigma_{\text{along}} = \frac{v \times \sigma_{\Delta t}}{\sqrt{2}}
\approx \frac{170 \times 0.3}{\sqrt{2}} \approx 36~\text{mm}
\end{equation}

Both resolutions are well within the $\sim$1--5~cm required by the
track extrapolation uncertainty.

\textbf{Timing.} The coincidence of two SiPM signals gives timing
resolution $\sigma_t \approx \sigma_{t,\text{SiPM}}/\sqrt{2}
\approx 0.7$~ns --- far better than the $\sim$10~ns needed for
bunch crossing identification.

\subsubsection{Operational advantages}

The physicist emphasized operational simplicity as a decisive
criterion for a subsystem covering $>$1000~m$^2$ that must operate
reliably for 15~years:

\begin{table}[htbp]
\centering
\caption{Operational comparison of scintillator bars and gaseous
detectors ($\mu$-RWELL or similar) for the muon system.}
\label{tab:muon_operations}
\begin{tabular}{lll}
\toprule
Aspect & Scintillator bars & Gaseous detectors \\
\midrule
Gas system & None & Required (mixing, \\
           &      & distribution, monitoring) \\
\addlinespace
High voltage & $\sim$30--50~V (SiPM) & $\sim$3--4~kV \\
\addlinespace
Environmental & Low (solid state) & Moderate (gain depends \\
sensitivity   &                   & on $T$, $P$, humidity) \\
\addlinespace
Aging & Minimal (scintillator & Gas aging possible \\
      & stable for decades)   & over 15~years \\
\addlinespace
Maintenance & Individual SiPM & Gas system servicing, \\
            & replacement      & HV monitoring \\
\addlinespace
Commissioning & Straightforward & Gas conditioning, \\
              &                 & HV training \\
\bottomrule
\end{tabular}
\end{table}

The absence of a gas system is perhaps the single most important
operational advantage. A gas system serving $\sim$1000--1400~m$^2$
of gaseous detectors requires continuous gas mixing, distribution,
purity monitoring, flow control, and potentially gas recovery (for
expensive or greenhouse gases). This infrastructure must be
maintained throughout the 15-year program, including during shutdown
periods. A scintillator system has no equivalent requirement ---
once installed, it operates with minimal intervention.

\subsubsection{Cost}

For $\sim$1000--1400~m$^2$ total instrumented area with bars of
$\sim$30~mm width:
\begin{itemize}[itemsep=2pt]
\item Bars per station (two orthogonal layers):
  $\sim$12,000.
\item Total for 3--4 stations: $\sim$36,000--48,000 bars.
\item Cost per bar (scintillator + fiber + 2~SiPMs):
  $\sim$\$15--40.
\item Total bar cost: $\sim$\$1--1.5M.
\item Including electronics, cables, support structures:
  $\sim$\$3--5M total.
\end{itemize}

This is a very modest cost for a major detector subsystem ---
comparable to or less than a gaseous detector system of similar
area, and far less than the multi-million-channel SiPM systems
required for the HCAL.

\subsubsection{Precedents}

The Belle~II experiment is upgrading its muon system
(replacing RPCs that suffered from aging and efficiency loss) with
exactly this technology, in a radiation environment similar to
FCC-ee. Their evaluation of multiple technology options and selection
of scintillator bars validates the choice for similar conditions.

The Mu2e cosmic ray veto demonstrates the technology at large scale,
with experience in industrial scintillator extrusion, mass
production of bars with embedded fibers, SiPM quality control, and
system integration directly applicable to FCC-ee.

\subsection{Integration with Return Yoke}
\label{sec:evo_muon_integration}

The muon detector stations are interleaved with iron absorber layers
in the return yoke outside the HCAL. The total absorber depth before
the outermost muon station determines the hadron punch-through
rejection:

\begin{table}[htbp]
\centering
\caption{Absorber depth and approximate hadron punch-through
probability for the muon system.}
\label{tab:muon_absorber}
\begin{tabular}{lSS}
\toprule
Component & {Depth ($\lambda_I$)} &
{Cumulative ($\lambda_I$)} \\
\midrule
HCAL (steel + scintillator) & 6.0 & 6.0 \\
Solenoid coil & 0.4 & 6.4 \\
Return yoke layers & {1--2} & {7.4--8.4} \\
\midrule
\multicolumn{2}{l}{Punch-through probability at
8~$\lambda_I$:} & {$\sim$0.03\%} \\
\bottomrule
\end{tabular}
\end{table}

With the HCAL providing 6~$\lambda_I$, the solenoid coil adding
$\sim$0.4~$\lambda_I$, and 1--2~$\lambda_I$ of iron in the return
yoke between muon stations, the total absorber depth is
$\sim$7.5--8.5~$\lambda_I$. The hadron punch-through probability at
this depth is $<$0.1\%, ensuring clean muon identification.

As discussed in Section~\ref{sec:solenoid_placement}, with the
solenoid inside the HCAL, the return yoke thickness is driven
primarily by muon physics needs rather than magnetic flux
requirements (the HCAL steel carries a substantial fraction of the
return flux). The yoke iron layers are optimized for muon
identification: thick enough for adequate hadron absorption, thin
enough to keep the overall detector compact, with 3--4 gaps for
muon detector stations at strategic depths.

A typical station arrangement might be:
\begin{itemize}[itemsep=2pt]
\item Station~1: immediately outside the HCAL/coil, after
  $\sim$6.4~$\lambda_I$.
\item Station~2: after $\sim$10~cm additional iron
  ($\sim$0.6~$\lambda_I$ additional).
\item Station~3: after $\sim$10~cm more iron
  ($\sim$1.2~$\lambda_I$ cumulative additional).
\item Station~4 (optional): after final iron layer
  ($\sim$1.8~$\lambda_I$ cumulative additional).
\end{itemize}

The first station is primarily for muon identification (anything
reaching it through 6.4~$\lambda_I$ is very likely a muon). The
outer stations provide redundancy, improved purity through
multi-station coincidence, and a crude standalone momentum
measurement from the deflection in the magnetized iron (useful as a
cross-check, though the tracker measurement is far superior).

\subsection{Summary of Muon System Evolution}
\label{sec:evo_muon_summary}

The muon system change from $\mu$-RWELL to scintillator bars is
perhaps the clearest example in this study of a design choice driven
by properly matching technology to requirements:

\begin{enumerate}[itemsep=2pt]
\item The FCC-ee muon system needs are fundamentally different from
  LHC: identification rather than measurement, in a benign rate and
  radiation environment.

\item The $\mu$-RWELL's impressive capabilities (100~$\mu$m
  resolution, MHz/cm$^2$ rate handling, radiation hardness) are
  entirely wasted at FCC-ee --- the experiment does not need and
  cannot benefit from any of them.

\item Scintillator bars meet every actual requirement (efficiency,
  position resolution, timing) while offering decisive advantages
  in operational simplicity, reliability, and cost.

\item The Belle~II and Mu2e precedents demonstrate the technology
  at the relevant scale and in comparable operating conditions.
\end{enumerate}

The lesson extends beyond the muon system: in detector design,
selecting the most advanced available technology is not always the
best choice. The right technology is the one that meets the actual
requirements with the greatest margin of reliability and operational
simplicity, particularly for subsystems that must function
unattended over a decade or more.

\section{Luminosity Monitor}
\label{sec:evo_luminosity}

The luminosity monitor technology --- a precision silicon-tungsten
sampling calorimeter --- did not change from CL1a to CL2a. However,
the discussion substantially deepened the understanding of the
requirements, revealing that the principal challenges are
metrological and operational rather than detector-physical. The
physicist's questions exposed a circular logic in one proposed
monitoring approach and clarified the distinct requirements for
absolute and relative luminosity determination.

\subsection{The Measurement Method}
\label{sec:evo_lumi_method}

Luminosity at FCC-ee is measured by counting small-angle Bhabha
scattering ($e^+e^- \to e^+e^-$) events within a precisely defined
angular acceptance:
\begin{equation}
\mathcal{L} = \frac{N_{\text{Bhabha}}}
{\sigma_{\text{Bhabha}}(\theta_{\min}, \theta_{\max})}
\end{equation}

The Bhabha cross-section at small angles is large, precisely
calculable in QED (dominated by $t$-channel photon exchange), and
falls steeply with angle:
$d\sigma/d\theta \propto 1/\theta^3$.

This steep dependence means the accepted cross-section is dominated
by events near the inner acceptance edge $\theta_{\min}$, and a
small error in $\theta_{\min}$ produces a large error in the
luminosity:
\begin{equation}
\frac{\delta\sigma}{\sigma} \approx
2\frac{\delta\theta_{\min}}{\theta_{\min}}
\label{eq:lumi_sensitivity}
\end{equation}

The precision requirements are:
\begin{itemize}[itemsep=2pt]
\item \textbf{Absolute luminosity:} $10^{-4}$, needed for the
  Z-pole cross-section measurements, $N_\nu$ determination, and
  electroweak precision program.
\item \textbf{Relative luminosity:} $10^{-5}$, needed for the
  Z~lineshape scan and W/top threshold scans, where the shape of
  the cross-section vs.\ energy encodes the particle mass and width.
\end{itemize}

\subsection{Position and Stability Requirements}
\label{sec:evo_lumi_position}

The physicist pressed for a detailed accounting of the geometric
precision requirements, distinguishing between knowledge (how well
you know where the detector is) and stability (how well it stays
there).

\subsubsection{Geometric parameters}

For a luminometer at distance $z \approx 2500$~mm from the
interaction point with inner acceptance at
$\theta_{\min} \approx 60$~mrad ($r_{\min} \approx 150$~mm),
the sensitivity of the luminosity to various geometric parameters
differs significantly:

\begin{table}[htbp]
\centering
\caption{Geometric precision requirements for the luminosity
monitor, derived from the $10^{-4}$ absolute and $10^{-5}$ relative
luminosity targets using
Equation~\ref{eq:lumi_sensitivity} and its analogs for other
geometric parameters.}
\label{tab:lumi_geometry}
\begin{tabular}{lSSl}
\toprule
Parameter & {For $10^{-4}$} & {For $10^{-5}$} & Difficulty \\
\midrule
Inner edge radius & 7.5 {~$\mu$m} & 0.75 {~$\mu$m}
                  & Very demanding \\
$z$ position      & 125 {~$\mu$m} & 12.5 {~$\mu$m}
                  & Moderate \\
Transverse offset & 1.5 {~mm} & 0.5 {~mm}
                  & Relaxed \\
Tilt              & 0.6 {~mrad} & 0.2 {~mrad}
                  & Relaxed \\
\bottomrule
\end{tabular}
\end{table}

The inner edge radius dominates --- both its absolute knowledge
(for $10^{-4}$) and its stability over the duration of the
lineshape scan (for $10^{-5}$). The transverse offset and tilt
enter only at second order (due to azimuthal symmetry cancellation)
and are comparatively relaxed.

The $z$ position requirement is also relatively relaxed because it
enters through the ratio $\delta z/z$, which benefits from the large
lever arm ($z \approx 2500$~mm).

\subsubsection{Absolute position knowledge}

The inner edge of the silicon sensors can be defined by
photolithography to $\sim$1~$\mu$m precision. The sensor position
on its mounting structure can be surveyed to $\sim$few~$\mu$m using
coordinate measuring machines. The as-built inner radius can
therefore be known to $\sim$3--5~$\mu$m, meeting the 7.5~$\mu$m
requirement.

However, the relevant quantity is the inner edge position relative
to the \textbf{beam axis}, not to a mechanical reference. The beam
position at the luminometer ($z = 2500$~mm from the IP) depends on
both the beam position and angle at the interaction point:
\begin{equation}
x_{\text{beam}}(z_{\text{lumi}}) = x_{\text{beam}}(0) +
x'_{\text{beam}} \times z_{\text{lumi}}
\end{equation}

A beam angle uncertainty of 1~$\mu$rad gives a position uncertainty
of 2.5~$\mu$m at the luminometer. The beam angle must therefore be
known to $\sim$3~$\mu$rad for the absolute measurement, requiring
precise beam position monitors (BPMs) near the interaction point and
careful cross-calibration between the BPM reference frame and the
luminometer reference frame.

\subsubsection{Stability for relative luminosity}

The $10^{-5}$ relative requirement demands that the effective inner
edge radius be stable to $\sim$0.75~$\mu$m over the duration of the
lineshape scan, which may span months to years as different energy
points are measured in separate running periods.

This is an extraordinary stability requirement. Its feasibility
depends on the thermal environment, mechanical design, and
operational strategy.

\subsection{Thermal and Mechanical Stability}
\label{sec:evo_lumi_stability}

\subsubsection{Thermal expansion}

The physicist identified temperature variations as a potentially
limiting factor. The coefficient of thermal expansion (CTE)
determines how temperature changes translate to position shifts:

\begin{table}[htbp]
\centering
\caption{Thermal expansion of the luminometer inner edge
($r_{\min} = 150$~mm) for different support materials.}
\label{tab:lumi_thermal}
\begin{tabular}{lSS}
\toprule
Support material & {CTE ($\mu$m/m/$^\circ$C)}
                 & {$\delta r$ per $^\circ$C ($\mu$m)} \\
\midrule
Aluminum & 23  & 3.45 \\
Steel    & 12  & 1.80 \\
Invar    & 1.3 & 0.20 \\
Carbon fiber (axial) & {1--2} & {0.15--0.30} \\
\bottomrule
\end{tabular}
\end{table}

For an aluminum support structure, a temperature change of just
$\sim$2$^\circ$C gives a 7~$\mu$m shift --- consuming the full
$10^{-4}$ tolerance. For the $10^{-5}$ relative requirement
(0.75~$\mu$m), the tolerance tightens to $\sim$0.2$^\circ$C in
aluminum.

With Invar or carbon fiber support ($\delta r \approx
0.2$~$\mu$m/$^\circ$C), a temperature variation of
$\sim$4$^\circ$C is tolerable for $10^{-5}$. This is achievable
with reasonable environmental control.

The relevant thermal path includes not just the luminometer body but
the \textbf{entire mechanical chain} from the beam pipe (which
defines the beam axis reference) to the luminometer inner edge.
Every element in this chain contributes thermal expansion. This
argues strongly for:
\begin{itemize}[itemsep=2pt]
\item Low-CTE materials (Invar, carbon fiber) for all
  positioning-critical elements.
\item Monolithic construction where possible, minimizing joints that
  can exhibit creep or relaxation.
\item Active temperature monitoring with precision sensors
  ($\sim$10~mK resolution) for offline correction of residual
  thermal expansion.
\end{itemize}

\subsubsection{Vibration}

The machine-detector interface region contains several vibration
sources: cryogenic compressors and cold heads, cooling water flow,
vacuum pumps (during commissioning), and ground motion.

For a symmetric vibration of amplitude $A$ about the nominal
position, the first-order effect on the acceptance cancels by
symmetry, and the residual second-order effect is:
\begin{equation}
\frac{\delta\sigma}{\sigma} \sim
\left(\frac{A}{r_{\min}}\right)^2
\end{equation}

Even for $A = 50$~$\mu$m (a large vibration amplitude):
$\delta\sigma/\sigma \sim (50/150{,}000)^2 \sim 10^{-7}$ --- 
completely negligible.

The more serious concern is relative vibration between the
luminometer and the beam axis, where the two vibrate independently.
This is mitigated by mounting the luminometer as rigidly as possible
to the beam pipe or to a common support structure with the BPMs,
so that they move together. The machine-detector interface design
should treat the beam pipe, BPMs, and luminometer support as a
single mechanical system.

\subsection{Monitoring the Acceptance}
\label{sec:evo_lumi_monitoring}

The AI initially proposed ``continuous Bhabha monitoring to track
the effective acceptance in real time.'' The physicist identified a
fundamental flaw in this reasoning.

\subsubsection{The circular logic}

The luminometer measures luminosity by counting Bhabha events within
its acceptance. If the acceptance shifts and the Bhabha count
changes, the luminometer cannot distinguish ``luminosity changed''
from ``acceptance changed'' --- it measures only the product
$\mathcal{L} \times A_{\text{eff}}$. Using the same Bhabha events
to simultaneously determine the luminosity and monitor the
acceptance is circular.

Certain observables within the luminometer data are sensitive to
specific types of geometry changes independent of luminosity: the
azimuthal distribution develops a $\cos\phi$ modulation if the beam
shifts transversely, and an up-down asymmetry reveals vertical
offsets. However, these observables detect transverse displacements
and tilts --- parameters for which the tolerance is already relaxed
(Table~\ref{tab:lumi_geometry}). They are \textbf{insensitive to
the critical parameter}: a pure radial shift of the inner acceptance
edge, because the detector pads shift with the edge and the
distribution measured in detector coordinates is unchanged.

\subsubsection{What actually works}

The acceptance must be monitored through means independent of the
luminosity measurement itself:

\textbf{Hardware position monitoring.} Capacitive position sensors
or laser interferometric systems can measure the luminometer position
relative to the beam pipe (or a stable reference) with sub-$\mu$m
precision, continuously. These directly track the quantity of
interest --- the geometric relationship between detector and beam
axis --- without relying on physics events.

\textbf{Beam position monitoring.} BPMs near the interaction point
provide the beam orbit with $\sim$$\mu$m precision and $\sim$$\mu$rad
angular knowledge. Combined with hardware position monitoring of the
luminometer, this gives the effective beam-luminometer geometry.

\textbf{Wide-angle Bhabha cross-check.} Bhabha scattering at wide
angles ($\theta > 200$~mrad), detected in the main electromagnetic
calorimeter, provides a completely independent luminosity measurement
with different systematic uncertainties. The angular acceptance is
defined by the main tracker and ECAL rather than the forward
luminometer, so systematic shifts in the luminometer acceptance do
not affect it.

However, the statistical precision of wide-angle Bhabhas
is limited. Using a Bhabha cross-section of 10--20~nb
typical for wide-angle selections and the Z-pole
luminosity of $\sim$2$\times$10$^{36}$~cm$^{-2}$s$^{-1}$,
the rate is $\sim$20--40~Hz, giving $\sim$2--3 million
events per day. The statistical precision per month is:
\begin{equation}
\frac{1}{\sqrt{7 \times 10^7}} \approx 1.2 \times 10^{-4}
\end{equation}
This is adequate as a cross-check at the $10^{-4}$ level
but \textbf{insufficient for $10^{-5}$ monitoring} on
practical timescales. Wide-angle Bhabhas serve as an
independent validation, not a primary monitoring tool.

\subsection{Operational Strategy for Relative Luminosity}
\label{sec:evo_lumi_operations}

Given that hardware monitoring at the $10^{-5}$ level is extremely
demanding and physics-based monitoring is statistics-limited, the
operational strategy becomes a critical component of achieving
$10^{-5}$ relative precision.

\subsubsection{Interleaved energy scans}

The most powerful approach is to alternate between energy points on
short timescales (hours) rather than running at one energy for weeks
before switching. Any slow systematic drift (thermal expansion,
mechanical creep, beam orbit evolution) is then common to all energy
points within each short cycle and cancels in the cross-section
ratios.

The residual systematic is the drift within a single cycle (hours),
which is far easier to control than drift over months:
\begin{itemize}[itemsep=2pt]
\item Thermal drifts over hours are small and smooth, easily tracked
  by temperature sensors.
\item Mechanical creep over hours is negligible for a properly
  designed structure.
\item Beam orbit variations over hours are tracked by BPMs.
\end{itemize}

The interleaving strategy converts a long-term stability problem
(0.75~$\mu$m over months) into a short-term stability problem
(0.75~$\mu$m over hours), which is far more tractable.

\subsubsection{Redundancy}

Two identical luminometers on opposite sides of the interaction point
provide independent measurements. Agreement between the two gives
confidence in the result; disagreement reveals systematic problems.
Since the two luminometers are at different $z$ positions (and
different $\phi$ orientations relative to the crossing plane), they
have different sensitivities to beam geometry changes and different
thermal/mechanical environments, making them genuinely complementary.

\subsubsection{Integrated monitoring approach}

The $10^{-5}$ relative luminosity is ultimately achieved through a
combination of:
\begin{enumerate}[itemsep=2pt]
\item \textbf{Intrinsically stable mechanical design} --- low-CTE
  materials, monolithic construction, kinematic mounting.
\item \textbf{Continuous hardware monitoring} --- capacitive sensors
  or interferometry tracking the luminometer-beam geometry at
  sub-$\mu$m precision.
\item \textbf{Interleaved energy scans} --- converting long-term
  drifts to short-term variations that cancel in cross-section
  ratios.
\item \textbf{Independent cross-checks} --- wide-angle Bhabhas in
  the main detector, redundant luminometers on both sides of the IP.
\end{enumerate}

No single technique achieves $10^{-5}$ alone --- the combination is
essential.

\subsection{Detector Design}
\label{sec:evo_lumi_detector}

The silicon-tungsten sampling calorimeter design is well-matched to
the requirements:

\begin{itemize}[itemsep=2pt]
\item \textbf{$\sim$20 layers} of tungsten ($\sim$1~$X_0 = 3.5$~mm
  each) interleaved with silicon pad sensors.
\item \textbf{Total depth $\sim$20~$X_0$} ($\sim$100--120~mm),
  adequate for full containment of Bhabha electrons up to the
  beam energy.
\item \textbf{Fine radial segmentation} near the inner edge
  ($\sim$1~mm pad pitch), enabling precise measurement of the
  Bhabha radial distribution for acceptance studies.
\item \textbf{Azimuthal segmentation} for monitoring beam-detector
  alignment through the $\phi$ distribution of Bhabha events.
\item \textbf{Multiple radial rings} providing redundant acceptance
  definitions at different $\theta$ values.
\end{itemize}

The silicon sensor edges, defined by photolithography, provide
the most precise available definition of the physical acceptance
boundary. The tungsten absorber's short radiation length and small
Moli\`ere radius ($\sim$9~mm) keep the calorimeter compact and the
showers narrow, maximizing the position resolution for individual
Bhabha electrons.

\subsection{Integration with the Machine-Detector Interface}
\label{sec:evo_lumi_mdi}

The luminometer design must be developed jointly with the accelerator
team because:
\begin{itemize}[itemsep=2pt]
\item The beam pipe geometry and inner radius at $z = 2500$~mm
  constrain the luminometer inner acceptance.
\item The $\sim$30~mrad crossing angle affects the acceptance
  geometry and the Bhabha event topology.
\item Final focus elements create synchrotron radiation backgrounds
  that must be shielded from the luminometer.
\item The mechanical support path from beam pipe to luminometer
  determines the thermal and vibrational stability.
\item Access for maintenance, alignment surveys, and hardware
  monitoring must be planned from the start within the congested
  MDI region.
\end{itemize}

The luminometer is among the most accelerator-coupled detector
elements, and its design cannot be finalized independently of the
interaction region layout.

\subsection{Summary of Luminosity Monitor Discussion}
\label{sec:evo_lumi_summary}

The luminosity monitor technology choice did not change, but the
discussion produced several important clarifications:

\begin{enumerate}[itemsep=2pt]
\item The \textbf{position requirements were quantified}: 7.5~$\mu$m
  for absolute luminosity, 0.75~$\mu$m stability for relative
  luminosity. The inner edge radius is the critical parameter;
  transverse offsets and tilts are comparatively relaxed.

\item \textbf{Thermal stability} was identified as a potentially
  limiting factor, driving the choice of low-CTE support materials
  (Invar, carbon fiber) and active temperature monitoring.

\item The proposal for \textbf{continuous Bhabha-based acceptance
  monitoring was found to be circular} --- the luminometer cannot
  monitor its own acceptance using the same events it uses to
  measure luminosity. Hardware monitoring (capacitive sensors,
  interferometry) is the correct approach.

\item The \textbf{operational strategy} (interleaved energy scans,
  redundant luminometers, independent wide-angle cross-checks) was
  recognized as equally important as the hardware design for
  achieving $10^{-5}$ relative precision.

\item The design must be \textbf{co-developed with the accelerator
  MDI team}, as the luminometer is among the most
  accelerator-coupled detector elements.
\end{enumerate}

The luminosity monitor discussion illustrates that for some
subsystems, the detector technology is not the primary challenge ---
the metrology, environmental control, and operational strategy are
equally or more important. A perfect detector with inadequate
mounting stability or suboptimal scan strategy would fail to achieve
the physics goals regardless of its intrinsic performance.

% ======================================================================
\chapter{The Revised Integrated Detector}
\label{ch:revised}
% ======================================================================

\noindent\textit{This chapter presents the revised detector concepts
CL2a and CL2b, incorporating all the changes that emerged from the
subsystem examination of Chapter~\ref{ch:evolution}. The primary
concept CL2a is described in detail, including a self-consistent
geometric layout. The complementary concept CL2b is presented more
briefly, emphasizing where it differs from CL2a.}

\section{CL2a: Detector Layout}
\label{sec:cl2a_layout}

The CL2a detector is built from the inside out, with each
subsystem's dimensions following from the previous one. The
self-consistent radial layout in the barrel region is summarized in
Table~\ref{tab:cl2a_radial} and shown schematically in
Figure~\ref{fig:cl2a_layout}.

\begin{figure}[htbp]
\centering
\fbox{\parbox{0.85\textwidth}{\centering\vspace{5cm}
[Schematic cross-section of CL2a --- to be inserted]
\vspace{5cm}}}
\caption{Schematic radial cross-section of the CL2a detector
concept (not to scale). The dimensions correspond to
Table~\ref{tab:cl2a_radial}.}
\label{fig:cl2a_layout}
\end{figure}

\begin{table}[htbp]
\centering
\caption{CL2a radial layout in the barrel region at $z = 0$. The
$R_{\text{in}}$ and $R_{\text{out}}$ columns give the inner and
outer radial extent of each subsystem. Material budgets are for a
radial track at normal incidence.}
\label{tab:cl2a_radial}
\begin{tabular}{llSSl}
\toprule
Subsystem & Technology
  & {$R_{\text{in}}$} & {$R_{\text{out}}$}
  & Material \\
  & & {(mm)} & {(mm)} & \\
\midrule
Beam pipe & Be, locally thinned
  & 10 & 12 & 0.08\% $X_0$ \\
\addlinespace
Vertex/inner Si & MAPS, 65~nm,
  & 12 & 300 & $<$0.2\% $X_0$ \\
  & 6 curved layers & & & (total) \\
\addlinespace
Inner wall & Carbon fiber
  & 340 & 345 & $\sim$0.1\% $X_0$ \\
\addlinespace
Main tracker & Drift chamber,
  & 345 & 2000 & $\sim$0.2\% $X_0$ \\
  & He:iC$_4$H$_{10}$ 90:10 & & & (gas + wires) \\
\addlinespace
Outer wrapper & AC-LGAD,
  & 2000 & 2005 & 0.4--1.0\% $X_0$ \\
  & single layer & & & \\
\addlinespace
ECAL & PbWO$_4$ crystal,
  & 2050 & 2250 & 22~$X_0$ \\
  & 2 segments & & & \\
\addlinespace
Solenoid & SC, 3~T design,
  & 2300 & 2450 & $\sim$0.4~$\lambda_I$ \\
  & variable 1.5--3~T & & & \\
\addlinespace
HCAL & Scint.-steel tiles,
  & 2500 & 3750 & $\sim$6~$\lambda_I$ \\
  & $\sim$50 layers & & & \\
\addlinespace
Return yoke + & Steel + scint. bars,
  & 3750 & 4400 & $\sim$1.5~$\lambda_I$ \\
muon system & 3--4 stations & & & \\
\bottomrule
\end{tabular}
\end{table}

Several design choices in this layout deserve comment:

\subsubsection{Gaps between subsystems}

Small radial gaps are allowed between subsystems for mechanical
clearance, cable routing, and thermal isolation:
\begin{itemize}[itemsep=2pt]
\item Between outer wrapper and ECAL ($\sim$45~mm): space for
  wrapper services, ECAL front-face support, and a small air gap.
\item Between ECAL and solenoid ($\sim$50~mm): mechanical clearance
  and thermal isolation between the room-temperature ECAL and the
  cryogenic solenoid.
\item Between solenoid and HCAL ($\sim$50~mm): clearance for coil
  cryostat and HCAL front-face support.
\end{itemize}

These gaps contribute dead material (cables, support structures) that
is not included in the idealized material budgets above. In practice,
the gaps should be minimized and any services routed along $z$
(parallel to the beam) rather than radially where possible.

\subsubsection{HCAL depth}

The HCAL consists of $\sim$50 layers, compared with $\sim$50 layers in CL1a, of 20~mm steel absorber plates
interleaved with 3~mm scintillator tiles, with $\sim$2~mm gaps for
tile wrapping and SiPM mounting. The total pitch per layer is
$\sim$25~mm, giving a total radial depth of $\sim$1250~mm.

The steel thickness totals $\sim$1000~mm, providing $\sim$6~$\lambda_I$
($\lambda_I^{\text{steel}} \approx 167$~mm). Combined with the
solenoid coil ($\sim$0.4~$\lambda_I$) and the crystal ECAL
($\sim$1~$\lambda_I$ nuclear), the total absorber before the
muon system is $\sim$7.4~$\lambda_I$.

The HCAL steel serves simultaneously as calorimeter absorber and
magnetic flux return (Section~\ref{sec:evo_solenoid}),
carrying a substantial fraction of the solenoid return flux.

\subsubsection{Muon system depth}

The return yoke outside the HCAL consists of 3--4 iron layers
($\sim$100~mm each) interleaved with scintillator bar muon stations.
The total additional absorber depth is $\sim$1.5~$\lambda_I$,
bringing the cumulative depth before the outermost muon station to
$\sim$8.9~$\lambda_I$ and reducing the hadron punch-through
probability to $<$0.05\%.

\subsubsection{Overall dimensions}

The CL2a barrel detector has an outer radius of $\sim$4.4~m. The
barrel half-length is determined by the angular coverage requirement
and the longest subsystem:
\begin{itemize}[itemsep=2pt]
\item Drift chamber active volume: $|z| < 2000$~mm.
\item Drift chamber stay system: extends to
  $|z| \approx 2500$--2800~mm.
\item ECAL barrel: $|z| \lesssim 2500$~mm.
\item HCAL barrel: $|z| \lesssim 2800$~mm.
\item Return yoke/muon: $|z| \lesssim 3200$~mm.
\end{itemize}

The total barrel half-length is $\sim$3.0--3.2~m, giving an overall
barrel length of $\sim$6.0--6.4~m. Endcap systems (tracking disks,
ECAL endcap, HCAL endcap, and muon endcap stations) extend the
total detector length to $\sim$10--12~m.

\subsection{Longitudinal Layout}
\label{sec:cl2a_longitudinal}

The longitudinal layout is constrained by the drift chamber stay
system, which consumes $\sim$500--800~mm per end between the active
wire volume and the outer cylinder. This sets the minimum $z$ extent
of the barrel calorimetry and determines where endcap systems must
begin:

\begin{table}[htbp]
\centering
\caption{CL2a longitudinal layout (half-lengths from IP).}
\label{tab:cl2a_longitudinal}
\begin{tabular}{lSl}
\toprule
Subsystem & {$|z|_{\text{max}}$ (mm)} & Notes \\
\midrule
Vertex/inner Si & {$\sim$150--200} & per layer \\
\addlinespace
DC active volume & 2000 & Wire length \\
\addlinespace
DC stay system & {2500--2800} & Drives barrel \\
               &              & length \\
\addlinespace
ECAL barrel & {$\sim$2500} & Matches DC \\
            &              & envelope \\
\addlinespace
HCAL barrel & {$\sim$2800} & Outside stays \\
\addlinespace
Muon barrel & {$\sim$3200} & Outermost \\
\addlinespace
Luminometer & {$\sim$2500} & In MDI region \\
\bottomrule
\end{tabular}
\end{table}

The polar angle coverage of the barrel tracking system extends to:
\begin{equation}
\theta_{\min}^{\text{barrel}} \approx
\arctan\left(\frac{R_{\text{inner}}^{\text{DC}}}{z_{\text{active}}^{\text{DC}}}\right)
= \arctan\left(\frac{345}{2000}\right) \approx 9.8^\circ
\end{equation}
corresponding to $|\cos\theta| \lesssim 0.985$ --- excellent forward
coverage.

\section{CL2a: Technology Summary}
\label{sec:cl2a_technology}

Table~\ref{tab:cl2a_summary} summarizes the CL2a technology choices
alongside those of CL1a, highlighting what changed and what remained.

\begin{table}[htbp]
\centering
\caption{Technology comparison between CL1a (initial) and CL2a
(revised). Changes are indicated in bold.}
\label{tab:cl2a_summary}
\begin{tabular}{lll}
\toprule
Subsystem & CL1a & CL2a \\
\midrule
Beam pipe & Be, thinned & Be, thinned (unchanged) \\
\addlinespace
Vertex + & Separate vertex & \textbf{Unified MAPS,} \\
inner Si & + inner Si & \textbf{6 layers} \\
\addlinespace
Main tracker & Drift chamber & Drift chamber \\
             & (He-based)    & (unchanged) \\
\addlinespace
Outer wrapper & Strips + LGAD & \textbf{Single AC-LGAD} \\
              & (two layers)  & \textbf{layer} \\
\addlinespace
ECAL & Noble liquid (LAr) & \textbf{PbWO$_4$ crystal,} \\
     &                    & \textbf{2 segments,} \\
     &                    & \textbf{dSiPM (counting mode)} \\
\addlinespace
Solenoid & 2~T, outside HCAL & \textbf{1.5--3~T variable,} \\
         &                    & \textbf{inside HCAL} \\
\addlinespace
HCAL & Scint.-steel tiles & Scint.-steel tiles \\
     &                    & \textbf{(as flux return)} \\
\addlinespace
Muon & $\mu$-RWELL & \textbf{Scintillator bars,} \\
     &             & \textbf{WLS fiber + SiPM} \\
\addlinespace
Luminometer & Si-W calorimeter & Si-W calorimeter \\
            &                  & (unchanged) \\
\bottomrule
\end{tabular}
\end{table}

\section{CL2a: Performance Targets}
\label{sec:cl2a_performance}

Table~\ref{tab:cl2a_performance} summarizes the expected performance
of the CL2a concept for key physics observables.

\begin{table}[htbp]
\centering
\caption{CL2a expected performance for key observables.
Values marked with $^*$ are estimates based on scaling arguments
rather than full simulation.}
\label{tab:cl2a_performance}
\begin{tabular}{lll}
\toprule
Observable & Target & Driving subsystem \\
\midrule
Impact parameter & $\sim$3~$\mu$m $\oplus$
  & Vertex detector \\
$\sigma_{d_0}$ & $\sim$10~$\mu$m/($p\sin^{3/2}\theta$)
  & + beam pipe \\
\addlinespace
Momentum & $\sigma(p_T)/p_T^2 \lesssim$
  & Tracker + \\
resolution & $2 \times 10^{-5}$~GeV$^{-1}$
  & outer wrapper + $B$ \\
\addlinespace
EM energy & $\sim$3\%/$\sqrt{E}$
  & Crystal ECAL \\
resolution & $\oplus$ $\sim$0.3\%
  & \\
\addlinespace
Jet energy$^*$ & $\sim$3.5--4.5\% at 45~GeV
  & Tracker + ECAL + \\
(PFA)     & $\sim$5--6\% at 100~GeV
  & HCAL + $B$ field \\
\addlinespace
$\pi/K$ & Up to $\sim$3~GeV (TOF,
  & Outer wrapper + \\
separation & at $3\sigma$ level)
  & main tracker \\
           & + higher ($dE/dx$)
  & \\
\addlinespace
Muon ID & $>$95\% for $p > 3$~GeV
  & Absorber depth + \\
efficiency &
  & muon stations \\
\addlinespace
Luminosity & $10^{-4}$ (absolute)
  & Luminometer + \\
           & $10^{-5}$ (relative)
  & scan strategy \\
\bottomrule
\end{tabular}
\end{table}

The jet energy resolution numbers are estimated from the
decomposition of Equation~\ref{eq:pfa_decomposition} with the
crystal ECAL and 2--3~T variable field, and carry significant
uncertainty. A proper evaluation requires full simulation of the
CL2a geometry with realistic material budget, particle flow
reconstruction, and the specific magnetic field configuration at
each energy point. This is identified as a priority R\&D activity.

\section{Interplay Between Subsystem Choices}
\label{sec:interplay}

One of the principal findings of this study is that subsystem choices
are strongly coupled --- a change in one subsystem propagates
through the design and influences optimal choices elsewhere. Several
of these couplings emerged through the discussion:

\begin{enumerate}[itemsep=4pt]
\item \textbf{ECAL technology influences the HCAL
  strategy.} The separate deep crystal ECAL measures
  all incoming photons before they reach the HCAL,
  reducing cluster multiplicity and simplifying the
  pattern recognition task. This reduces the performance
  difference between tile and dual-readout HCAL options,
  and makes the HCAL intrinsic resolution less critical
  relative to the particle flow confusion term.

\item \textbf{ECAL crystal choice drives readout R\&D.}
  PbWO$_4$'s low light yield is simultaneously what
  makes SiPM readout conceivable (avoiding the
  catastrophic saturation that higher-yield crystals
  would produce) and what limits the photostatistical
  contribution to energy resolution. The readout
  technology choice (counting-mode dSiPM vs.\ analog
  SiPM) is a direct consequence of the crystal
  selection.

\item \textbf{ECAL technology influences PFA performance assessment.}
  The commonly cited PFA jet resolution numbers assume Si-W ECAL
  granularity. With a crystal ECAL, the photon energy term improves
  while the confusion term changes. The net effect is roughly
  neutral at Z-pole energies but must be properly evaluated with
  full simulation.

\item \textbf{Solenoid placement drives cost, flux
  return, and muon system design.} Moving the solenoid
  inside the HCAL reduces cost by $\sim$4$\times$,
  enables the HCAL steel to serve as flux return
  (substantially reducing the dedicated return yoke mass
  and cost), and allows the muon absorber to be optimized
  for physics rather than magnetics. The cost is
  $\sim$0.4~$\lambda_I$ of dead material, whose impact
  is mitigated by the particle flow framework: the coil
  material primarily affects the neutral hadron component
  ($\sim$10--15\% of jet energy), and the confusion term
  dominates jet resolution at all FCC-ee energies.

\item \textbf{Variable magnetic field couples to all tracking and
  PFA performance.} The field strength affects momentum resolution,
  Lorentz angle in the drift chamber, particle flow confusion term,
  and the accelerator beam optics. The variable field strategy
  (1.5--3~T matched to beam energy) optimizes all these
  simultaneously.

\item \textbf{Tracker gas choice couples to material budget and
  vertex performance.} The drift chamber's helium-based gas
  ($\sim$0.03\% $X_0$) preserves the ultra-low material budget
  achieved by the MAPS vertex detector. An argon-based TPC gas
  ($\sim$1.5\% $X_0$) would partially negate the vertex detector's
  material investment, particularly for low-momentum tracks at the
  Z~pole.

\item \textbf{Outer wrapper material affects ECAL performance.}
  Every fraction of $X_0$ in the outer wrapper increases the photon
  conversion probability before the ECAL. The single-layer AC-LGAD
  design ($\sim$0.4--1.0\% $X_0$) was chosen partly to minimize
  this impact; the two-layer design of CL1a would have roughly
  doubled it.
\end{enumerate}

These couplings argue against optimizing each subsystem independently
--- a practice that can lead to locally optimal but globally
suboptimal designs. The integrated approach pursued in this study,
where subsystem discussions explicitly considered the impact on
neighboring systems, proved essential for arriving at a coherent
overall concept.

\section{CL2b: The Revised Complementary Detector}
\label{sec:cl2b}

The complementary detector CL2b is updated from CL1b to reflect the
insights from the subsystem examination. The guiding principle
remains maximal complementarity with CL2a: different technologies
for the major subsystems, providing independent systematic checks
on all major physics measurements.

\begin{table}[htbp]
\centering
\caption{CL2a and CL2b compared. Shared elements are listed below
the dividing line. Changes from the initial CL1a/CL1b pairing are
indicated in italics.}
\label{tab:cl2a_cl2b}
\begin{tabular}{lll}
\toprule
Subsystem & CL2a & CL2b \\
\midrule
Main tracker & Drift chamber (He) & \textit{TPC (Ar-based)} \\
\addlinespace
ECAL & PbWO$_4$ crystal & \textit{Si-W, high-} \\
     &                  & \textit{granularity PFA} \\
\addlinespace
HCAL & Scint.-steel tiles & Dual-readout fibers \\
\addlinespace
\midrule
\multicolumn{3}{l}{\textit{Shared elements:}} \\
\addlinespace
Beam pipe      & \multicolumn{2}{l}{Be, locally thinned} \\
Vertex/inner Si & \multicolumn{2}{l}{Unified MAPS, 6 layers} \\
Outer wrapper  & \multicolumn{2}{l}{AC-LGAD, single layer} \\
Solenoid       & \multicolumn{2}{l}{\textit{Variable 1.5--3~T,
                 inside HCAL}} \\
Muon system    & \multicolumn{2}{l}{\textit{Scintillator bars}} \\
Luminometer    & \multicolumn{2}{l}{Si-W calorimeter} \\
\bottomrule
\end{tabular}
\end{table}

The key complementary choices and their motivation:

\begin{itemize}[itemsep=4pt]
\item \textbf{TPC main tracker.} The TPC provides true 3D space
  points, superior $z$ resolution, and unambiguous pattern
  recognition --- a fundamentally different tracking approach from
  the drift chamber. The argon-based gas enables Ramsauer-Townsend
  diffusion suppression for competitive spatial resolution at the
  cost of higher material budget ($\sim$1.5\% $X_0$ vs.\
  $\sim$0.03\% $X_0$). The CEPC TPC studies and ILD-FCCee concept
  provide a strong foundation.

  The different systematic profiles --- drift chamber calibration
  (time-to-distance, wire positions) vs.\ TPC calibration (drift
  velocity, space charge) --- give genuinely independent
  cross-checks on tracking performance.

\item \textbf{Si-W electromagnetic calorimeter.} With
  5$\times$5~mm$^2$ pads and $\sim$30 longitudinal layers, the Si-W
  ECAL is specifically optimized for particle flow, providing fine
  3D shower imaging that the crystal ECAL cannot match. 
  This gives the best possible particle flow performance
(at the cost of substantially worse intrinsic photon
energy resolution than the crystal's
$\sim$3\%/$\sqrt{E}$, as expected for any sampling
calorimeter), complementing CL2a's approach of superior
energy resolution with adequate pattern recognition.

  The comparison of jet energy resolution between CL2a (crystal ECAL
  + tiles) and CL2b (Si-W ECAL + dual-readout) across the full
  FCC-ee energy range would be scientifically valuable data for
  future collider detector design.

\item \textbf{Dual-readout fiber HCAL.} The fundamentally different
  approach to hadronic energy measurement --- event-by-event
  $f_{\text{em}}$ correction vs.\ particle flow avoidance ---
  provides independent systematic checks on all jet energy
  measurements. As argued in Section~\ref{sec:evo_hcal_assessment},
  this is perhaps the strongest case for two-detector
  complementarity.

  The dual-readout HCAL integrates more naturally with the Si-W ECAL
  in CL2b than it would with the crystal ECAL in CL2a, because the
  Si-W ECAL provides the fine granularity needed for particle flow
  of the charged component while the dual-readout HCAL provides
  robust standalone hadronic measurement for the neutral component.
\end{itemize}

The shared elements were updated from CL1b to reflect the
conclusions of Chapter~\ref{ch:evolution}:
\begin{itemize}[itemsep=2pt]
\item The vertex and inner silicon are unified (same as CL2a).
\item The solenoid is inside the HCAL with variable field (same as
  CL2a).
\item The muon system uses scintillator bars (same as CL2a).
\end{itemize}

These shared elements were identified as clearly optimal regardless
of the overall detector philosophy, so there is no benefit to making
them different between the two detectors. Complementarity is most
valuable where the technology choice genuinely affects the systematic
profile of the physics measurements --- the tracker, ECAL, and HCAL.

Note that CL2b with a dual-readout fiber HCAL and solenoid inside
would require the fiber calorimeter to use a ferromagnetic absorber
(steel) rather than the copper or brass typically used in RD52
prototypes, so that the HCAL steel can serve as flux return. This
has not been studied in detail and represents an R\&D item specific
to CL2b. If a non-magnetic absorber proves necessary for the
dual-readout concept, the solenoid placement in CL2b might need to
differ from CL2a --- an acceptable departure from the shared
baseline if required by the physics.

\section{Summary of Changes from Initial to Revised Concepts}
\label{sec:changes_summary}

Table~\ref{tab:all_changes} provides a consolidated summary of all
changes between the initial concepts (CL1a/CL1b) and the revised
concepts (CL2a/CL2b), with the primary reason for each change.

\begin{table}[htbp]
\centering
\caption{Summary of all changes from CL1a to CL2a, with the
primary reason for each. Changes are numbered for reference.}
\label{tab:all_changes}
\begin{tabular}{clll}
\toprule
\# & Subsystem & Change & Primary reason \\
\midrule
1 & Vertex + & Separate $\to$ & No physical basis for \\
  & inner Si & unified & separation at FCC-ee \\
\addlinespace
2 & Outer    & Two layers $\to$ & Material budget \\
  & wrapper  & single AC-LGAD & reduction \\
\addlinespace
3 & ECAL & Noble liquid $\to$ & Cryostat dead material \\
  &      & PbWO$_4$ crystal & eliminates noble liquid \\
  &      &                  & resolution advantage \\
\addlinespace
\addlinespace
4 & Solenoid & 2~T fixed $\to$ & Match field to beam \\
  & field   & 1.5--3~T variable & energy; accelerator \\
  &         &                   & coupling at Z pole \\
\addlinespace
5 & Solenoid & Outside $\to$ & $\sim$4$\times$ cost \\
  & placement & inside HCAL & reduction; HCAL as \\
  &           &             & flux return \\
\addlinespace
6 & Muon & $\mu$-RWELL $\to$ & Match technology \\
  & system & scintillator bars & to actual (modest) \\
  &        &                   & requirements \\
\bottomrule
\end{tabular}
\end{table}

\section{Unresolved Questions and R\&D Priorities}
\label{sec:open_questions}

Several important questions remain open. These are identified as
priorities for future R\&D, simulation studies, or design
optimization:

\begin{enumerate}[itemsep=4pt]
\item \textbf{Main tracker technology.} The drift chamber vs.\ TPC
  choice is the least resolved major decision. A direct simulation
  comparison under FCC-ee conditions --- including realistic material
  budgets (He vs.\ Ar gas), pattern recognition in complex Z-pole
  events, the impact of gas material on vertexing at low momenta,
  and $z$-resolution effects on acceptance determination --- would
  be highly valuable.

\item \textbf{Cluster counting performance.} The practical
  resolution of $dN/dx$ cluster counting, accounting for delta ray
  contamination, diffusion, left-right folding, and electronics
  bandwidth, needs quantitative determination through detailed
  simulation with a realistic cluster-finding algorithm. The gas
  optimization (potentially favoring lower primary ionization rate
  or lower drift velocity than conventionally assumed) deserves
  systematic study.

\item \textbf{SiPM readout for crystal ECAL.} The
  preferred counting-mode dSiPM approach requires
  development of a device matched to PbWO$_4$
  scintillation properties (cell pitch, recharge time,
  counter depth, integration window) and test beam
  validation across the FCC-ee energy range. The analog
  SiPM fallback, with saturation curve correction, also
  requires test beam characterization to establish the
  achievable energy resolution.
  
\item \textbf{AC-LGAD multi-hit performance.} The outer wrapper's
  ability to serve position, $z$, and timing functions with a single
  AC-LGAD layer depends on the multi-hit disentangling performance,
  which needs systematic beam test characterization under realistic
  conditions.

\item \textbf{Crystal ECAL particle flow performance.} Full
  simulation of jet energy resolution with the CL2a geometry
  (crystal ECAL, variable magnetic field, scintillator-steel HCAL)
  is needed to validate or correct the estimates from scaling
  arguments used in this report.

\item \textbf{Solenoid coil thickness.} Achieving $\lesssim$150~mm
  coil thickness at 3~T design field requires detailed engineering
  study, potentially including high-temperature superconductor (HTS)
  options.

\item \textbf{Dual-readout HCAL engineering for CL2b.}
  The dual-readout fiber concept has been demonstrated
  with copper/brass absorber. For CL2b (with the
  solenoid inside), steel absorber is required for flux
  return. While there is no fundamental physics obstacle
  --- the dual-readout correction applies regardless of
  absorber material --- the performance with steel has
  not been validated, and fiber routing at full $4\pi$
  scale remains an unsolved engineering challenge
  regardless of absorber choice.
  
\item \textbf{Luminometer stability.} Demonstration of sub-$\mu$m
  position stability in a realistic MDI thermal and vibrational
  environment, using low-CTE materials and hardware position
  monitoring.

\item \textbf{Air cooling of MAPS vertex detector.} Validation of
  air cooling for the ultra-thin curved MAPS shells at
  representative power density, with measurement of vibration and
  thermal uniformity.

\item \textbf{Sensor inside beam pipe.} Feasibility R\&D for the
  upgrade option: UHV compatibility of thinned MAPS, beam background
  simulation, beam impedance studies.
\end{enumerate}

% ======================================================================
\chapter{Reflections on Human-AI Collaboration}
\label{ch:reflections}
% ======================================================================

\noindent\textit{This chapter reflects on the process of using
human-AI dialogue to explore detector design. It is jointly authored:
the AI drafted the assessment of its own contributions and
limitations, while observations about the human role draw on the
physicist's perspective as communicated during the dialogue. Both
contributors reviewed the final text.}

\section{What the AI Brought to the Discussion}
\label{sec:what_ai_brought}

The AI assistant contributed several capabilities that proved useful
for the detector design exploration:

\subsection{Rapid Survey of Technology Options}

For each subsystem, the AI could quickly lay out the available
technology options, their demonstrated performance parameters,
and their relative advantages and disadvantages. This provided a
starting framework for discussion that would otherwise require
extensive literature review. Examples include the comparison of
noble liquid vs.\ crystal vs.\ Si-W ECAL options
(Section~\ref{sec:evo_ecal}), the LGAD variant comparison for the
outer wrapper (Section~\ref{sec:evo_lgad_options}), and the drift
chamber vs.\ TPC assessment (Section~\ref{sec:evo_tracker_choice}).

While these surveys are not a substitute for expert knowledge ---
the AI's information has a training cutoff and may miss the most
recent developments --- they provide a useful starting point that
can be corrected and refined through dialogue.

\subsection{Quantitative Estimates on Demand}

Throughout the discussion, the AI provided back-of-envelope
calculations for quantities such as:
\begin{itemize}[itemsep=2pt]
\item Wire sag in the drift chamber
  (Section~\ref{sec:evo_dc_mechanical}).
\item Spoke material budget and azimuthal filling fraction
  (Section~\ref{sec:evo_dc_mechanical}).
\item Primary ionization statistics and cluster counting resolution
  (Section~\ref{sec:evo_cluster_counting}).
\item Cryostat material budget and its impact on effective energy
  resolution (Section~\ref{sec:evo_cryostat}).
\item SiPM photoelectron counts and saturation levels
  (Section~\ref{sec:evo_sipm}).
\item Luminometer position requirements from the Bhabha angular
  sensitivity (Section~\ref{sec:evo_lumi_position}).
\item Solenoid cost scaling and flux return calculations
  (Section~\ref{sec:evo_solenoid}).
\end{itemize}

These estimates, while approximate, served to make the discussion
quantitative rather than purely qualitative. They enabled rapid
assessment of whether a proposed effect was negligible, important,
or dominant --- a critical capability for design trade-off
discussions. Several estimates were corrected through the dialogue
(e.g., the factor-of-two from left-right folding in cluster
counting, Section~\ref{sec:evo_cluster_counting}), illustrating both
the value and the limitations of rapid analytical estimates.

\subsection{Systematic Comparison Frameworks}

The AI organized comparisons into structured tables and systematic
assessments, making it easier to see the trade-offs clearly. The
technology comparison tables throughout Chapter~\ref{ch:evolution}
(e.g., Tables~\ref{tab:tracker_comparison},
\ref{tab:effective_resolution}, \ref{tab:muon_requirements},
\ref{tab:muon_operations}) helped organize the discussion and
identify the decisive factors for each choice.

\subsection{Willingness to Revise Positions}

When presented with compelling arguments, the AI revised its
positions --- sometimes substantially. The ECAL technology change
(noble liquid to crystal), the solenoid placement change (outside to
inside HCAL), and the muon system technology change ($\mu$-RWELL to
scintillator bars) all involved the AI acknowledging that its initial
reasoning was flawed or incomplete and adopting the alternative.

This willingness to revise is essential for productive collaboration.
An AI that defended its initial positions regardless of contrary
evidence would be far less useful as a design exploration partner.

\subsection{Documentation and Synthesis}

The AI's ability to maintain context over a long
conversation had both technical and practical benefits.
During the discussion, it enabled cross-referencing
between subsystem decisions --- for example, recognizing
that the solenoid placement inside the HCAL has
consequences for the muon system absorber design, or
that the crystal ECAL choice affects the HCAL's pattern
recognition demands. This contributed directly to the
system-level integration discussed in
Section~\ref{sec:interplay}.

As a practical matter, the AI's ability to produce
structured written output from the dialogue --- this
report --- is a significant contribution. The synthesis
of Chapter~\ref{ch:revised} draws together conclusions
from dozens of individual technical discussions into a
coherent integrated concept, a task that benefits from
the AI's ability to organize large amounts of discussed
material.\footnote{In practice, the conversation
eventually exceeded the platform's context limits and
became inaccessible, necessitating a separate editing
session. The context maintenance capability is real and
valuable but bounded by platform constraints.}

\section{What Required Human Guidance}
\label{sec:human_guidance}

The physicist's contributions were qualitatively different from the
AI's and proved essential for arriving at sound conclusions. Several
categories of human contribution are identifiable:

\subsection{Practical Experience and Engineering Judgment}

The physicist brought knowledge that is not well-represented in
published literature --- the kind of understanding that comes from
having built, operated, and debugged real detector systems. Examples
include:
\begin{itemize}[itemsep=2pt]
\item The observation that amplitude-based discrimination of primary
  clusters from delta ray secondaries is undermined by avalanche
  fluctuations and space charge effects
  (Section~\ref{sec:evo_cluster_counting}) --- effects that are
  known to experts but rarely discussed in papers promoting cluster
  counting.
\item The recognition that $\mu$-RWELL's impressive specifications
  are entirely irrelevant for FCC-ee's benign muon system
  environment (Section~\ref{sec:evo_muon}).
\item The identification of temperature stability as a potentially
  limiting factor for the luminometer
  (Section~\ref{sec:evo_lumi_stability}).
\item The practical assessment of digital SiPM power
  requirements, informed by consultation with subject
  matter experts outside this collaboration,
  distinguishing between counting-only mode (negligible
  power) and timing mode (uncertain, with experts unable
  to commit to estimates without specific circuit
  design) (Section~\ref{sec:evo_sipm}).
\end{itemize}

\subsection{Asking the Right Question}

Several of the most consequential shifts in the design were
triggered not by the AI's analysis but by a single well-chosen
question from the physicist:
\begin{itemize}[itemsep=2pt]
\item ``Why did you separate the vertex detector and inner silicon
  tracker?'' --- leading to the unified MAPS subsystem
  (Section~\ref{sec:evo_vertex_silicon}).
\item ``How thick is the cryostat?'' --- leading to the abandonment
  of noble liquid ECAL in favor of crystals
  (Section~\ref{sec:evo_cryostat}).
\item ``Why not liquid krypton instead of argon?'' --- exposing an
  unjustified inherited assumption
  (Section~\ref{sec:evo_lar_to_lkr}).
\item ``The solenoid cost scales as $(R_{\text{out}}/R_{\text{in}})^3$
  --- is it worth it?'' --- triggering the solenoid placement
  change (Section~\ref{sec:evo_solenoid}).
\item ``Can the HCAL serve as flux return?'' --- strengthening the
  case for inside placement
  (Section~\ref{sec:evo_hcal_flux}).
\item ``The crystal ECAL filters out the EM component --- doesn't
  that simplify the HCAL's job?'' --- reframing the HCAL technology
  comparison (Section~\ref{sec:evo_hcal_assessment}).
\item ``How does the luminometer monitor its own acceptance using
  the same events it measures?'' --- exposing a logical circularity
  (Section~\ref{sec:evo_lumi_monitoring}).
\end{itemize}

In each case, the question itself contained the key physical insight.
The AI's role was to work through the implications, but the
direction came from the human.

\subsection{Recognizing Inappropriate Technology Choices}

A recurring pattern was the physicist identifying cases where the AI
selected technology based on impressiveness rather than suitability:
\begin{itemize}[itemsep=2pt]
\item $\mu$-RWELL for the muon system: excellent technology,
  wrong application.
\item The tendency to favor the most advanced LGAD variant without
  fully accounting for maturity and practical complications.
\item The initial preference for noble liquid ECAL based on its
  prominent role in the ATLAS detector, without critically examining
  whether the same arguments apply at FCC-ee.
\end{itemize}

This reflects a systematic tendency of the AI: when trained on
literature that emphasizes novel and high-performing technologies,
the AI naturally gravitates toward those technologies even when
simpler, more mature alternatives are better matched to the actual
requirements. The human's role in correcting this tendency was
essential.

\subsection{System-Level Thinking}

The physicist consistently thought about how choices in one subsystem
affect others --- the kind of system-level reasoning that experienced
detector designers develop over careers:
\begin{itemize}[itemsep=2pt]
\item The separate deep ECAL simplifying the HCAL's
  pattern recognition task.
\item The solenoid placement driving flux return, cost, and muon
  system design simultaneously.
\item The outer wrapper material affecting ECAL performance.
\item The tracker gas choice affecting vertex detector performance
  at low momenta.
\end{itemize}

While the AI could follow and extend these connections
once pointed out, in this study the initial
identification of the coupling came from the physicist
rather than the AI. Whether this reflects a fundamental
limitation of current AI systems or a consequence of the
challenge-and-response mode adopted here is not clear
from a single study.

\section{Observations on AI Reasoning Patterns}
\label{sec:ai_reasoning}

\noindent\textit{This section presents the physicist's
observations on patterns in AI reasoning that emerged
during the collaboration.}

\subsection{Knowledge Without Connections}

A striking pattern emerged early in the discussion. When
examining the TPC option for the main tracker, the
physicist asked the AI to explain the low diffusion
achievable in argon-based TPCs. The AI identified
magnetic field suppression of transverse diffusion but
did not mention the Ramsauer-Townsend effect --- the
minimum in the electron-argon scattering cross-section
near 0.5~eV that enables ``cool'' electron transport,
which is a primary motivation for argon-based TPC gases.

When the physicist raised the Ramsauer-Townsend effect
explicitly, the AI immediately responded with a correct
and detailed explanation of the mechanism and its
implications for TPC spatial resolution. The knowledge
was present but the connection to TPC diffusion was not
made until prompted.

This pattern --- possessing correct individual facts
without making the cross-domain connections that
experienced physicists find intuitive --- recurred
throughout the collaboration. It suggests that while AI
has broad factual coverage, the associative links between
concepts that constitute physical intuition are weaker
or latent, requiring external activation.

\subsection{Exploration Bounded by the Questioner}

The challenge-and-response mode that proved productive
for this study also creates a systematic bias: the
design space explored is limited to what the physicist
thought to ask about. Issues the physicist did not raise
went unexamined.

This is a methodological limitation rather than purely
an AI limitation. However, it differs from what one might
expect with a second human expert, who would bring
different experience and different blind spots, and might
independently raise concerns the first physicist had not
considered. The AI, operating primarily in response mode,
tends to share the questioner's blind spots rather than
compensating for them. When the physicist did not
challenge an aspect of the design, the AI rarely
volunteered concerns unprompted.

A more proactive AI --- one that systematically
questioned its own proposals in the way the physicist
questioned them --- would partially address this
limitation. Whether current or near-future AI systems
can be configured to operate in this more self-critical
mode is an open question.

\subsection{Tendency Toward Deference}

When the physicist challenged AI positions on technical
matters, the AI rarely maintained its original stance.
While this produced efficient convergence, it may not
always produce the best outcome. Many design questions
do not have clear right or wrong answers --- they
involve trade-offs where reasonable people can disagree.
In such cases, productive dialogue benefits from both
parties articulating and defending their positions,
forcing each to sharpen their reasoning. An experienced
human collaborator would sometimes push back ---
``I hear your concern, but I still think this approach
is better because...'' --- even when the question is
not fully resolvable. This resistance, largely absent
from the AI's behavior in this study, would strengthen
the final product regardless of which position
ultimately prevails.

\subsection{Recency and Visibility Bias}

The AI appeared to weight technologies more heavily when
they are frequently discussed in recent literature,
independent of their actual suitability. The IDEA-style
drift chamber, extensively discussed in the FCC-ee
context, was presented as the starting point for the
main tracker; conventional drift chamber designs with
decades of successful operation received less initial
attention. Similarly, $\mu$-RWELL was presented as the
muon system technology, possibly influenced by its
prominence in recent CMS upgrade literature, despite
being dramatically overqualified for FCC-ee's benign
environment. In both cases, less recently published but
equally or more suitable alternatives required the
physicist's intervention to surface.

This bias likely reflects the structure of the training
data: recent papers and conference talks dominate, and
technologies that generate active publication streams
receive disproportionate weight in the AI's
``prior.'' Technologies that are mature, well-understood,
and no longer subjects of active R\&D publication may be
underweighted despite being excellent solutions.

\section{Limitations of the Approach}
\label{sec:limitations}

Several important limitations should be acknowledged:

\subsection{Knowledge Cutoff}

The AI's training data has a cutoff date, and it does not have
access to the most recent results from ongoing R\&D programs,
conference presentations, or preprints. During the discussion, the
physicist mentioned recent CEPC workshop results that the AI could
not access or verify. For a field that evolves rapidly, this is a
significant limitation. The AI was transparent about this limitation
when it arose, which is the appropriate response --- fabricating
knowledge of recent results would be far worse than acknowledging
ignorance.

It should be noted that this limitation is partly a consequence of
the interaction mode used in this study. Tool-augmented
configurations, in which the AI can search the web or query
databases during the conversation, would mitigate the knowledge
cutoff problem by allowing the AI to access current information
on demand. Such capabilities are available in some AI platforms
and are developing rapidly. A future study of this kind would
likely benefit from such tools, reducing the burden on the human
contributor to supply current information and enabling more
thorough literature coverage.

\subsection{Approximate Quantitative Estimates}

The calculations throughout this report are analytical estimates
and scaling arguments, not results of detailed Monte Carlo
simulation or finite element analysis. They are useful for
establishing orders of magnitude and identifying dominant effects,
but they cannot substitute for proper simulation. Several
conclusions --- particularly the jet energy resolution estimates
with the crystal ECAL and the cluster counting performance
assessment --- require validation through full simulation before
being used to guide actual detector construction decisions.

\subsection{Technology Enthusiasm Bias}

As noted in Section~\ref{sec:human_guidance}, and
related to the recency bias discussed in
Section~\ref{sec:ai_reasoning}, the AI showed a
systematic tendency toward selecting the most
technically impressive available technology. This likely
reflects the training data: papers and conference talks
naturally emphasize novel, high-performing technologies,
creating a bias in the AI's ``prior'' toward such
technologies. The correction --- matching technology to
requirements rather than selecting the most advanced
option --- required human intervention in several
instances.

It should be noted that this is not a uniquely AI
limitation. Human physicists are susceptible to
analogous biases: attraction to ``shiny new things''
on one hand, or over-reliance on familiar tools on the
other. The difference is in the mechanism --- the AI's
bias is shaped by literature frequency, while a human's
may be shaped by personal experience, institutional
culture, or involvement in specific R\&D programs ---
but the resulting design errors can be similar.

\subsection{Limited Access to Unpublished Knowledge}

Much of the practical knowledge that experienced
detector physicists possess is never published: lessons
learned from commissioning, failure modes discovered in
operation, calibration difficulties, and engineering
compromises that proved consequential. The AI has access
only to what appears in the published literature (and
whatever fraction of informal knowledge appears in talks
and proceedings). This creates a systematic gap that
cannot be closed by better training data alone --- some
knowledge exists only in the experience of
practitioners.

Again, this limitation is not unique to AI. Newcomers
to the field face a similar challenge: armed with
published scaling laws and performance projections, they
may push a design argument to its logical conclusion
without recognizing that at some point a different
consideration --- mechanical, thermal, operational ---
supersedes the original scaling. The knowledge of where
these boundaries lie is precisely what distinguishes
experienced practitioners, and it is acquired through
years of hands-on work rather than literature study.
The AI, in this sense, resembles an unusually
well-read but inexperienced newcomer.

\subsection{No Independent Validation}

The AI cannot independently verify its own reasoning or catch its
own errors in the way that a second human expert might. The examples
of the physicist correcting the AI's mistakes --- the factor-of-two
in cluster counting, the circular logic in luminometer monitoring,
the technology mismatches --- suggest that unchecked AI analysis
would contain significant errors. The value of the AI in this study
was realized only in combination with expert human review.

\section{Assessment}
\label{sec:assessment}

The human-AI collaboration proved productive for detector design
exploration, with each contributor bringing distinct and
complementary capabilities:
\begin{itemize}[itemsep=2pt]
\item The AI provided breadth: rapid survey of options, quantitative
  estimates, systematic organization, and the ability to follow
  chains of reasoning across many technical domains.
\item The human provided depth: practical experience, physical
  intuition, system-level thinking, and the critical judgment to
  recognize when the AI's reasoning was flawed or its technology
  choice was inappropriate.
\end{itemize}

The collaboration was most productive when operating in a
challenge-and-response mode: the AI presented an analysis, the
human identified weaknesses or asked penetrating questions, and the
AI revised in response. This iterative process converged toward
conclusions that neither contributor would likely have reached
independently --- the AI would not have questioned its own initial
assumptions, and the human would have spent considerably more time
assembling the quantitative framework for each comparison.

The subsequent editing of this report --- itself a
human-AI collaboration --- revealed additional issues
not caught during the original dialogue: inconsistent
numbers, unsupported quantitative claims, and framing
that did not survive closer scrutiny. This reinforces
the point that AI-generated analysis benefits from
multiple rounds of review, not just real-time correction
during the conversation.

The approach has clear limitations: it cannot replace detailed
simulation, engineering design, or the collective expertise of a
large collaboration. But as a tool for structured exploration of
design space, rapid evaluation of trade-offs, and documentation of
the reasoning behind design choices, it has demonstrable value.

Whether the specific detector concepts developed here (CL2a and
CL2b) prove to be close to the eventual FCC-ee detector design is
less important than whether the \emph{process} of systematic
examination, challenge, and revision leads to well-reasoned and
well-documented design choices. On that criterion, we believe this
exploration has been a useful exercise.

% ======================================================================
\chapter{Summary}
\label{ch:summary}
% ======================================================================

This report has presented an exploration of detector design for the
FCC-ee $e^+e^-$ collider, conducted through an extended dialogue
between a physicist and an AI assistant. Starting from initial
detector concepts (CL1a and CL1b), each subsystem was examined in
detail, leading to revised concepts (CL2a and CL2b) that differ
from the starting point in several significant ways.

The principal conclusions are:

\begin{enumerate}[itemsep=4pt]

\item \textbf{Vertex and inner silicon tracker:} A unified MAPS
  subsystem in 65~nm CMOS, using curved self-supporting shells with
  air cooling, extending continuously from the beam pipe
  ($R \approx 12$~mm) to the main tracker entrance
  ($R \approx 300$~mm). Total material budget $<$0.2\% $X_0$ for
  six layers. The separation between ``vertex detector'' and ``inner
  tracker'' found in LHC-era designs has no physical basis at
  FCC-ee.

\item \textbf{Main tracker:} A drift chamber with He-based gas is
  retained for CL2a, primarily for its ultra-low material budget.
  Cluster counting ($dN/dx$) performance is likely more modest than
  projected, with practical effects (delta ray contamination,
  diffusion, electronics bandwidth) degrading the resolution
  significantly from the theoretical $1/\sqrt{N}$ floor. A TPC
  is recommended for the complementary detector CL2b.

\item \textbf{Electromagnetic calorimeter:} PbWO$_4$ crystals
  replace the initial noble liquid choice. 
  The decisive factor is the cryostat dead material
($\sim$10--15\%~$X_0$), which degrades the noble
liquid calorimeter performance through photon
conversions, non-Gaussian tails, and constant term
contributions, while crystals achieve
$\sim$3\%/$\sqrt{E}$ with negligible dead material.
The absence of radiation damage at FCC-ee and the
abundant clean calibration samples across all FCC-ee
running points remove the historical weaknesses of
crystal calorimetry.
  Dual-readout (\v{C}erenkov
  filtering) in the ECAL was considered but not adopted, as it
  provides no benefit for electromagnetic showers while degrading
  the scintillation light collection.

\item \textbf{Solenoid:} Variable field (1.5--3~T, matched to beam
  energy) with inside-HCAL placement. The variable field naturally
  aligns physics needs with accelerator constraints. The inside
  placement reduces cost by $\sim$4$\times$ and enables the HCAL
  steel to serve as magnetic flux return.

\item \textbf{Hadronic calorimeter:} Scintillator-steel tiles for
  CL2a, with dual-readout fibers strongly recommended for CL2b. 
  The separate deep ECAL simplifies the HCAL's pattern
recognition task, and the particle flow confusion term
dominates jet resolution over the HCAL intrinsic
resolution.
  The HCAL technology
  choice is identified as the strongest case for two-detector
  complementarity.

\item \textbf{Muon system:} Scintillator bars with WLS fiber and
  dual-end SiPM readout replace $\mu$-RWELL. This matches the
  technology to the actual (modest) requirements of muon
  identification at FCC-ee, with decisive advantages in operational
  simplicity, reliability, and cost.

\item \textbf{Luminosity monitor:} Si-W calorimeter (unchanged),
  with the understanding that the dominant challenges are metrological
  (sub-$\mu$m position knowledge and stability) and operational
  (interleaved energy scans for $10^{-5}$ relative precision) rather
  than detector-physical.

\end{enumerate}

Several cross-cutting themes emerged from the study:

\begin{itemize}[itemsep=4pt]
\item \textbf{Practical considerations matter.} The cryostat dead
  material, the solenoid cost scaling, 
  the SiPM dynamic range challenge at FCC-ee energies, 
  and the operational complexity of gaseous
  muon detectors all proved decisive for technology choices. Abstract
  performance comparisons that neglect these practicalities can lead
  to suboptimal designs.

\item \textbf{Subsystem choices are coupled.} The ECAL technology
  choice influences the HCAL strategy; the solenoid placement
  affects cost, flux return, and the muon system; the tracker gas
  choice affects vertex performance. Optimizing subsystems
  independently can yield a globally suboptimal detector.

\item \textbf{FCC-ee's benign environment is an opportunity.} The
  absence of radiation damage enables technologies (crystals, MAPS
  at 20~$\mu$m thickness, air cooling, scintillator-based muon
  system) that are impractical at hadron colliders. Designs inherited
  from LHC-era thinking do not necessarily transfer to FCC-ee.

\item \textbf{The magnetic field is a key variable.}
  The variable field strategy (adapting field strength to beam
  energy) optimizes PFA performance, momentum
  resolution, and accelerator compatibility
  simultaneously. The field strength is a significant
  driver of jet energy resolution through its effect on
  the confusion term.

  \item \textbf{Two-detector complementarity is valuable.}
  The most difficult technology choices (drift chamber
  vs.\ TPC, crystal vs.\ Si-W ECAL, tiles vs.\
  dual-readout HCAL) may not have clear winners --- they
  involve genuine trade-offs between different
  performance characteristics. Having both approaches
  deployed simultaneously provides independent systematic
  checks, complementary performance, and risk mitigation.
  \end{itemize}

The human-AI collaboration proved productive for this exploration.
The AI contributed rapid technology surveys, quantitative estimates,
and systematic organization. The human contributed practical
experience, system-level thinking, and the critical questions that
redirected the analysis at key points. The most consequential design
changes were triggered by human questions rather than AI analysis,
but the AI's ability to rapidly work through the implications of
each question made the iterative exploration efficient.

The detector concepts presented here --- CL2a and CL2b --- are
not proposed as final designs. They are the output of a structured
design exploration that has identified the key trade-offs, quantified
the dominant effects, and documented the reasoning behind each
choice. Several conclusions require validation through detailed
simulation, and a number of R\&D items have been identified. The
value of this exercise lies in the systematic process of examination
and revision, which we hope contributes usefully to the ongoing
discussion within the FCC-ee detector community.

% ======================================================================
\chapter*{Declaration of Generative AI in the Research Process}
\addcontentsline{toc}{chapter}{Declaration of Generative AI
  in the Research Process}
\label{ch:declaration}
% ======================================================================

This work was developed through an extended, structured dialogue
between the author and Claude (Anthropic, \texttt{claude-opus-4},
accessed via the Anthropic web interface), a large language model
AI assistant. The nature and extent of the AI's contribution goes
substantially beyond manuscript preparation and is integral to
the research content:

\begin{enumerate}[itemsep=4pt]
\item \textbf{Conceptual development.} The AI generated the initial
  detector concepts (CL1a and CL1b), including the rationale for
  each subsystem technology choice and the criteria for integrated
  detector design. These served as the starting point for the
  collaborative exploration.

\item \textbf{Technical analysis.} The AI performed quantitative
  estimates throughout the study, including material budget
  calculations, force balance analyses, ionization statistics,
  energy resolution projections, cost scaling arguments, and
  magnetic flux calculations. These estimates are analytical
  approximations, not results of dedicated simulation.

\item \textbf{Iterative design evolution.} Through iterative
  dialogue, the author posed questions and challenges based on
  practical experience and physics insight. The AI revised its
  analyses and design choices in response, leading to the
  substantially revised concepts (CL2a and CL2b). The key
  questions and insights that drove the major design changes
  originated with the human author, as documented in
  Chapter~\ref{ch:reflections}.

\item \textbf{Manuscript drafting.} The AI drafted the majority of
  the manuscript text, which the author reviewed, edited, and in
  some cases substantially revised. 

\item \textbf{Literature survey.} The AI provided surveys of
  available technologies and their published performance
  parameters. These surveys reflect the AI's training data,
  which has a knowledge cutoff, and may not include the most
  recent results. The author verified key references and added
  references known to be missing.
\end{enumerate}

The AI's contributions were subject to the following limitations,
discussed in detail in Chapter~\ref{ch:reflections}:
\begin{itemize}[itemsep=2pt]
\item Quantitative estimates are approximate, based on analytical
  scaling arguments rather than detailed simulation.
\item The AI exhibited a systematic tendency toward selecting
  technically impressive technologies over simpler alternatives
  better matched to the actual requirements; this was corrected
  through human review.
\item The AI's knowledge has a training cutoff and does not include
  the most recent conference results or unpublished developments.
\item Several errors in the AI's reasoning were identified and
  corrected by the author during the dialogue (e.g., a factor-of-two
  error in cluster counting time-domain density, circular logic in
  luminometer acceptance monitoring).
\end{itemize}

All AI-assisted analysis and conclusions were critically reviewed
by the author. The author takes full responsibility for the
scientific content, conclusions, and any remaining errors in this
publication.

\bibliographystyle{unsrtnat}
\bibliography{references}

\end{document}